\documentclass{LMCS}

\usepackage{epic, eepic, epsf, epsfig, pst-node,pstricks,pst-grad}
\usepackage{gastex,color,empheq}
\usepackage{amsmath,amssymb,latexsym}
\usepackage{xspace}
\usepackage{dsfont}
\usepackage{cite}
\usepackage[mathscr]{euscript}
\usepackage{enumerate,hyperref}

\def\doi{4 (4:16) 2008}
\lmcsheading%
{\doi}
{1--35}
{}
{}
{Sep.~\phantom{0}3, 2007}
{Dec.~24, 2008}
{}   
\begin{document}


\newcommand{\qedclaim}{\qed}

\newcommand{\redc}[1]{#1}
\newcommand{\greenc}[1]{#1}
\newcommand{\whitec}[1]{\textcolor{white}{#1}}

\newcommand{\oa}{{\overline{a}}}
\newcommand{\ob}{{\overline{b}}}

\renewcommand{\uplus}{\mathrel{\mathaccent\cdot\cup}}

\newcommand{\rightone}{\mathrel{\curvearrowright}}
\newcommand{\leftone}{\mathrel{\curvearrowleft}}
\newcommand{\righttwo}{\mathrel{\rotatebox[origin=cc]{180}{$\curvearrowleft$}}}
\newcommand{\lefttwo}{\mathrel{\rotatebox[origin=cc]{180}{$\curvearrowright$}}}
\newcommand{\rightthree}{\mathrel{\stackrel{3}{\curvearrowright}}}
\newcommand{\leftthree}{\mathrel{\stackrel{3}{\curvearrowleft}}}

\newcommand{\rightright}{\mathrel{\leadsto}}
\newcommand{\rightleft}{\mathrel{\rotatebox[origin=cc]{90}{\,$\circlearrowright$\,}}}
\newcommand{\leftleft}{\mathrel{\reflectbox{$\leadsto$}}}
\newcommand{\leftright}{\mathrel{\rotatebox[origin=cc]{-90}{\,$\circlearrowleft$\,}}}

\newcommand{\rightk}{\mathrel{\stackrel{\stack}{\curvearrowright}}}
\newcommand{\leftk}{\mathrel{\stackrel{\stack}{\curvearrowleft}}}

\newcommand{\lra}{\longrightarrow}
\newcommand{\crSigma}{\widetilde{\Sigma}}
\newcommand{\ncrSigma}{\widehat{\Sigma}}
\newcommand{\cSigma}{\Sigma_c}
\newcommand{\rSigma}{\Sigma_r}
\newcommand{\intSigma}{\Sigma_{\mathit{int}}}

\newcommand{\signNW}{\tau_{\crSigma}}
\newcommand{\signG}{\tau_{\mathit{Grids}}}

\newcommand{\Init}{Q_I}

\newcommand{\hinv}{h'}

\newcommand{\NW}{\mathbb{NW}}
\newcommand{\infNW}{\mathbb{NW}^\omega}
\newcommand{\Grids}{\mathbb{G}}

\newcommand{\LSigma}{{\bf{\Sigma}}}

\newcommand{\Contents}{\mathit{Cont}}

\newcommand{\dom}{\mathrm{dom}}
\newcommand{\codom}{\mathrm{co}\textup{-}\mathrm{dom}}

\newcommand{\dist}{d}
\newcommand{\radius}{r}

\newcommand{\succord}{\mathord{\lessdot}}
\newcommand{\succrel}{\mathrel{\lessdot}}
\renewcommand{\succ}[2]{#1 \mathrel{\lessdot} #2}
\renewcommand{\epsilon}{\varepsilon}

\renewcommand{\mod}{~\mathrm{mod}~}

\newcommand{\modit}{\mathit{mod}}

\newcommand{\Sph}[3]{#1\text{-}\mathrm{Sph}(#2,#3)}

\newcommand{\simulate}[4]{#3 \mathrel{\equiv^{#1}_{#2}} #4}
\newcommand{\wsimulate}[4]{#3 \mathrel{\sqsubseteq^{#1}_{#2}} #4}
\newcommand{\notwsimulate}[4]{#3 \mathrel{\not\sqsubseteq^{#1}_{#2}} #4}
\newcommand{\edge}[1]{\mathrel{\leftrightarrow}_{#1}}
\newcommand{\redge}[1]{\mathrel{\rightarrow}_{#1}}

\newcommand{\Nat}{N}
\newcommand{\infDom}{\N_{+}}
\newcommand{\posN}{\N_{+}}

\newcommand{\sphere}{S}
\newcommand{\esphere}{E}

\newcommand{\call}{\mu^{-1}}
\newcommand{\ret}{\mu}

\newcommand{\cdelta}{\delta_c}
\newcommand{\rdelta}{\delta_r}
\newcommand{\intdelta}{\delta_{\mathit{int}}}
\newcommand{\crdelta}{\langle\cdelta,\rdelta,\intdelta\rangle}

\newcommand{\stack}{s}
\newcommand{\nstack}{K}

\newcommand{\Lang}{\mathcal{L}}

\renewcommand{\phi}{\varphi}

\newcommand{\nword}{W}
\newcommand{\str}{\mathrm{string}}
\newcommand{\nested}{\mathrm{nested}}

\newcommand{\Aut}{\mathcal{A}}
\newcommand{\PA}{\mathcal{A}}
\newcommand{\WA}{\mathcal{B}}

\newcommand{\MVPA}{\textsc{Mvpa}\xspace}
\newcommand{\MNWA}{\textsc{Mnwa}\xspace}
\newcommand{\GMNWA}{\textsc{GMnwa}\xspace}

\newcommand{\MSO}{\textup{MSO}}
\newcommand{\FO}{\textup{FO}}
\newcommand{\EMSO}{\textup{EMSO}}

\newcommand{\tVPA}{2\textsc{vpa}\xspace}
\newcommand{\tNWA}{2\textsc{nwa}\xspace}

\newcommand{\isom}{\cong}

\newcommand{\N}{\mathds{N}}

\newcommand{\gdown}{\mathord{\mathrm{succ}_1}}
\newcommand{\gright}{\mathord{\mathrm{succ}_2}}
\newcommand{\succone}{\mathord{\mathrm{succ}_1}}
\newcommand{\succtwo}{\mathord{\mathrm{succ}_2}}
\newcommand{\succk}{\mathord{\mathrm{succ}_k}}

\newcommand{\foequiv}[1]{\mathrel{\equiv}_{#1}}
\newcommand{\inffoequiv}[1]{\mathrel{\equiv}_{#1}^\infty}
\newcommand{\threquiv}[2]{\mathrel{\leftrightarrows}_{#1,#2}}
\newcommand{\infthrequiv}[2]{\mathrel{\leftrightarrows}_{#1,#2}^\infty}

\newtheorem{myclaim}[thm]{Claim}{\bfseries}{\rmfamily}

\renewcommand{\arraystretch}{1.4}

\newcommand{\state}{\sphere}
\newcommand{\cstate}{\sphere_c}

\newcommand{\Buchi}{B{\"u}chi\xspace}

\newcommand{\AllSpheres}{\mathit{Spheres}}
\newcommand{\Spheres}{\mathit{Spheres}}

\newcommand{\AlleSpheres}{\mathit{eSpheres}}

\newcommand{\core}{\mathit{core}}
\newcommand{\espheremap}{\mathit{esphere}}
\newcommand{\labeling}{\mathit{label}}
\newcommand{\inst}{\mathit{col}}
\newcommand{\col}{\mathit{col}}
\newcommand{\const}{\#\mathit{Col}}

\newcommand{\State}{\mathcal{E}}
\newcommand{\cState}{\mathcal{E}_c}

\newcommand{\tW}{{\widetilde{W}}}
\newcommand{\tn}{\widetilde{n}}
\newcommand{\tsuccord}{\widetilde{\succord}}
\newcommand{\tmu}{\widetilde{\mu}}
\newcommand{\tlambda}{\widetilde{\lambda}}
\newcommand{\trho}{\widetilde{\rho}}

\newcommand{\maxN}{\mathit{maxSize}(r)}

\newcommand{\Dir}{\Delta}

\newcommand{\Wleadsto}[4]{#3 \mathrel{{\xhookrightarrow{#2~}}_{#1}} #4}
\newcommand{\Wleadstoeq}[4]{#3 \mathrel{{\xRightarrow{~#2~}}_{#1}} #4}
\newcommand{\longWleadstoeq}[4]{#3 \mathrel{{\xRightarrow{#2~}}_{#1}} #4}

\newcommand{\eWleadsto}[5]{#3 \mathrel{{\xhookrightarrow[{#5}]{#2}}}
#4}

\newcommand{\Paths}{\mathit{Paths}}

\title[On the Expressive Power of 2-Stack Visibly Pushdown Automata]{On
  the Expressive Power of\\2-Stack Visibly Pushdown Automata}

\author[B.~Bollig]{Benedikt Bollig}
 \address{LSV, ENS Cachan, CNRS ---
   61, avenue du Pr{\'e}sident Wilson,
   94235 Cachan Cedex, France}
\email{bollig@lsv.ens-cachan.fr}



\begin{abstract}
  Visibly pushdown automata are input-driven pushdown automata that recognize
  some non-regular context-free languages while preserving the nice closure
  and decidability properties of finite automata. Visibly pushdown automata
  with multiple stacks have been considered recently by La Torre, Madhusudan,
  and Parlato, who exploit the concept of visibility further to obtain a rich
  automata class that can even express properties beyond the class of
  context-free languages. At the same time, their automata are closed under
  boolean operations, have a decidable emptiness and inclusion problem, and
  enjoy a logical characterization in terms of a monadic second-order logic
  over words with an additional nesting structure. These results require a
  restricted version of visibly pushdown automata with multiple stacks whose
  behavior can be split up into a fixed number of phases.

  In this paper, we consider 2-stack visibly pushdown automata (i.e., visibly
  pushdown automata with two stacks) in their unrestricted form. We show that
  they are expressively equivalent to the existential fragment of monadic
  second-order logic. Furthermore, it turns out that monadic second-order
  quantifier alternation forms an infinite hierarchy wrt.\ words with multiple
  nestings. Combining these results, we conclude that 2-stack visibly pushdown
  automata are not closed under complementation.

  Finally, we discuss the expressive power of \Buchi 2-stack visibly pushdown
  automata running on infinite (nested) words. Extending the logic by an
  infinity quantifier, we can likewise establish equivalence to existential
  monadic second-order logic.
\end{abstract}

\maketitle



\section{Introduction}\label{sec:intro}

The notion of a regular word language has ever played an important r\^ole in
computer science, as it constitutes a robust concept that enjoys manifold
representations in terms of finite automata, regular expressions, monadic
second-order logic, etc. Generalizing regular languages towards richer classes
and more expressive formalisms is often accompanied by the loss of robustness
and decidability properties. It is, for example, well-known that the class of
context-free languages, represented by pushdown automata, is not closed under
complementation and that universality, equivalence, and inclusion are
undecidable problems \cite{Hopcroft2000}.

\emph{Visibly pushdown languages} have been introduced by Alur and Madhusudan
to overcome this deficiency while subsuming many interesting and useful
context-free properties \cite{AM2004}. Visibly pushdown languages are
represented by special pushdown automata whose stack operations are driven by
the input. More precisely, the underlying alphabet of possible actions is
partitioned into (1) call, (2) return, and (3) internal actions, which, when
reading an action, indicates if (1) a stack symbol is pushed on the stack, (2)
a stack symbol is read and popped from the stack, or (3) the stack is not
touched at all, respectively. Such a partition gives rise to a
\emph{call-return alphabet}. Though this limits the expressive power of
pushdown automata, the such defined class of visibly pushdown languages is
rich enough to model various interesting non-regular properties for program
analysis. Even more, this class preserves some important closure properties of
regular languages, such as the closure under boolean operations, and it
exhibits decidable problems, such as inclusion, that are undecidable in the
context of general pushdown automata. Last but not least, the visibly pushdown
languages are captured by a monadic second-order logic that makes use of a
binary nesting predicate. Such a logic is suitable in the context of
visibility, as the nesting structure of a word is uniquely determined,
regardless of a particular run of the pushdown automaton. The logical
characterization smoothly extends the classical theory of regular languages
\cite{Buechi:60,Elgot1961}. For context-free languages, quantification over
\emph{matchings}, which are not implicitly given when we do not have
visibility, is necessary to obtain a logical characterization
\cite{Schwentick94}.

Visibly pushdown automata with multiple stacks have been considered recently
and independently by La Torre, Madhusudan, and Parlato \cite{Madhusudan2007},
as well as Carotenuto, Murano, and Peron \cite{Murano2007}. The aim of these
papers is to exploit the concept of visibility further to obtain even richer
classes of non-regular languages while preserving important closure properties
and decidability of verification-related problems such as emptiness and
inclusion.

In \cite{Madhusudan2007}, the authors consider visibly pushdown automata with
arbitrarily many stacks. To retain the nice properties of visibly pushdown
automata with only one stack, the idea is to restrict the domain, i.e., the
possible inputs, to those words that can be divided into at most $k$ phases
for a predefined $k$. In every phase, pop actions correspond to one and the
same stack. These restricted visibly pushdown automata have a decidable
emptiness problem, which is shown by a reduction to the emptiness problem for
finite tree automata, and are closed under union, intersection, and
complementation (wrt.\ the domain of $k$-phase words). Moreover, a word
language is recognizable if, and only if, it can be defined in monadic
second-order logic where the usual logic over words is expanded by a matching
predicate that matches a push with its corresponding pop event. As mentioned
above, such a matching is unique wrt.\ the underlying call-return alphabet.
The only negative result in this regard is that multi-stack visibly pushdown
automata cannot be determinized.

The paper \cite{Murano2007} considers visibly pushdown automata with two
stacks and call-return alphabets that appear more general than those of
\cite{Madhusudan2007}: Any stack is associated with a partition of one and the
same alphabet into call, return, and local transitions so that an action might
be both a call action for the first stack and, at the same time, a return
action for the second. In this way, both stacks can be worked on
simultaneously. Note that, if we restrict to the alphabets of
\cite{Madhusudan2007} where the stack alphabets are disjoint, the models from
\cite{Murano2007} and \cite{Madhusudan2007} coincide. Carotenuto et al.\ show
that the emptiness problem of their model is undecidable. Their approach to
gain decidability is to exclude simultaneous pop operations by introducing an
ordering constraint on stacks, which is inspired by \cite{multi96} (see also
\cite{ABH-dlt08}). More precisely, a pop operation on the second stack is only
possible if the first stack is empty. Under these restrictions, the emptiness
problem turns out to be decidable in polynomial time (note that the number of
stacks is fixed).\footnote{In \cite{Murano2007}, the authors argue that
  2-stack visibly pushdown automata without restriction are closed under
  complementation, but their proof makes use of the incorrect assumption that
  these automata are determinizable. In fact, \mbox{2-stack} visibly pushdown
  automata can in general not be determinized \cite{Madhusudan2007}. In the
  present paper, we show that 2-stack visibly pushdown automata are actually
  not closed under complementation.}

In this paper, we consider 2-stack visibly pushdown automata (i.e., visibly
pushdown automata with two stacks) where each action is exclusive to one of
the stacks, unless we deal with an internal action, which does not affect the
stacks at all. Thus, we adopt the model of \cite{Madhusudan2007}, though we
have to restrict to two stacks for our main results. One of these results
states that the corresponding language class is precisely characterized by the
existential fragment of monadic second-order logic where a first-order kernel
is preceded by a block of existentially quantified second-order variables. In
a second step, we show that the full monadic second-order logic is strictly
more expressive than its existential fragment so that we conclude that 2-stack
visibly pushdown automata are not closed under complementation. Note that our
model has an undecidable emptiness problem, as can be easily seen.

The key technique in our proofs is to consider words over call-return
alphabets as relational structures, called \emph{nested words} \cite{AlurM06}.
Nested words augment ordinary words with a nesting relation that, as the
logical atomic predicate mentioned above, relates push with corresponding pop
events. More precisely, we consider a nested word to be a graph whose nodes
are labeled with actions and are related in terms of a matching and an
immediate-predecessor relation. We thus deal with structures of bounded
degree: every node has at most two incoming edges (one from the immediate
predecessor and one from a push event if we deal with a pop event operating on
the non-empty stack) and, similarly, at most two outgoing edges. As there is a
one-to-one correspondence between words and their nested counterpart, we may
consider nested-word automata \cite{AlurM06}, which are equivalent to visibly
pushdown automata but operate on the enriched word structures. There have been
several notions of automata on graphs and partial orders
\cite{Tho-automata-ttp,ThoPOMIV96} that are similar to nested-word automata
and have one idea in common: the state that is taken after executing some
event depends on the states that have been visited in neighboring events. Such
defined automata may likewise operate on models for concurrent-systems
executions such as Mazurkiewicz traces \cite{Droste2000} and message sequence
charts \cite{BolligJournal}. In the framework of nested-word automata, to
determine the state after executing a pop operation, we therefore have to
consider both the state of the immediate-predecessor position and the state
that had been reached after the execution of the corresponding push event. To
obtain a logical characterization of nested-word automata over two stacks, we
adopt a technique from \cite{BolligJournal}: for a natural number $r$, we
compute a nested-word automaton $\WA_\radius$ that computes the \emph{sphere}
of radius $\radius$ around any event $i$, i.e., the restriction of the input
word to those events that have distance at most $\radius$ from $i$. Once we
have this automaton, we can apply Hanf's Theorem, which states that
satisfaction of a given first-order formula depends on the number of these
local spheres counted up to a threshold that depends on the quantifier-nesting
depth of the formula \cite{Hanf1965}. This finally leads us to a logical
characterization of 2-stack visibly pushdown automata in terms of existential
monadic second-order logic. Note that our construction of $\WA_\radius$ is
close to the nontrivial technique applied in \cite{BolligJournal}. In the
context of nested words, however, the correctness proof is more complicated.
The fact that we deal with two stacks only is crucial, and the construction
fails as soon as a third stack comes into play.

Then, we exploit the concept of nested words to show that full monadic
second-order logic is more expressive than its existential fragment. This is
done by a first-order interpretation of nested words over two stacks into
grids, for which the analogous result has been known \cite{MST02}.

An extension of Hanf's Theorem has been established to cope with infinite
structures \cite{LSV:06:11}. This allows us to apply the automaton
$\WA_\radius$ to also obtain a logical characterization of the canonical
extension of 2-stack visibly pushdown automata towards \Buchi automata running
on infinite words.

\paragraph{Outline of the paper} In Section~\ref{sec:MVPA}, we introduce
multi-stack visibly pushdown automata, running on words, as well as
multi-stack nested-word automata, which operate on nested words. We establish
expressive equivalence of these two models. Section~\ref{sec:MSO} recalls
monadic second-order logic over relational structures and, in particular,
nested words. There, we also state Hanf's Theorem, which provides a normal
form of first-order definable properties in terms of spheres. The construction
of the sphere automaton $\WA_\radius$, which is, to some extent, the core
contribution of this paper, is the subject of Section~\ref{sec:sphereaut}. By
means of this automaton, we can show expressive equivalence of 2-stack visibly
pushdown automata and existential monadic second-order logic
(Section~\ref{subsect:maintheorem}). Section~\ref{sec:grids} establishes the
gap between this fragment and the full logic, from which we conclude that
2-stack visibly pushdown automata cannot be complemented in general. By
slightly modifying our logic, we obtain, in Section~\ref{sec:buechi}, a
characterization of \Buchi 2-stack visibly pushdown automata, running on
infinite words. We conclude with Section~\ref{sec:openproblems} stating some
related open problems.


\section{Multi-Stack Visibly Pushdown Automata}\label{sec:MVPA}

The set $\{0,1,2,\ldots\}$ of natural numbers is denoted by $\N$, the set
$\{1,2,\ldots\}$ of positive natural numbers by $\infDom$. We call any finite
set an \emph{alphabet}. For a set $\Sigma$, we denote by $\Sigma^\ast$,
$\Sigma^+$, and $\Sigma^\omega$ the sets of finite, nonempty finite, and
infinite strings over $\Sigma$, respectively.\footnote{From now on, to avoid
  confusion with nested words, we use the term ``string'' rather than ``word''
  if we deal with elements from $\Sigma^\ast \cup \Sigma^\omega$.} The empty
string is denoted by $\varepsilon$. For a natural number $n \in \N$, we let
$[n]$ stand for the set $\{1,\ldots,n\}$ (i.e., $[0]$ is the empty set). In
this paper, we will identify isomorphic structures and we use $\isom$ to
denote isomorphism.

Let $\nstack \ge 1$ be a positive natural number. A ($\nstack$-stack)
\emph{call-return alphabet} is a collection
$\langle\{(\cSigma^\stack,\rSigma^\stack)\}_{\stack \in
  [\nstack]},\intSigma\rangle$ of pairwise disjoint alphabets.
Intuitively, $\cSigma^\stack$ contains the actions that call the stack
$\stack$, $\rSigma^\stack$ is the set of returns of stack $\stack$, and
$\intSigma$ is a set of internal actions, which do not involve any stack
operation.

We fix $\nstack \ge 1$ and a $\nstack$-stack call-return alphabet
$\crSigma=\langle\{(\cSigma^\stack,\rSigma^\stack)\}_{\stack \in
  [\nstack]},\intSigma\rangle$. Moreover, we set $\cSigma = \bigcup_{\stack
  \in [\nstack]} \cSigma^\stack$, $\rSigma = \bigcup_{\stack \in [\nstack]}
\rSigma^\stack$, and $\Sigma = \cSigma \cup \rSigma \cup \intSigma$.

\subsection{Multi-Stack Visibly Pushdown Automata}

\begin{defi}
  A \emph{multi-stack visibly pushdown automaton} (\MVPA) over $\crSigma$ is a
  tuple $\PA=(Q,\Gamma,\delta,\Init,F)$ where
\begin{enumerate}[$\bullet$]
\item $Q$ is its finite set of \emph{states},
\item $\Init \subseteq Q$ is the set of \emph{initial states},
\item $F \subseteq Q$ is the set of \emph{final states},
\item $\Gamma$ is the finite \emph{stack alphabet} containing a special symbol
  $\bot$ that will represent the empty stack, and
\item $\delta$ provides the \emph{transitions} in terms of a triple $\crdelta$
  with
\[\begin{array}{rcl}
  \cdelta & \!\!\mathrel{\subseteq}\!\! & Q \times \cSigma \times (\Gamma \setminus
  \{\bot\}) \times Q,\\ \rdelta & \!\!\mathrel{\subseteq}\!\! & Q \times \rSigma \times
  \Gamma \times Q,~\text{and}\\ \intdelta & \!\!\mathrel{\subseteq}\!\! & Q \times
  \intSigma \times Q~.\end{array}\]
\end{enumerate}

A \emph{2-stack visibly pushdown automaton} ($\tVPA$) is an \MVPA that is
defined over a 2-stack alphabet (i.e., $K=2$).
\end{defi}

A transition $(q,a,A,q') \in \cdelta$, say with $a \in \cSigma^\stack$, is a
push transition meaning that, being in state $q$, the automaton can read $a$,
push the symbol $A \in \Gamma \setminus \{\bot\}$ onto the $\stack$-th stack,
and go over to state $q'$. A transition $(q,a,A,q') \in \rdelta$, say with $a
\in \rSigma^\stack$, allows us to pop $A \neq \bot$ from the $\stack$-th stack
when reading $a$, while the control changes from state $q$ to state $q'$. If,
however, $A = \bot$, then the stack is not touched, i.e., $\bot$ is never
popped. Finally, a transition $(q,a,q') \in \intdelta$ is applied when reading
internal actions $a \in \intSigma$. They do not involve any stack operation
and, actually, do not even allow us to read from the stack.

Let us formalize the behavior of the \MVPA $\PA$. A \emph{stack contents} is a
nonempty finite sequence from $\Contents={(\Gamma \setminus \{\bot\})}^\ast
\cdot \{\bot\}$. The leftmost symbol is thus the top symbol of the stack
contents. A configuration of $\PA$ consists of a state and a stack contents
for every stack. Hence, it is an element of $Q \times \Contents^{[K]}$.
Consider a string $w = a_1 \ldots a_n \in \Sigma^+$. A \emph{run} of $\PA$ on
$w$ is a sequence $\rho = (q_0,\sigma_0^1, \ldots, \sigma_0^\nstack) \ldots
(q_n,\sigma_n^1, \ldots, \sigma_n^\nstack) \in (Q \times \Contents^{[K]})^+$
such that $q_0 \in \Init$, $\sigma_0^\stack = \bot$ for each stack $\stack \in
[\nstack]$, and, for all $i \in \{1,\ldots,n\}$, the following hold:
\begin{enumerate}[\hbox to6 pt{\hfill}]
\item\noindent{\hskip-11 pt\bf [Push]:}\ If $a_i \in \cSigma^\stack$ for $\stack \in [\nstack]$, then
  there is a stack symbol $A \in \Gamma \setminus \{\bot\}$ such that
  $(q_{i-1},a_i,A,q_i) \in \cdelta$, $\sigma_i^\stack = A \cdot
  \sigma_{i-1}^\stack$, and $\sigma_i^{\stack'} = \sigma_{i-1}^{\stack'}$ for
  every $\stack' \in [\nstack] \setminus \{\stack\}$.
\item\noindent{\hskip-11 pt\bf [Pop]:}\ If $a_i \in \rSigma^\stack$ for $\stack \in [\nstack]$, then
  there is a stack symbol $A \in \Gamma$ such that $(q_{i-1},a_i,A,q_i) \in
  \rdelta$, $\sigma_i^{\stack'} = \sigma_{i-1}^{\stack'}$ for every $\stack' \in
  [\nstack] \setminus \{\stack\}$, and either $A \neq \bot$ and
  $\sigma_{i-1}^\stack = A \cdot \sigma_i^\stack$, or $A = \bot$ and
  $\sigma_{i-1}^\stack = \sigma_i^\stack = \bot$.
\item\noindent{\hskip-11 pt\bf [Internal]:}\ If $a_i \in \intSigma$, then $(q_{i-1},a_i,q_i) \in
  \intdelta$, and $\sigma_i^\stack = \sigma_{i-1}^\stack$ for every $\stack
  \in [\nstack]$.
\end{enumerate}
The run $\rho$ is accepting if $q_n \in F$. A string $w \in \Sigma^+$ is
accepted by $\PA$ if there is an accepting run of $\PA$ on $w$. The set of
accepted strings forms the (string) language of $\PA$, which is a subset of
$\Sigma^+$ and denoted by $L(\PA)$.\footnote{To simplify the presentation, the
  empty word $\varepsilon$ is excluded from the domain.}

\begin{exa}\label{ex:mvpa}
  There is no \MVPA that recognizes the context-sensitive language
  $\{a^nb^nc^n \mid n \ge 1\}$, no matter which call-return alphabet we chose.
  Note that, however, with the more general notion of a call-return alphabet
  from \cite{Murano2007}, it is possible to recognize this language by means
  of two stacks. Now consider the 2-stack call-return alphabet $\crSigma$
  given by $\cSigma^1 = \{ a \}$, $\rSigma^1 = \{ \oa \}$, $\cSigma^2 =
  \{b\}$, $\rSigma^2 = \{ \ob \}$, and $\intSigma = \emptyset$. The language
  $L=\{( a b)^n \oa ^{n+1} \ob ^{n+1} \mid n \ge 1\}$ can be recognized by
  some \tVPA over $\crSigma$, even by the restricted model of 2-phase \tVPA
  from \cite{Madhusudan2007}, as every word from $L$ can be split into at most
  two return phases. In the following, we define a \tVPA
  $\PA=(\{q_0,\ldots,q_4\},\{\$,\bot\},\delta,\{q_0\},\{q_0\})$ over
  $\crSigma$ such that $L(\PA) = L^+$, which is no longer divisible into a
  bounded number of return phases. The transition relation $\delta$ is given
  as follows (a graphical illustration is provided in Figure~\ref{fig:mvpa}):
  \[\hspace{-2em}\begin{array}{rlcrl}
    \cdelta: & (q_0, a ,\$,q_2) & ~~~~~~ & \rdelta: & (q_3, \oa ,\$,q_3) \\
    & (q_2, b,\$,q_1) & & & (q_3, \oa ,\bot,q_4)\\
    & (q_1, a ,\$,q_2) & & & (q_4, \ob ,\$,q_4)\\
    & (q_2, b,\$,q_3) & & & (q_4, \ob ,\bot,q_0)
  \end{array}\]
  The idea is that the finite-state control ensures that an input word matches
  the regular expression $(( a b)^+ \oa ^+ \ob ^+)^+$. To guarantee that, in
  any iteration, the number of $a$ is by one less than the number of $\oa$,
  any push action $a$ stores a stack symbol $\$$ in stack $1$, which can then
  be removed by the corresponding pop action $\oa$ unless the symbol $\bot$ is
  discovered. We do the same for $b$ and $\ob$ on stack $2$.
\end{exa}

\begin{figure}[h]
  \centering
\scalebox{1}{
\begin{picture}(20,30)(0,-3)
\gasset{Nframe=y,Nw=5.5,Nh=5.5,Nmr=8,ilength=4} 

\node[Nmarks=ir](q0)(0,20){$q_0$}
\node(q1)(40,20){$q_1$}
\node(q2)(20,20){$q_2$}
\node(q3)(20,0){$q_3$}
\node(q4)(0,0){$q_4$}

\drawloop[loopCW=n,loopdiam=5,loopangle=0,ELside=r](q3){$\oa,\$$}
\drawloop[loopCW=y,loopdiam=5,loopangle=180,ELside=l](q4){$\ob,\$$}

\drawedge[ELside=l](q0,q2){$a,\$$}
\drawedge[ELside=l](q2,q3){$b,\$$}
\drawedge[ELside=l](q3,q4){$\oa,\bot$}
\drawedge[ELside=l](q4,q0){$\ob,\bot$}

\drawedge[curvedepth=2,ELside=l](q2,q1){$b,\$$}
\drawedge[curvedepth=2,ELside=l](q1,q2){$a,\$$}

\end{picture}
}
\caption{A \tVPA \label{fig:mvpa}}
\end{figure}


\subsection{Nested Words and Multi-Stack Nested-Word Automata}


We will now see how strings over symbols from the call-return alphabet
$\crSigma$ can be represented by relational structures. Basically, to a
string, we add a binary predicate that combines push with corresponding pop
events.
Let $\stack \in [\nstack]$. A string $w \in \Sigma^\ast$ is called
$\stack$-\emph{well formed} if it is generated by the context-free grammar
\begin{align*}
  A ::= &~ aAb ~\mid~ AA ~\mid~ \varepsilon ~\mid~ c
\end{align*}
where $a \in \cSigma^\stack$, $b \in \rSigma^\stack$, and $c \in \Sigma \setminus
(\cSigma^\stack \mathrel{\cup} \rSigma^\stack)$.

\begin{defi}
  A \emph{nested word} over $\crSigma$ is a structure
  $([n],\succord,\mu,\lambda)$ where $n \in \posN$ (we call the elements from
  $[n]$ \emph{positions}, \emph{nodes}, or \emph{events}), $\succord=\{(i,i+1)
  \mid i \in [n-1]\}$, $\lambda: [n] \rightarrow \Sigma$, and $\mu =
  \bigcup_{\stack \in [\nstack]} \mu^\stack \subseteq [n] \times [n]$ where,
  for every $\stack \in [\nstack]$ and $(i,j) \in [n] \times [n]$, $(i,j) \in
  \mu^\stack$ iff $i < j$, $\lambda(i) \in \cSigma^\stack$, $\lambda(j) \in
  \rSigma^\stack$, and $\lambda(i+1) \ldots \lambda(j-1)$ is $\stack$-well
  formed.
\end{defi}
\noindent
The set of nested words over $\crSigma$ is denoted by $\NW(\crSigma)$.

\begin{figure}[h]
\begin{center}
  \scalebox{1}{
\begin{picture}(110,38)(-50,-19)
\gasset{Nh=4,Nw=4,Nadjustdist=1,AHangle=35,AHLength=1.0,AHlength=0.4,Nframe=n,Nfill=n,linewidth=0.08}
\unitlength=0.27em

\node(W)(0.5,0){$\redc{a} \lra \greenc{b} \lra \redc{a} \lra {\greenc{b}} \lra
  \redc{\oa} \lra \redc{\oa} \lra \redc{\oa} \lra \greenc{\ob} \lra \greenc{\ob}
  \lra \greenc{\ob}~~~~~~~~$}

\node(A)(-45,-0.6){}
\node(B)(6,0.2){}
\drawbcedge(A,-35,17,B,-4,17){}

\node(A)(-25,-0.6){}
\node(B)(-4,0.2){}
\drawbcedge(A,-21,10,B,-10,10){}

\node[Nframe=n](A)(-35,-0.7){}
\node[Nframe=n](B)(35,-0.6){}
\drawbcedge(A,-30,-20,B,30,-20){}

\node[Nframe=n](A)(-16,-0.9){}
\node[Nframe=n](B)(26,-0.8){}
\drawbcedge(A,-8,-13,B,20,-13){}

\gasset{Nframe=n,Nfill=n}

\node(Pos)(-44,16){\footnotesize\textcolor{black}{1}}
\node(Pos)(-34.5,16){\footnotesize\textcolor{black}{2}}
\node(Pos)(-24.5,16){\footnotesize\textcolor{black}{3}}
\node(Pos)(-15,16){\footnotesize\textcolor{black}{4}}
\node(Pos)(-5,16){\footnotesize\textcolor{black}{5}}
\node(Pos)(5,16){\footnotesize\textcolor{black}{6}}
\node(Pos)(15,16){\footnotesize\textcolor{black}{7}}
\node(Pos)(25,16){\footnotesize\textcolor{black}{8}}
\node(Pos)(34.5,16){\footnotesize\textcolor{black}{9}}
\node(Pos)(44,16){\footnotesize\textcolor{black}{10}}

\end{picture}









}
\caption{A nested word\label{fig:exaNW}}
\end{center}
\end{figure}
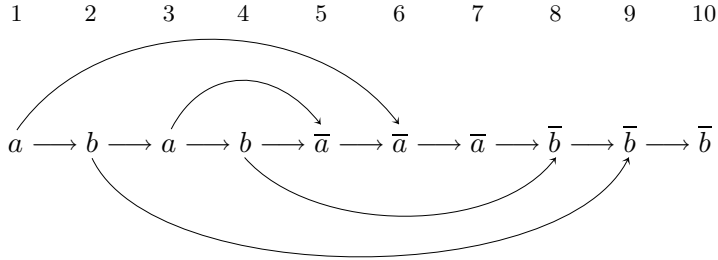

Figure~\ref{fig:exaNW} depicts a nested word over a 2-stack call-return
alphabet. Throughout the paper, we take advantage of the fact that nested
words over a 2-stack call-return alphabet can be written as a string with one
type of \emph{stack edges} above the string and the other below the string,
where the first type concerns the first stack and the other type concerns the
second stack. In the 2-stack case, the edges do not intersect.

Note that a nested word needs not be \emph{well-matched}. It might have
pending calls, i.e., calls without matching return, as well as pending
returns, i.e., returns that do not have a matching call. Therefore, the
relations $\mu$ and its inverse $\mu^{-1}$ can be seen as partial maps $[n]
\dashrightarrow [n]$, in the obvious manner. Moreover, observe that, given
nested words $\nword=([n],\succord,\mu,\lambda)$ and
$\nword'=([n'],\succord',\mu',\lambda')$, $n=n' \mathrel{\wedge}
\lambda=\lambda'$ implies $\nword = \nword'$. It is therefore justified to
represent $\nword$ as the string $\str(\nword) := \lambda(1) \ldots \lambda(n)
\in \Sigma^+$. This naturally extends to sets $\Lang$ of nested words and we
set $\str(\Lang) := \{\str(W) \mid W \in \Lang\}$. Vice versa, given a string
$w \in \Sigma^+$, there is precisely one nested word $\nword$ over $\crSigma$
such that $\str(\nword) = w$. This unique nested word is denoted $\nested(w)$.
For $L \subseteq \Sigma^+$, we let $\nested(L) := \{\nested(w) \mid w \in
L\}$.


\begin{exa}\label{ex:exaNW}
  Consider the 2-stack call-return alphabet $\crSigma$ from
  Example~\ref{ex:mvpa}, which was given by $\cSigma^1 = \{ a \}$, $\rSigma^1
  = \{ \oa \}$, $\cSigma^2 = \{ b\}$, $\rSigma^2 = \{ \ob \}$, and $\intSigma
  = \emptyset$. Figure~\ref{fig:exaNW} depicts a nested word
  $W=([n],\succord,\mu,\lambda)$ over $\crSigma$ with $n=10$. The straight
  arrows represent $\succord$, the curved arrows capture $\mu$ (those above
  the horizontal correspond to the first stack). For example, $(2,9) \in \mu$.
  Thus, $\mu(2)$ and $\mu^{-1}(9)$ are defined, whereas both $\mu^{-1}(7)$ and
  $\mu^{-1}(10)$ are not. In terms of visibly pushdown automata, this means
  that positions 7 and 10 are employed when the first/second stack is empty,
  respectively. Observe that $W=\nested( a\, b\, a\, b\, \oa\, \oa\, \oa\,
  \ob\, \ob\, \ob)$ and $\str(W)= a\, b\, a\, b\, \oa\, \oa\, \oa\, \ob\,
  \ob\, \ob $.
\end{exa}

We now turn to an automata model that is suited to nested words and, to some
extent, is equivalent to \MVPA. Our model is an extension of nested-word
automata for one stack, which has been considered in \cite{AlurM06}, to
multiple stacks. We also extend the model of \cite{AlurM06} by \emph{calling
  states}. If the state that is reached after executing some action $a$ is a
calling state, then the corresponding run is accepting only if this $a$ is a
call with a matching return (i.e., it is not pending). We will later see that
this concept does not increase the expressive power of our automata but turns
out to be a convenient tool when we translate logical formulas into automata.

\begin{defi}
  A \emph{generalized multi-stack nested-word automaton} (generalized \MNWA)
  over $\crSigma$ is a tuple $\WA=(Q,\delta,\Init,F,C)$ where
\begin{enumerate}[$\bullet$]
\item $Q$ is the finite set of \emph{states},
\item $\Init \subseteq Q$ is the set of \emph{initial states},
\item $F \subseteq Q$ is the set of \emph{final states},
\item $C \subseteq Q$ is a set of \emph{calling states}, and
\item $\delta$ is a pair $\langle \delta_1,\delta_2 \rangle$ of relations
  $\delta_1 \mathrel{\subseteq} Q \mathrel{\times} \Sigma \mathrel{\times} Q$
  and $\delta_2 \mathrel{\subseteq} Q \mathrel{\times} Q \mathrel{\times}
  \rSigma \mathrel{\times} Q$, which contain the \emph{transitions}.
\end{enumerate}
We call $\WA$ a \emph{multi-stack nested-word automaton} (\MNWA) if $C =
\emptyset$.

A \emph{(generalized) 2-stack nested-word automaton} ((generalized) $\tNWA$)
is a (generalized, respectively) $\MNWA$ that is defined over a 2-stack
alphabet (i.e., $K=2$).
\end{defi}

Intuitively, $\delta_1$ contains all the local and push transitions, as well
as all the pop transitions that act on an empty stack (i.e., in terms of
nested words and nested-word automata, those transitions that perform an
action from $\rSigma$ that is not matched by a corresponding calling action).
A run of $\WA$ on a nested word $\nword=([n],\succord,\mu,\lambda)$ over
$\crSigma$ is a mapping $\rho: [n] \rightarrow Q$ such that
$(q,\lambda(1),\rho(1)) \in \delta_1$ for some $q \in \Init$, and, for all $i
\in \{2,\ldots,n\}$, we have
 \[\left\{
 \begin{array}{rll}
   (\rho(\mu^{-1}(i)),\rho(i-1),\lambda(i),\rho(i)) & \!\!\!\in \delta_2 &~~ \text{~if~}
   \mu^{-1}(i) \text{~is~defined}\\
   (\rho(i-1),\lambda(i),\rho(i)) & \!\!\!\in \delta_1 &~~ \text{~otherwise}
 \end{array}
\right.
 \]
 The run $\rho$ is accepting if $\rho(n) \in F$ and, for all $i \in [n]$ with
 $\rho(i) \in C$, $\mu(i)$ is defined. The language of $\WA$, denoted by
 $\Lang(\WA)$, is the set of nested words from $\NW(\crSigma)$ that allow for
 an accepting run of $\WA$.


 Recall that there is a one-to-one correspondence between strings and nested
 words. We let therefore $\Lang(\PA)$ with $\PA$ an $\MVPA$ stand for the set
 $\nested(L(\PA))$.

\begin{exa}\label{ex:mnwa}
  Consider again the 2-stack call-return alphabet $\crSigma$ given by
  $\cSigma^1 = \{ a \}$, $\rSigma^1 = \{ \oa \}$, $\cSigma^2 = \{b\}$,
  $\rSigma^2 = \{ \ob \}$, and $\intSigma = \emptyset$. In
  Example~\ref{ex:mvpa}, we have seen that, for $L=\{( a b )^n \oa ^{n+1} \ob
  ^{n+1} \mid n \ge 1\}$, the iteration $L^+$ is the language of some \tVPA
  over $\crSigma$. We can also specify a \tNWA
  $\WA=(\{q_0,\ldots,q_4\},\delta,\{q_0\},\{q_0\},\emptyset)$ over $\crSigma$
  such that $\Lang(\WA) = \nested(L^+)$. Note that $\Lang(\WA)$ will contain,
  for example, the nested word that is depicted in Figure~\ref{fig:exaNW}. The
  transition relation $\delta$ is given as follows:
  \[\hspace{-2em}\begin{array}{rlcrl}
    \delta_1: & (q_0, a ,q_2) & ~~~~~~ & \delta_2: & (q_2,q_3, \oa ,q_3) \\
    & (q_2, b ,q_1) & & & (q_3,q_4, \ob ,q_4)\\
    & (q_1, a ,q_2) & & & (q_1,q_4, \ob ,q_4)\\
    & (q_2, b ,q_3)\\
    & (q_3, \oa ,q_4)\\
    & (q_4, \ob ,q_0)
  \end{array}\]
  Similarly to Example~\ref{ex:mvpa}, the finite-state control will ensure the
  general regular structure of a word without explicit ``counting''. This
  counting is then implicitly done by the relation $\delta_2$, which requires
  a matching call for a return. A graphical description of $\WA$ is given in
  Figure~\ref{fig:mnwa}. Hereby, a return transition with an adjoining set of
  states indicates that one state of this set must have been reached right
  after executing the corresponding call (in particular, the return must not
  be pending), whereas the remaining return transitions, $(q_3,\oa,q_4)$ and
  $(q_4,\ob,q_0)$, apply only to pending returns.
\end{exa}

\begin{figure}[h]
  \centering
\scalebox{1}{
\begin{picture}(20,28)(0,-3)
\gasset{Nframe=y,Nw=5.5,Nh=5.5,Nmr=8,ilength=4} 

\node[Nmarks=ir](q0)(0,20){$q_0$}
\node(q1)(40,20){$q_1$}
\node(q2)(20,20){$q_2$}
\node(q3)(20,0){$q_3$}
\node(q4)(0,0){$q_4$}

\drawloop[loopCW=n,loopdiam=5,loopangle=0,ELside=r](q3){$\{q_2\},\oa$}
\drawloop[loopCW=y,loopdiam=5,loopangle=180,ELside=l](q4){$\{q_1,q_3\},\ob$}

\drawedge[ELside=l](q0,q2){$a$}
\drawedge[ELside=l](q2,q3){$b$}
\drawedge[ELside=l](q3,q4){$\oa$}
\drawedge[ELside=l](q4,q0){$\ob$}

\drawedge[curvedepth=2,ELside=l](q2,q1){$b$}
\drawedge[curvedepth=2,ELside=l](q1,q2){$a$}

\end{picture}
}
\caption{A \tNWA \label{fig:mnwa}}
\end{figure}

A general technique for a reduction from \MVPA to \MNWA and vice versa can be
found below (Lemma~\ref{lem:PAWA}).

We can show that the use of calling states does not increase the
expressiveness of \MNWA. Note that, however, the concept of calling states
will turn out to be helpful when building the sphere automaton in
Section~\ref{sec:sphereaut}.

\begin{lem}\label{lem:generalized}
  For every generalized \MNWA $\WA$ over $\crSigma$, there is an \MNWA $\WA'$
  over $\crSigma$ such that $\Lang(\WA') = \Lang(\WA)$.
\end{lem}

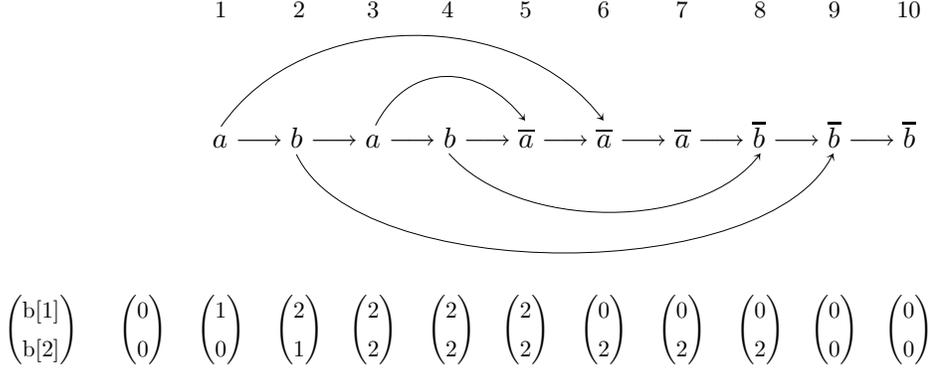
\begin{figure}[h]
\begin{center}
{
  \scalebox{1}{
\begin{picture}(120,50)(-65,-32)
\gasset{Nh=4,Nw=4,Nadjustdist=1,AHangle=35,AHLength=1.0,AHlength=0.4,Nframe=n,Nfill=n,linewidth=0.08}
\unitlength=0.27em

\node(W)(0.5,0){$\redc{a} \lra \greenc{b} \lra \redc{a} \lra {\greenc{b}} \lra
  \redc{\oa} \lra \redc{\oa} \lra \redc{\oa} \lra \greenc{\ob} \lra \greenc{\ob}
  \lra \greenc{\ob}~~~~~~~~$}

\node(A)(-45,-0.6){}
\node(B)(6,0.2){}
\drawbcedge(A,-35,17,B,-4,17){}

\node(A)(-25,-0.6){}
\node(B)(-4,0.2){}
\drawbcedge(A,-21,10,B,-10,10){}

\node[Nframe=n](A)(-35,-0.7){}
\node[Nframe=n](B)(35,-0.6){}
\drawbcedge(A,-30,-20,B,30,-20){}

\node[Nframe=n](A)(-16,-0.9){}
\node[Nframe=n](B)(26,-0.8){}
\drawbcedge(A,-8,-13,B,20,-13){}

\gasset{Nframe=n,Nfill=n}

\node(Pos)(-44,16){\footnotesize\textcolor{black}{1}}
\node(Pos)(-34,16){\footnotesize\textcolor{black}{2}}
\node(Pos)(-24.5,16){\footnotesize\textcolor{black}{3}}
\node(Pos)(-15,16){\footnotesize\textcolor{black}{4}}
\node(Pos)(-5,16){\footnotesize\textcolor{black}{5}}
\node(Pos)(5,16){\footnotesize\textcolor{black}{6}}
\node(Pos)(15,16){\footnotesize\textcolor{black}{7}}
\node(Pos)(25,16){\footnotesize\textcolor{black}{8}}
\node(Pos)(34.5,16){\footnotesize\textcolor{black}{9}}
\node(Pos)(44,16){\footnotesize\textcolor{black}{10}}

\node(Pos)(-67,-25){\scalebox{0.8}{$\begin{pmatrix}\textup{b}[1]\\\textup{b}[2]\end{pmatrix}$}}
\node(Pos)(-54,-25){\scalebox{0.8}{$\begin{pmatrix}0\\0\end{pmatrix}$}}
\node(Pos)(-44,-25){\scalebox{0.8}{$\begin{pmatrix}1\\0\end{pmatrix}$}}
\node(Pos)(-34,-25){\scalebox{0.8}{$\begin{pmatrix}2\\1\end{pmatrix}$}}
\node(Pos)(-24.5,-25){\scalebox{0.8}{$\begin{pmatrix}2\\2\end{pmatrix}$}}
\node(Pos)(-14.5,-25){\scalebox{0.8}{$\begin{pmatrix}2\\2\end{pmatrix}$}}
\node(Pos)(-5,-25){\scalebox{0.8}{$\begin{pmatrix}2\\2\end{pmatrix}$}}
\node(Pos)(5,-25){\scalebox{0.8}{$\begin{pmatrix}0\\2\end{pmatrix}$}}
\node(Pos)(15,-25){\scalebox{0.8}{$\begin{pmatrix}0\\2\end{pmatrix}$}}
\node(Pos)(25,-25){\scalebox{0.8}{$\begin{pmatrix}0\\2\end{pmatrix}$}}
\node(Pos)(34.5,-25){\scalebox{0.8}{$\begin{pmatrix}0\\0\end{pmatrix}$}}
\node(Pos)(44,-25){\scalebox{0.8}{$\begin{pmatrix}0\\0\end{pmatrix}$}}

\end{picture}

}
}
\caption{The flag construction\label{fig:flag}}
\end{center}
\end{figure}

\begin{proof}
  In the construction of an $\MNWA$, we exploit the following property of a
  nested word $\nword=([n],\succord,\mu,\lambda)$: given $(i,j) \in \mu$, say,
  with $\lambda(i) \in \cSigma^\stack$, $\mu(i')$ is defined for all $i' \in
  \{i+1,\ldots,j-1\}$ satisfying $\lambda(i') \in \cSigma^\stack$. Basically,
  $\WA'$ will simulate $\WA$. In addition, whenever a calling state is
  assigned to a position labeled with an element from $\cSigma^\stack$, we
  will set a flag $\overline{\textup{b}}[\stack]=1$, which can only be
  resolved and turn into a final state ($\overline{\textup{b}}[\stack]=0$)
  when a matching return position has been found. As any interim call position
  that concerns stack $\stack$ is matched anyway, the flags
  $\overline{\textup{b}}[\stack]$ in that interval are set to $2$. Thus, while
  a flag is $1$ or $2$, there is still some unmatched calling position. Hence,
  a final state requires every flag to equal $0$, which also designates the
  initial state.

  Let us become more precise and let $\WA=(Q,\delta,\Init,F,C)$ be a
  generalized \MNWA. We determine the
  \MNWA~$\WA'=(Q',\delta',\Init',F',\emptyset)$ by $Q' = Q \times
  \{0,1,2\}^{[\nstack]}$, $\Init' = \Init \times \{(0)_{\stack \in
    [\nstack]}\}$, $F' = F \times \{(0)_{\stack \in [\nstack]}\}$, and
  $\delta' = \langle \delta_1' , \delta_2' \rangle$ where
  \begin{enumerate}[$\bullet$]
  \item $\delta_1'$ is the set of triples
    $((q,\overline{\textup{b}}),a,(q',\overline{\textup{b}}')) \in Q' \times \Sigma \times Q'$
    such that $(q,a,q') \in \delta_1$, $q'\in C$ implies $a \in \cSigma$, and,
    for every $\stack \in [\nstack]$,
  \[\overline{\textup{b}}'[\stack] = \left\{
    \begin{array}{cl}
      2 & ~~ \text{~if~} \overline{\textup{b}}[\stack] \in \{1,2\}\\
      1 & ~~ \text{~if~} \overline{\textup{b}}[\stack]  = 0 \text{~and~} a \in \cSigma^\stack
      \text{~and~} q' \in C\\
      0 & ~~ \text{~otherwise}
 \end{array}
\right.
\]
\item $\delta_2'$ is the set of quadruples
  $((p,\overline{\textup{c}}),(q,\overline{\textup{b}}),a,(q',\overline{\textup{b}}'))
  \in Q' \times Q' \times \rSigma \times Q'$ such that $(p,q,a,q') \in
  \delta_2$, $q' \not\in C$, and, for every $\stack \in [\nstack]$,
  \[\overline{\textup{b}}'[\stack] = \left\{
    \begin{array}{cl}
      0 & ~~ \text{~if~} \overline{\textup{c}}[\stack]  = 1\\
      \overline{\textup{b}}[\stack] & ~~ \text{~otherwise}
 \end{array}
\right.
\]
\end{enumerate}
In fact, we can show that $\Lang(\WA) = \Lang(\WA')$.


Note that the flag assignments depend deterministically on the input word and
the states assigned to the positions. Let $W = ([n],\succord,\mu,\lambda)$ be
a nested word over $\crSigma$.

Suppose $\rho$ to be an accepting run of $\WA$ on $W$ and let $\widehat{\rho}:
[n] \rightarrow \{0,1,2\}^{[K]}$ be the unique supplement of $\rho$ according
to the flag construction. To verify that $(\rho,\widehat{\rho})$ is indeed an
accepting run of $\WA'$ on $W$, we need to show that
$\widehat{\rho}(n)[\stack] = 0$ for all $\stack \in [K]$. So let $\stack \in
[K]$. If there is no $i \in [n]$ such that $\lambda(i) \in \cSigma^\stack$ and
$\rho(i) \in C$, then we clearly have $\widehat{\rho}(n)[\stack] = 0$, as the
flag for stack $\stack$ never changes its value during the run. If the flag
changes its value from $0$ to $1$, then this happens at a position $i \in [n]$
such that $\lambda(i) \in \cSigma^\stack$ and $\rho(i) \in C$. As $\rho$ is an
accepting run of $\WA$ on $W$, there is $j \in [n]$ such that $(i,j) \in \mu$.
By construction of $\WA'$, $\widehat{\rho}(i)[\stack] = 1$,
$\widehat{\rho}(i')[\stack] = 2$ for all $i' \in \{i + 1,\ldots,j-1\}$, and
$\widehat{\rho}(j)[\stack] = 0$. Thus, we finally have
$\widehat{\rho}(n)[\stack] = 0$.

Conversely, let $\rho: [n] \rightarrow Q$ and $\widehat{\rho}: [n] \rightarrow
\{0,1,2\}^{[K]}$ be mappings such that $(\rho,\widehat{\rho})$ is an
accepting run of $\WA'$ on $W$. Clearly, $\rho$ is a run of $\WA$ on $W$. So
let us verify that it is accepting. First, observe that $\rho(n) \in F$. So
suppose $i \in [n]$ such that $\rho(i)$ is a calling state. According to the
construction of $\WA'$, $\lambda(i) \in \cSigma^\stack$ for some $\stack$.
Moreover, we have $\widehat{\rho}(i)[\stack] = \{1,2\}$. As
$\widehat{\rho}(n)[\stack] = 0$, there must be $i' \le i$ and $j' > i$ such
that $\lambda(i') \in \cSigma^\stack$ and $(i',j') \in \mu$. This implies that
$\mu(i)$ is indeed defined so that we can conclude that $\rho$ is an accepting
run of $\WA$ on $W$.
\end{proof}
The flag construction from the previous proof is illustrated in
Figure~\ref{fig:flag}, where we assume a run on the nested word such that
every state associated with a symbol from $\{a,b\}$ is a calling state.

\begin{lem}\label{lem:PAWA}
  Let $\Lang \mathrel{\subseteq} \NW(\crSigma)$ be a set of nested words over
  $\crSigma$. The following are equivalent:
\begin{enumerate}[\em(1)]
\item There is an \MVPA $\PA$ over $\crSigma$ such that $\Lang(\PA) = \Lang$.
\item There is an \MNWA $\WA$ over $\crSigma$ such that $\Lang(\WA) = \Lang$.
\end{enumerate}
\end{lem}

\begin{proof}
  Given an $\MVPA$ $\PA=(Q,\Gamma,\delta,\Init,F)$, we define an $\MNWA$
  $\WA=(Q',\delta',\Init',F',\emptyset)$ with $\Lang(\PA) = \Lang(\WA)$ as
  follows: $Q' = Q \times \Gamma$, $\Init' = \Init \times \{\bot\}$, $F' = F
  \times \Gamma$, and $\delta' = \langle \delta_1' , \delta_2' \rangle$ where
  \begin{enumerate}[$\bullet$]
  \item $\delta_1'$ is the set of triples $((q,A),a,(q',A')) \in Q' \times
    \Sigma \times Q'$ such that $(q,a,A',q') \in \cdelta$, $(q,a,q') \in
    \intdelta$, or $(q,a,\bot,q') \in \rdelta$, and
  \item $\delta_2'$ is the set of quadruples $((p,B),(q,A),a,(q',A')) \in Q'
    \times Q' \times \Sigma \times Q'$ such that $(q,a,B,q') \in \rdelta$.
  \end{enumerate}
  The idea is that the stack symbol associated with a transition is
  incorporated into the state of the \MNWA. When an internal or unmatched
  return action is performed, then we may chose an arbitrary stack symbol, as
  it will not be reconsidered later in the run.

  For the converse direction, let $\WA=(Q,\delta,\Init,F,\emptyset)$ be an
  \MNWA. Consider the \MVPA $\PA=(Q,Q \uplus \{\bot\},\delta',\Init,F)$ where
  $\delta' = \langle \cdelta' , \rdelta', \intdelta' \rangle$ is given by
  \begin{enumerate}[$\bullet$]
  \item $\cdelta' = \{(q,a,q',q') \mid (q,a,q') \in \delta_1 \mathrel{\cap} (Q
    \times \cSigma \times Q)\}$,
  \item $\intdelta' = \delta_1 \mathrel{\cap} (Q \times \intSigma \times Q)$,
    and
  \item $\rdelta'$ is the set of tuples $(q,a,A,q') \in Q \times \rSigma
    \times \Gamma \times Q$ such that either $(q,a,q') \in \delta_1$ and $A =
    \bot$, or $(A,q,a,q') \in \delta_2$.
\end{enumerate}
Here, we need to ensure that, when $\PA$ performs a matched return action, we
can access the state that $\WA$ has associated with the corresponding call. To
this aim, $\PA$ just pushes the state onto the stack so that it becomes
accessible when the corresponding return is executed. It is straightforward to
show that $\Lang(\PA) = \Lang(\WA)$.
\end{proof}


\section{Monadic Second-Order Logic and Hanf's Theorem}\label{sec:MSO}

\subsection{Monadic Second-Order Logic over Relational Structures}

We fix supplies of first-order variables $x,y,\ldots$ and second-order
variables $X,Y,\ldots$. Let $\tau$ be a function-free signature. The set
$\MSO(\tau)$ of \emph{monadic second-order} (MSO) formulas over $\tau$ is
given by the following grammar:
\begin{align*}
  \phi ::= &~ P(x_1,\ldots,x_m) ~\mid~ x_1 = x_2 ~\mid~ x \in X ~\mid~ \neg
  \phi ~\mid~ \phi_1 \vee \phi_2 ~\mid~ \exists x \phi ~\mid~ \exists X \phi
\end{align*}
Hereby, $m \ge 1$, $P \in \tau$ is an $m$-ary predicate symbol, the $x_k$ and
$x$ are first-order variables, and $X$ is a second-order variable.
%
Moreover, we will make use of the usual abbreviations such as $\phi_1
\mathrel{\wedge} \phi_2$ for $\neg(\neg \phi_1 \vee \neg\phi_2)$, $\phi_1
\rightarrow \phi_2$ for $\neg \phi_1 \vee \phi_2$, etc. Given a
$\tau$-structure $\mathfrak{A}$ with universe $A$, a formula
$\phi(x_1,\ldots,x_m,X_1,\ldots,X_n) \in \MSO(\tau)$ with free variables in
$\{x_1,\ldots,x_m,X_1,\ldots,X_n\}$, $(u_1,\ldots,u_m) \in A^m$, and
$(U_1,\ldots,U_n) \in {(2^A)}^n$, we write, as usual, $\mathfrak{A} \models
\phi[u_1,\ldots,u_m,U_1,\ldots,U_n]$ if $\mathfrak{A}$ satisfies $\phi$ when
assigning $(u_1,\ldots,u_m)$ to $(x_1,\ldots,x_m)$ and $(U_1,\ldots,U_n)$ to
$(X_1,\ldots,X_n)$.

Let us identify some important fragments of $\MSO(\tau)$. The set $\FO(\tau)$
of \emph{first order} (FO) formulas over $\tau$ comprises those formulas from
$\MSO(\tau)$ that do not contain any second-order quantifier. Furthermore, an
\emph{existential} MSO (EMSO) formula is of the form $\exists X_1 \ldots
\exists X_n \phi$ with $\phi \in \FO(\tau)$. The corresponding class of
formulas is denoted $\EMSO(\tau)$. More generally, given $m \ge 1$, we denote
by $\LSigma_m(\tau)$ the set of formulas of the form $\exists \overline{X_1}
\forall \overline{X_2} \ldots \exists / \forall \overline{X_m} \phi$ where
$\phi \in \FO(\tau)$ and the $\overline{X_k}$ are blocks of second-order
variables, possibly empty or of different length.

We will later make use of the notion of \emph{definability} relative to a
class of structures. Let $\mathcal{F} \subseteq \MSO(\tau)$ be a class of
formulas and $\Lang,\mathcal{C}$ be sets of $\tau$-structures. We say that
$\Lang$ is $\mathcal{F}$-\emph{definable relative to} $\mathcal{C}$ if there
is a sentence (i.e., a formula without any free variables) $\phi \in
\mathcal{F}$ such that $\Lang$ is the set of $\tau$-structures $\mathfrak{A}
\in \mathcal{C}$ such that $\mathfrak{A} \models \phi$.





\subsection{Hanf's Theorem for Nested Words, and Spheres}

We will now provide a signature that allows us to specify MSO properties of
nested words. Let $\crSigma$ be a call-return alphabet. We define
$\tau_{\crSigma}$ to be the signature $\{\lambda_a \mid a \in \Sigma\} \cup
\{\succord,\mu\}$ with $\lambda_a$ a unary and $\succord$ and $\mu$ binary
predicate symbols. We write the MSO formula $\lambda_a(x)$ as $\lambda(x)=a$
and the formula $\succord(x_1,x_2)$ as $x_1 \succrel x_2$. MSO formulas over
$\tau_{\crSigma}$ can be canonically interpreted over nested words
$([n],\succord,\mu,\lambda) \in \NW(\crSigma)$, as $\lambda$ can be seen as a
collection of unary relations $\lambda_a = \{i \in [n] \mid \lambda(i) = a\}$
where $a \in \Sigma$. Thus, nested words over $\crSigma$ are actually
$\tau_{\crSigma}$-structures. A sample MSO formula over $\tau_{\crSigma}$ such
that $\Sigma=\{a,b\}$ is $\forall x \forall y~\!(\lambda(x) = a
\mathrel{\wedge} \mu(x,y) \rightarrow \lambda(y)=b)$. It expresses that every
matching pair with a calling $a$ has a $b$-labeled return position. Given a
sentence $\phi \in \MSO(\signNW)$, we denote by $\Lang(\phi)$ the set of
nested words over $\crSigma$ that satisfy $\phi$, i.e., $\Lang(\phi) = \{W \in
\NW(\crSigma) \mid W \models \phi\}$.


Over nested words (more generally, structures of bounded degree), FO formulas
enjoy a normal form in terms of local formulas. A formula $\phi(x) \in
\FO(\signNW)$ with one free variable $x$ is said to be \emph{local} if there
is $r \in \N$ such that, in every subformula $\exists y \psi$ of $\phi$,
$\psi$ is of the form $(\dist(x,y) \le r) \wedge \chi$. Hereby, the formula
$\dist(x,y) \le r$ has the expected meaning and can be obtained inductively.
Informally, the truth of a local formula $\phi(x)$ depends only on the local
neighborhood around $x$.

Next, we state Hanf's locality theorem in terms of nested words. It actually
applies to general classes of structures of bounded degree.

\begin{thm}[Hanf \cite{Hanf1965}]\label{thm:hanf} Let $\phi \in
  \FO(\tau_{\crSigma})$ be a sentence. There is a positive Boolean combination
  $\psi$ of formulas of the form \[\exists^{=t}x\, \chi(x) \text{~~~and~~~}
  \exists^{>t}x\, \chi(x)\] where $t \in \N$ and $\chi \in
  \FO(\tau_{\crSigma})$ is local (with the obvious meaning of the quantifiers
  $\exists^{=t}$ and $\exists^{>t}$; note that there might occur different
  thresholds $t$ in $\psi$) such that, for every nested word $W \in
  \NW(\crSigma)$, we have
  \[W \models \phi \text{~~~iff~~~} W \models \psi.\] Moreover, $\psi$ can be
  computed effectively and in elementary time.
\end{thm}
For a comprehensive proof of this theorem, see, for example,
\cite{Tho97handbook,Libkin2004}. However, these proofs are not effective,
whereas the original proof by Hanf is effective. It is crucial to note that
Hanf's Theorem applies to the case of nested words as we deal with a class of
structures of bounded degree (see below for a formal definition). Indeed,
there is a uniform bound on the degree of nested words.

Let $\mathfrak{A}=(\Nat,\succord,\mu,\lambda,\ldots)$ and
$\mathfrak{A'}=(\Nat',\succord',\mu',\lambda',\ldots)$ be tuples such that
$(\Nat,\succord,\mu,\lambda)$ and $(\Nat',\succord',\mu',\lambda')$ are
$\signNW$-structures.
For $i,j \in N$ and $i',j' \in N'$, we write
$\wsimulate{\mathfrak{A}}{\mathfrak{A}'}{(i,j)}{(i',j')}$ if $\lambda(i) =
\lambda'(i')$, $\lambda(j) = \lambda'(j')$, $(i,j) \in \succord$ implies
$(i',j') \in \succord'$, and $(i,j) \in \mu$ implies $(i',j') \in \mu'$.
Theorem~\ref{thm:hanf} suggests that, over nested words, the validity of an FO
formula in a nested word depends on the local neighborhoods of the latter.
This leads to the notion of a \emph{sphere}, which will actually play a
central role in the remainder of this paper. A sphere of radius $r \in \N$
includes elements whose distance from a distinguished sphere center is bounded
by $r$. Given $i,j \in \Nat$, the \emph{distance} $\dist_\mathfrak{A}(i,j)$ of
$i$ and $j$ in $\mathfrak{A}$ is the minimal length of a path from $i$ to $j$
in the \emph{Gaifman graph} of $(\Nat,\succord,\mu,\lambda)$. The Gaifman
graph of $(\Nat,\succord,\mu,\lambda)$ is defined to be the undirected graph
$(N,\mathit{Arcs})$ where $(i,j) \in \mathit{Arcs}$ iff $(i,j) \in \succord
\mathrel{\cup} \mathord{\mu} \mathrel{\cup} \succord^{-1} \mathrel{\cup}
\mathord{\mu}^{-1}$ \cite{Libkin2004}. In particular, we have
$\dist_\mathfrak{A}(i,i) = 0$. If $\dist_\mathfrak{A}(i,j) = 1$, we also write
$i \edge{\mathfrak{A}} j$. We write $i \redge{\mathfrak{A}} j$ if $(i,j) \in
\succord \mathrel{\cup} \mu$.
The degree of a $\signNW$-structure is said to be \emph{bounded} by some
natural number $B$ if the degree of its Gaifman graph is bounded by $B$.
Observe that the degree of a nested word is bounded by $3$, which is therefore
a uniform bound for the class $\NW(\crSigma)$.

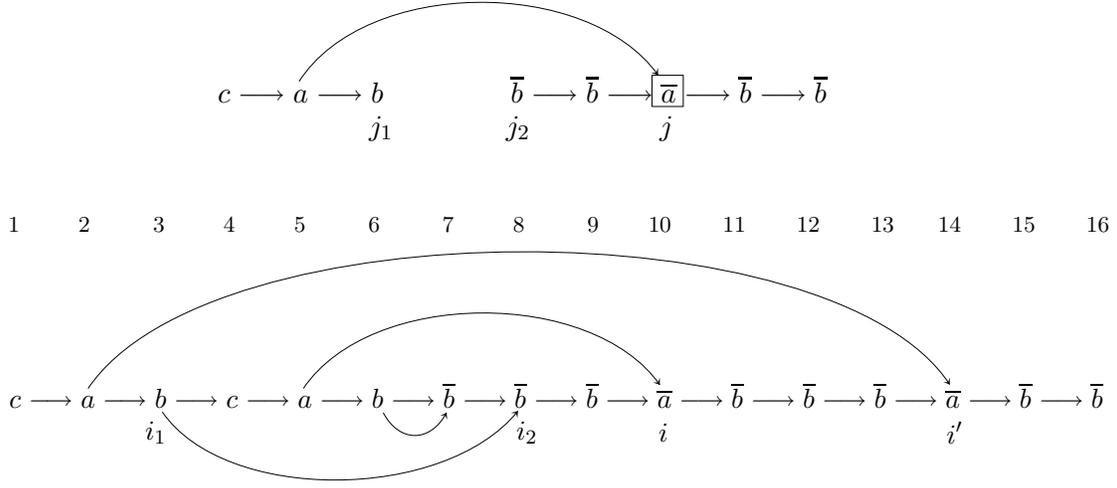
\begin{figure}[t]
\begin{center}
  \scalebox{1}{
\begin{picture}(99,25)(-45,-10)
\gasset{Nh=4,Nw=4,Nadjustdist=1,AHangle=35,AHLength=1.0,AHlength=0.4,Nframe=n,Nfill=n,linewidth=0.08}
\unitlength=0.27em

\node(W)(0,0){$c \lra a \lra b \hspace{4.4em} \overline{b} \lra \overline{b} \lra
\overline{a} \lra \overline{b} \lra \overline{b}$}

\node(A)(-29.5,-0.7){}
\node[Nframe=y,Nw=4,Nh=4,Nmr=0](B)(18.6,-0.2){}
\drawbcedge(A,-22,15,B,11,15){}

\node(A)(-18,-5){$j_1$}
\node(B)(-0.5,-5){$j_2$}
\node(C)(18.5,-5){$j$}

\gasset{Nframe=n,Nfill=n}

\end{picture}
}\vspace{4ex}\\
\scalebox{0.95}{
\begin{picture}(150,40)(-75,-15)
\gasset{Nh=4,Nw=4,Nadjustdist=1,AHangle=35,AHLength=1.0,AHlength=0.4,Nframe=n,Nfill=n,linewidth=0.08}
\unitlength=0.27em

\node(W)(0,0){$c \lra a \lra b \lra c \lra a \lra b \lra \overline{b} \lra \overline{b}
  \lra \overline{b} \lra \overline{a} \lra \overline{b} \lra \overline{b}
  \lra \overline{b} \lra \overline{a} \lra \overline{b}
  \lra \overline{b}$}

\node(A)(-64,-0.7){}
\node(B)(54,-0.2){}
\drawbcedge(A,-50,26,B,41,26){}

\node(A)(-35,-0.7){}
\node(B)(15,-0.2){}
\drawbcedge(A,-27,15,B,7,15){}

\node(A)(-24,-0.5){}
\node(B)(-14,-0.5){}
\drawbcedge(A,-22,-7,B,-16,-7){}

\node(A)(-54,-0.5){}
\node(B)(-4,-0.5){}
\drawbcedge(A,-47,-15,B,-13,-15){}

\node(A)(14.5,-5){$i$}
\node(B)(53.7,-5){$i'$}
\node(C)(-53.9,-5){$i_1$}
\node(C)(-3.9,-5){$i_2$}

\gasset{Nframe=n,Nfill=n}

\node(Pos)(-73,23){\footnotesize\textcolor{black}{1}}
\node(Pos)(-63.5,23){\footnotesize\textcolor{black}{2}}
\node(Pos)(-53.5,23){\footnotesize\textcolor{black}{3}}

\node(Pos)(-44,23){\footnotesize\textcolor{black}{4}}
\node(Pos)(-34.5,23){\footnotesize\textcolor{black}{5}}
\node(Pos)(-24.5,23){\footnotesize\textcolor{black}{6}}
\node(Pos)(-14.5,23){\footnotesize\textcolor{black}{7}}
\node(Pos)(-5,23){\footnotesize\textcolor{black}{8}}
\node(Pos)(5,23){\footnotesize\textcolor{black}{9}}
\node(Pos)(14,23){\footnotesize\textcolor{black}{10}}
\node(Pos)(24,23){\footnotesize\textcolor{black}{11}}
\node(Pos)(34,23){\footnotesize\textcolor{black}{12}}
\node(Pos)(44,23){\footnotesize\textcolor{black}{13}}
\node(Pos)(53,23){\footnotesize\textcolor{black}{14}}
\node(Pos)(63,23){\footnotesize\textcolor{black}{15}}
\node(Pos)(73,23){\footnotesize\textcolor{black}{16}}

\end{picture}
}
\caption{A $2$-shpere embedded into a nested word\label{fig:embedding}}
\end{center}
\end{figure}

Let $\mathfrak{B}=(\Nat,\succord,\mu,\lambda)$ be a $\signNW$-structure,
$\radius \in \N$, and $i \in \Nat$. The $r$-\emph{sphere} of $\mathfrak{B}$
around $i$, which we denote by $\Sph{r}{\mathfrak{B}}{i}$, is basically the
substructure of $\mathfrak{B}$ induced by the new universe $\{j \in \Nat \mid
\dist_\mathfrak{B}(i,j) \le \radius\}$, but extended by the constant $i$ as a
distinguished element, called the \emph{sphere center}.
Given an isomorphism type $\sphere$ of an $\radius$-sphere, we let
$|\mathfrak{B}|_\sphere := |\{i \in \Nat \mid S \isom
\Sph{r}{\mathfrak{B}}{i}\}|$ denote the number of points in $\mathfrak{B}$
that \emph{realize} $S$. For an example, consider Figure~\ref{fig:embedding},
showing a nested word $W$ and the 2-sphere of $W$ around $i=10$ where the
sphere center is marked as a rectangle. Note that $\Sph{2}{W}{10} \isom
\Sph{2}{W}{14}$ and $|W|_{\Sph{2}{W}{10}} = 2$.

We denote by $\Spheres_\radius(\crSigma)$ the set of (isomorphism types of)
$\radius$-spheres that arise from nested words over $\crSigma$, i.e.,
\[\AllSpheres_\radius(\crSigma) := \{\Sph{\radius}{W}{i} \mid W \in
\NW(\crSigma) \text{~and~} i \text{~is a node of } W\}\,.\] Note that
$\AllSpheres_\radius(\crSigma)$ is finite up to isomorphism, which is crucial
for the constructions in Section~\ref{sec:maintheorem}.

\section{2-Stack Visibly Pushdown Automata vs.\ Logic}\label{sec:EMSO}

\label{sec:maintheorem}



In this section, we focus on \tVPA. So let us fix a 2-stack call-return
alphabet
$\crSigma=\langle\{(\cSigma^1,\rSigma^1),(\cSigma^2,\rSigma^2)\},\intSigma\rangle$.

\subsection{The Main Result}

\label{subsect:maintheorem}
The key connection between FO logic and \tVPA/\tNWA~is provided by the
following proposition, which states the existence of an automaton that
computes the sphere around any node of a nested word.

\begin{prop}\label{prop:main}
  Let $\radius$ be any natural number. There are a generalized \tNWA
  $\WA_\radius=(Q,\delta,\Init,F,C)$ over $\crSigma$ and a mapping $\eta: Q
  \rightarrow \AllSpheres_\radius(\crSigma)$ such that
\begin{enumerate}[$\bullet$]
\item $\Lang(\WA_\radius) = \NW(\crSigma)$ (i.e., every nested word admits an
  accepting run of $\WA_\radius$), and
\item for every nested word $W \in \NW(\crSigma)$, every accepting run $\rho$
  of $\WA_\radius$ on $W$, and every node $i$ of $W$, we have $\eta(\rho(i))
  \isom \Sph{\radius}{W}{i}$.
\end{enumerate}
\end{prop}


Before we turn towards the proof of this statement, we will first show how
Proposition~\ref{prop:main} can be used to establish expressive equivalence of
\tVPA and EMSO logic.

\begin{lem}\label{lem:counting}
  Let $\radius,t \in \N$ and let $S \in \AllSpheres_\radius(\crSigma)$ be an
  $\radius$-sphere in some nested word over $\crSigma$. There are generalized
  $\tNWA$ $\WA^1$ and $\WA^2$ over $\crSigma$ such that $\Lang(\WA^1) = \{W
  \in \NW(\crSigma) \mid |W|_S = t\}$ and $\Lang(\WA^2) = \{W \in
  \NW(\crSigma) \mid |W|_S > t\}$.
\end{lem}

\begin{proof}
  In both cases, we start from the generalized $\tNWA$
  $\WA_\radius=(Q,\delta,\Init,F,C)$ and the mapping $\eta: Q \rightarrow
  \AllSpheres_\radius(\crSigma)$ from Proposition~\ref{prop:main}. For
  $k=1,2$, we obtain $\WA^k$ by extending the state space with a counter that,
  using $\eta$, counts the number of realizations of $S$ up to $t+1$. The new
  set of initial states is thus in both cases $\Init \times \{0\}$. However,
  the set of final states of $\WA^1$ is $F \times \{t\}$, the one of $\WA^2$
  is $F \times \{t+1\}$.
\end{proof}

We are now prepared to state the first main result of this paper.

\begin{thm}\label{thm:equiv}
  Let $\Lang \subseteq \NW(\crSigma)$ be a set of nested words over the
  2-stack call-return alphabet $\crSigma$. Then, the following are equivalent:
\begin{enumerate}[\em(1)]
\item There is a \tVPA $\PA$ over $\crSigma$ such that $\Lang(\PA) = \Lang$.
\item There is a sentence $\phi \in \EMSO(\tau_{\crSigma})$ such that
  $\Lang(\phi) = \Lang$.
\end{enumerate}
Both directions are effective. In particular, the \tVPA that we construct for
a given EMSO sentence can be computed in elementary time, and its size is
elementary in the size of the formula.
\end{thm}

\begin{proof}
  To prove $(1) \rightarrow (2)$, one can perform a standard construction of
  an EMSO formula from a \tNWA, where the latter can be extracted from the
  given \tVPA according to Lemma~\ref{lem:PAWA}. Basically, the formula
  ``guesses'' a possible run on the input word in terms of existentially
  quantified second-order variables and then verifies, in its first-order
  fragment, that we actually deal with a run that is accepting.

  So let us directly prove $(2) \rightarrow (1)$ and let $\phi = \exists X_1
  \ldots \exists X_m \psi(X_1,\ldots,X_m) \in \EMSO(\signNW)$ be a sentence
  with $\psi(X_1,\ldots,X_m) \in \FO(\signNW)$ (we suppose $m \ge 1$). We
  define a new 2-stack call-return alphabet
  \[\ncrSigma=\langle\{(\cSigma^1 \times 2^{[m]},\rSigma^1 \times
  2^{[m]}),(\cSigma^2 \times 2^{[m]},\rSigma^2 \times 2^{[m]})\},\intSigma
  \times 2^{[m]}\rangle\] where $2^{[m]}$ shall denote the powerset of $[m]$.
  From $\psi$, we obtain an FO formula $\psi'$ over $\tau_{\ncrSigma}$ by
  replacing each occurrence of $\lambda(x) = a$ with $\bigvee_{M \in 2^{[m]}}
  \lambda(x) = (a,M)$ and each occurrence of $x \in X_k$ with $\bigvee_{a \in
    \Sigma,~M \in 2^{[m]}} \lambda(x) = (a,M \cup \{k\})$. We set $\Lang
  \mathrel{\subseteq} \NW(\ncrSigma)$ to be the set of nested words that
  satisfy $\psi'$. From Hanf's Theorem (Theorem~\ref{thm:hanf}), we know that
  $\Lang$ is the language of a positive Boolean combination of formulas of the
  form $\exists^{=t}x\, \chi(x)$ and $\exists^{>t}x\, \chi(x)$ where $\chi$ is
  local. It is easy to see that the class of nested-word languages that are
  recognized by generalized \tNWA is closed under union and intersection.
  Thus, the validity of one such basic formula can be checked by a generalized
  \tNWA due to Lemma~\ref{lem:counting}. We deduce that there is a generalized
  \tNWA $\WA'$ over $\crSigma$ recognizing $\Lang$.

  Now, to check whether some nested word from $\NW(\crSigma)$ satisfies
  $\phi$, a generalized \tNWA $\WA$ with $\Lang(\WA) = \Lang(\phi)$ will guess
  an additional labeling for each node in terms of an element from $2^{[m]}$
  and then simulate $\WA'$. By Lemma~\ref{lem:generalized} and
  Lemma~\ref{lem:PAWA}, we finally obtain a \tVPA $\PA$ such that $\Lang(\PA)
  = \Lang(\phi)$.
\end{proof}

\subsection{Proof of Proposition~\ref{prop:main}}\label{sec:sphereaut}

We now turn to the proof of Proposition~\ref{prop:main}. In each state, the
generalized $\tNWA$ $\WA_\radius$ will guess the current sphere as well as
spheres of nodes nearby and the current position in these additional spheres.
Adding some global information allows us to locally check whether all the
guesses are correct. The rest of this section is devoted to the construction
of $\WA_\radius$ and a corresponding mapping $\eta$ to prove
Proposition~\ref{prop:main}.

\subsubsection{The Construction}

Recall that $\AllSpheres_\radius(\crSigma)$ denotes the set of all the
$\radius$-spheres that arise from nested words, i.e.,
$\AllSpheres_\radius(\crSigma) = \{\Sph{\radius}{W}{i} \mid W$ is a nested
word and $i$ is a position in $W\}$. An \emph{extended} $r$-sphere over
$\crSigma$ is a tuple $\esphere=(N,\succord,\mu,\lambda,\gamma,\alpha,\inst)$
where $\core(\esphere) := (N,\succord,\mu,\lambda,\gamma) \in
\AllSpheres_\radius(\crSigma)$ (in particular, $\gamma \in N$), $\alpha \in
\Nat$, and $\inst \in [\const]$ with $\const = 4 \cdot \maxN^2 + 1$ where
$\maxN$ is the maximal size of an $\radius$-sphere, i.e., $\maxN = \max\{|N|
\mid (N,\succord,\mu,\lambda,i) \in \AllSpheres_\radius(\crSigma)\}$. We say
that $\alpha$ is the \emph{active node} of $E$ and $\inst$ is its
\emph{color}. Strictly speaking,
$(N,\succord,\mu,\lambda,\gamma,\alpha,\inst)$ is not a mathematical
structure, as $\inst$ does not refer to an element of $N$. We introduced the
function $\core$ to extract a mathematical structure from an extended sphere,
which will allow us to deal with notions such as isomorphism.

Let $\AlleSpheres_\radius(\crSigma)$ denote the set of all the (isomorphism
classes of) extended spheres over $\crSigma$. For an extended sphere
$E=(N,\succord,\mu,\lambda,\gamma,\alpha,\inst)$ and an element $i \in N$, we
denote by $E[i]$ the extended sphere
$(N,\succord,\mu,\lambda,\gamma,i,\inst)$, i.e., the extended sphere that we
obtain by replacing the active node $\alpha$ with $i$.

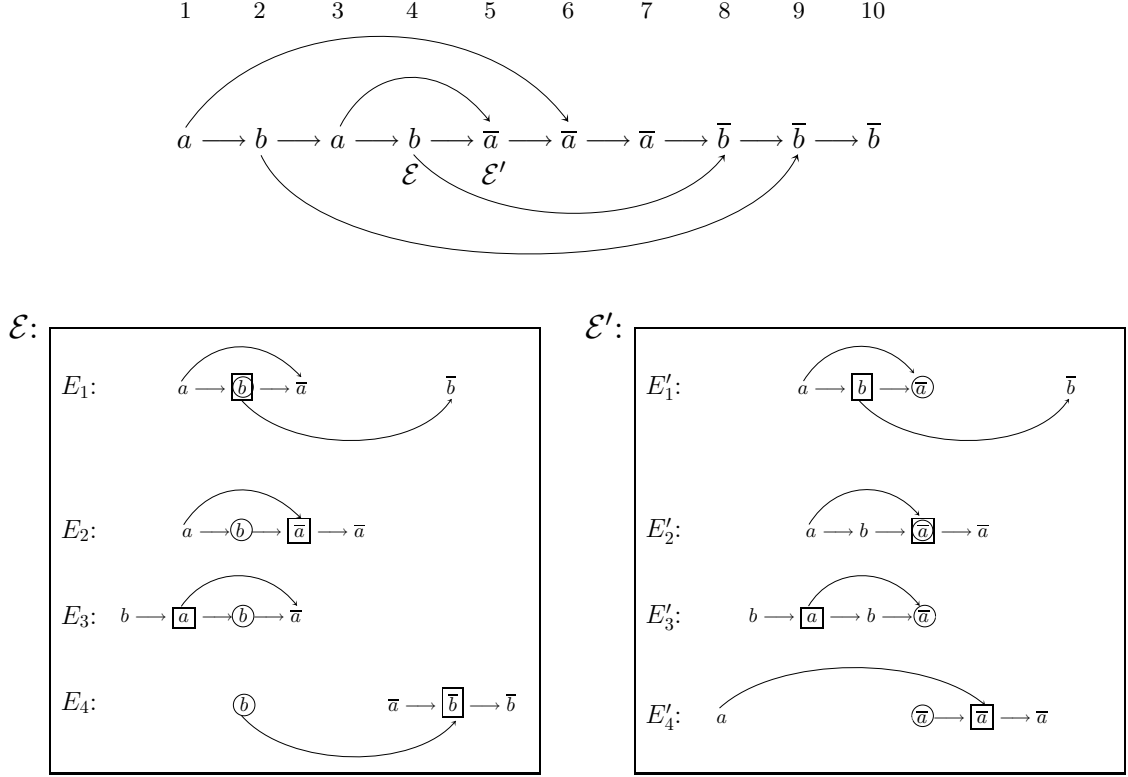
\begin{figure}
\begin{center}
  \scalebox{1}{
\begin{picture}(110,38)(-50,-19)
\gasset{Nh=4,Nw=4,Nadjustdist=1,AHangle=35,AHLength=1.0,AHlength=0.4,Nframe=n,Nfill=n,linewidth=0.08}
\unitlength=0.27em

\node(W)(0.5,0){$\redc{a} \lra \greenc{b} \lra \redc{a} \lra {\greenc{b}} \lra
  \redc{\oa} \lra \redc{\oa} \lra \redc{\oa} \lra \greenc{\ob} \lra \greenc{\ob}
  \lra \greenc{\ob}~~~~~~~~$}

\node(A)(-45,-0.6){}
\node(B)(6,0.2){}
\drawbcedge(A,-35,17,B,-4,17){}

\node(A)(-25,-0.6){}
\node(B)(-4,0.2){}
\drawbcedge(A,-21,10,B,-10,10){}

\node[Nframe=n](A)(-35,-0.7){}
\node[Nframe=n](B)(35,-0.6){}
\drawbcedge(A,-30,-20,B,30,-20){}

\node[Nframe=n](A)(-16,-0.9){}
\node[Nframe=n](B)(26,-0.8){}
\drawbcedge(A,-8,-13,B,20,-13){}

\gasset{Nframe=n,Nfill=n}

\node(Q)(-15.2,-5){$\State$}
\node(Q)(-4.5,-4.8){$\State'$}

\node(Pos)(-44,16){\footnotesize\textcolor{black}{1}}
\node(Pos)(-34.5,16){\footnotesize\textcolor{black}{2}}
\node(Pos)(-24.5,16){\footnotesize\textcolor{black}{3}}
\node(Pos)(-15,16){\footnotesize\textcolor{black}{4}}
\node(Pos)(-5,16){\footnotesize\textcolor{black}{5}}
\node(Pos)(5,16){\footnotesize\textcolor{black}{6}}
\node(Pos)(15,16){\footnotesize\textcolor{black}{7}}
\node(Pos)(25,16){\footnotesize\textcolor{black}{8}}
\node(Pos)(34.5,16){\footnotesize\textcolor{black}{9}}
\node(Pos)(44,16){\footnotesize\textcolor{black}{10}}

\end{picture}
}

\vspace{2ex}


\scalebox{1.15}{

$\State$:~
\fbox{
\begin{minipage}[t]{0.34\textwidth}
~\\
{
\scalebox{0.6}{
\begin{picture}(60,18)(-47,-9)
\gasset{Nh=4,Nw=4,Nadjustdist=1,AHangle=35,AHLength=1.0,AHlength=0.4,Nframe=n,Nfill=n,linewidth=0.08}
\unitlength=0.27em

\node(W)(-0.5,0){$\whitec{a \lra b \lra} a \lra {\greenc{\fbox{$b$}}} \lra
  \oa \whitec{\lra \oa \lra \redc{\oa} \lra} \ob \whitec{\lra \greenc{\ob}
  \lra \ob}~~~~~~~~$}

\node(A)(-27,-0.6){}
\node(B)(-2.5,0.4){}
\drawbcedge(A,-21,10,B,-10,10){}

\node[Nframe=n](A)(-16,-0.9){}
\node[Nframe=n](B)(25,-0.5){}
\drawbcedge(A,-8,-13,B,20,-13){}

\gasset{Nframe=n,Nfill=n}

\node[Nframe=n](E)(-45,0){\Large$E_1$:}

\end{picture}
}
}

\vspace{4ex}

\scalebox{0.6}{
\begin{picture}(60,16)(-47,-9)
\gasset{Nh=4,Nw=4,Nadjustdist=1,AHangle=35,AHLength=1.0,AHlength=0.4,Nframe=n,Nfill=n,linewidth=0.08}
\unitlength=0.27em

\node(W)(0.8,0){$\whitec{a \lra b \lra} a \lra \text{\greenc{$b$}} \lra
  \fbox{\greenc{$\oa$}} \lra \oa \whitec{\lra \redc{\oa} \lra \ob} \whitec{\lra \greenc{\ob}
  \lra \ob}~~~~~~~~$}

\node(A)(-26.5,-0.6){}
\node(B)(-2.5,0.8){}
\drawbcedge(A,-21,10,B,-10,10){}

\gasset{Nframe=n,Nfill=n}

\node[Nframe=n](E)(-45,0){\Large$E_2$:}

\end{picture}
}

\vspace{0ex}

\scalebox{0.6}{
\begin{picture}(60,16)(-47,-9)
\gasset{Nh=4,Nw=4,Nadjustdist=1,AHangle=35,AHLength=1.0,AHlength=0.4,Nframe=n,Nfill=n,linewidth=0.08}
\unitlength=0.27em

\node(W)(-1,0){$\whitec{a \lra} b \lra \fbox{$a$} \lra \text{\greenc{$b$}} \lra
  \greenc{\oa} \whitec{\lra \oa \lra \redc{\oa} \lra \ob} \whitec{\lra \greenc{\ob}
    \lra \ob}~~~~~~~~$}

\put(1,0){
\node(A)(-28,0.2){}
\node(B)(-4.5,0.4){}
\drawbcedge(A,-23,10,B,-10,10){}
}

\gasset{Nframe=n,Nfill=n}

\node[Nframe=n](E)(-45,0){\Large$E_3$:}

\end{picture}
}

\vspace{0ex}

\scalebox{0.6}{
\begin{picture}(60,16)(-47,-9)
\gasset{Nh=4,Nw=4,Nadjustdist=1,AHangle=35,AHLength=1.0,AHlength=0.4,Nframe=n,Nfill=n,linewidth=0.08}
\unitlength=0.27em

\node(W)(1,0){$\textcolor{white}{\whitec{a} \lra \whitec{b} \lra \whitec{a} \lra} \text{\greenc{$b$}} \whitec{\lra
  \whitec{\oa} \lra \whitec{\oa} \lra} \redc{\oa} \lra \fbox{\greenc{$\ob$}}  \lra \greenc{\ob}
  \whitec{\lra \whitec{\ob}}~~~~~~~~$}

\node[Nframe=n](A)(-16,-0.9){}
\node[Nframe=n](B)(26,-1.8){}
\drawbcedge(A,-8,-13,B,20,-13){}

\gasset{Nframe=n,Nfill=n}

\node[Nframe=n](E)(-45,0){\Large$E_4$:}

\end{picture}
}

\vspace{1ex}

\end{minipage}
}
\whitec{---}$\State'$:~
\fbox{
\begin{minipage}[t]{0.34\textwidth}
~\\
{
\scalebox{0.6}{
\begin{picture}(60,18)(-47,-9)
\gasset{Nh=4,Nw=4,Nadjustdist=1,AHangle=35,AHLength=1.0,AHlength=0.4,Nframe=n,Nfill=n,linewidth=0.08}
\unitlength=0.27em

\node(W)(6,0){$\whitec{a \lra b \lra} a \lra \greenc{\fbox{$b$}} \lra
  {\oa} \whitec{\lra \oa \lra \redc{\oa} \lra} \ob \whitec{\lra \greenc{\ob}
    \lra \ob}~~~~~~~~$}

\put(6,0){
\node(A)(-26,-0.6){}
\node(B)(-3.5,1){}
\drawbcedge(A,-21,10,B,-9,10){}
}

\put(6,0){
  \node[Nframe=n](A)(-16,-0.9){}
  \node[Nframe=n](B)(25,-0.5){}
  \drawbcedge(A,-8,-13,B,20,-13){}
}

\node[Nframe=y,Nh=3.8,Nw=3.8](E)(-123,0){}
\node[Nframe=y,Nh=4,Nw=4](E)(-123.3,-26.5){}
\node[Nframe=y,Nh=4,Nw=4](E)(-123,-42.5){}
\node[Nframe=y,Nh=4,Nw=4](E)(-122.8,-59){}

\node[Nframe=y,Nh=4,Nw=4](E)(3,0){}
\node[Nframe=y,Nh=3.8,Nw=3.8](E)(3,-26.6){}
\node[Nframe=y,Nh=4,Nw=4](E)(3.4,-42.5){}
\node[Nframe=y,Nh=4,Nw=4](E)(3,-61){}

\gasset{Nframe=n,Nfill=n}

\node[Nframe=n](E)(-45,0){\Large$E_1'$:}

\end{picture}
}
}

\vspace{4ex}

\scalebox{0.6}{
\begin{picture}(60,16)(-47,-9)
\gasset{Nh=4,Nw=4,Nadjustdist=1,AHangle=35,AHLength=1.0,AHlength=0.4,Nframe=n,Nfill=n,linewidth=0.08}
\unitlength=0.27em

\node(W)(8,0){$\whitec{a \lra b \lra} a \lra \greenc{b} \lra
  {\fbox{\greenc{$\oa$}}} \lra \oa \whitec{\lra \redc{\oa} \lra \ob} \whitec{\lra \greenc{\ob}
    \lra \ob}~~~~~~~~$}

\put(7,0){
\node(A)(-26,-0.6){}
\node(B)(-3.5,1.2){}
\drawbcedge(A,-21,10,B,-9,10){}
}

\gasset{Nframe=n,Nfill=n}

\node[Nframe=n](E)(-45,0){\Large$E_2'$:}

\end{picture}
}

\vspace{0ex}

\scalebox{0.6}{
\begin{picture}(60,16)(-47,-9)
\gasset{Nh=4,Nw=4,Nadjustdist=1,AHangle=35,AHLength=1.0,AHlength=0.4,Nframe=n,Nfill=n,linewidth=0.08}
\unitlength=0.27em

\node(W)(7,0){$\whitec{a \lra} b \lra \fbox{$a$} \lra \greenc{b} \lra
  {\greenc{\oa}} \whitec{\lra \oa \lra \redc{\oa} \lra \ob} \whitec{\lra \greenc{\ob}
  \lra \ob}~~~~~~~~$}

\put(8,0){
\node(A)(-27,0.2){}
\node(B)(-4.5,0.4){}
\drawbcedge(A,-23,10,B,-10,10){}
}

\gasset{Nframe=n,Nfill=n}

\node[Nframe=n](E)(-45,0){\Large$E_3'$:}

\end{picture}
}

\vspace{1ex}

\scalebox{0.6}{
\begin{picture}(60,16)(-47,-9)
\gasset{Nh=4,Nw=4,Nadjustdist=1,AHangle=35,AHLength=1.0,AHlength=0.4,Nframe=n,Nfill=n,linewidth=0.08}
\unitlength=0.27em

\node(W)(10,0){$a \whitec{\lra b \lra a \lra b \lra}
  {\oa} \lra \fbox{$\oa$} \lra \redc{\oa} \whitec{\lra \ob \lra \greenc{\ob}
    \lra \ob}~~~~~~~~$}

\put(-1,0){
\node(A)(-35,0.2){}
\node(B)(17,0.8){}
\drawbcedge(A,-25,12,B,7,12){}
}

\gasset{Nframe=n,Nfill=n}

\node[Nframe=n](E)(-45,0){\Large$E_4'$:}

\end{picture}
}
\end{minipage}
}
} 
\end{center}
\caption{A step of the sphere automaton\label{fig:sphereaut}}
\end{figure}

The idea of the construction of the generalized $\tNWA$ $\WA_\radius$ is the
following: A state $\State$ of $\WA_\radius$ is a set of extended spheres,
which reflect the ``environment'' of a node that $\State$ is assigned to. Now
suppose that, in a run of $\WA_\radius$ on a nested word $\tW =
([\tn],\tsuccord,\tmu,\tlambda)$, $\State$ is assigned to a position $i \in
[\tn]$ and contains $E=(N,\succord,\mu,\lambda,\gamma,\alpha,\inst)$. If the
run is accepting, this will mean that the environment of $i$ in $\tW$ looks
like the environment of $\alpha$ in $E$. In particular, $\State$ will contain
exactly one extended sphere $E=(N,\succord,\mu,\lambda,\gamma,\alpha,\inst)$
such that $\gamma$ and $\alpha$ coincide, meaning that $\Sph{\radius}{\tW}{i}
\isom (N,\succord,\mu,\lambda,\gamma)$. This is illustrated in
Figure~\ref{fig:sphereaut} depicting a nested word and a step of a run of the
sphere automaton for $r = 1$ on this word. States $\State$ and $\State'$ are
assigned to positions $4$ and $5$, respectively. Each state is a set of
extended spheres. For clarity, however, we will neglect colors in the example.
The sphere center is, as usual, depicted as a rectangle; the active node is
marked as a circle. Observe that each state contains precisely one extended
sphere in which the sphere center and the active node are identical. These are
$E_1 \in \State$, and, respectively, $E_2' \in \State'$. Indeed, $E_1$
corresponds to the $1$-sphere of the nested word around $4$, while $E_2'$
reflects the $1$-sphere around $5$.

Of course, $\WA_\radius$ has to locally guess the environment of a position.
But how can we ensure that a guess is correct? Obviously, we have to pass a
local guess to each neighboring position in $\tW$. So suppose again that a
state $\State$ containing $E=(N,\succord,\mu,\lambda,\gamma,\alpha,\inst)$ is
assigned to a node $i$ of $\tW$. As $\alpha$ shall correspond to $i$, we need
to ensure that $\lambda(\alpha) = \tlambda(i)$ (this will be taken care of by
item (2) in the definition of the transition relation below). Now suppose that
$\alpha$ has a $\succord$-successor $j \in N$, i.e., $\alpha \succrel j$.
Then, we have to guarantee that $i < \tn$. This is done by simply excluding
$\State$ from the set of final states (in Figure~\ref{fig:sphereaut}, neither
$\State$ nor $\State'$ are final states). Moreover, $j$ should correspond to
$i+1$, which is ensured by passing $E[j]$ to the state that will be assigned
to $i+1$ (see item (7); in Figure~\ref{fig:sphereaut}, $\State'$ must
therefore contain $E_1[j]$ where $j$ is the $\succrel$-successor of the active
node of $E_1$, and we actually have $E_1' \isom E_1[j]$). On the other hand,
if $i$ has a $\tsuccord$-successor, then $\alpha$ must have a
$\succord$-successor $j$ as well such that $E[j]$ belongs to the state that
will be assigned to $i+1$. Observe that this rule applies unless
$d_E(\gamma,\alpha) = \radius$, as then $i+1$ lies out of the area of
responsibility of $E$ (see item (5)). Similar requirements have to be
considered wrt.\ potential $\succord$-/$\tsuccord$-predecessors (see (3), (4),
and (6)), as well as wrt.\ the relations $\mu$ and $\tmu$ (see (3')--(7')).
One difficulty in our construction, however, is to guarantee the lack of an
edge. So assume the extended sphere $E$ is the one given by
Figure~\ref{fig:embedding} with $j_1$ as the active node. Let us neglect
colors for the moment. Suppose furthermore that $\tW$ is the nested word from
Figure~\ref{fig:embedding}, below the sphere. Then, an accepting run $\rho$ of
$\WA_\radius$ on $\tW$ will assign to $i_1$ a state that contains $E$ (modulo
some coloring). Moreover, the state assigned to $i$ will contain $E[j]$, where
the sphere center and the active node coincide. We observe that, in $E$, the
node $j_1$ is maximal. In particular, there is no $\mu$-edge between $j_1$ and
$j_2$. This should be reflected in $\tW$. A first idea to guarantee this might
be to just prevent $\rho(i_2)$ from containing the extended sphere $E[j_2]$
(note that $(i_1,i_2) \in \tmu$). This is, however, too restrictive. Actually,
$(\Sph{\radius}{\tW}{i},i_2)$ and $E[j_2]$ are isomorphic (neglecting the
coloring of $E$) so that $\rho(i_2)$ must contain $E[j_2]$. The solution is
already present in terms of the coloring of extended spheres. More precisely,
$\rho(i_2)$ is allowed to carry $E[j_2]$ as soon as it has a color that is
different from the color of the extended sphere $E[j_1]$ assigned to $i_1$.
Roughly speaking, there might be isomorphic spheres in $\tW$ that are
overlapping. To consider them simultaneously, they are thus equipped with
distinct colors.


The construction we obtain following the above ideas indeed allows us to
infer, from an accepting run assigning a state $\State$ to a node $i$, the
$\radius$-sphere around $i$. As mentioned above, we simply consider the
(unique up to isomorphism) extended sphere
$(N,\succord,\mu,\lambda,\gamma,\alpha,\inst)$ contained in $\State$ such that
$\gamma = \alpha$. Then, $(N,\succord,\mu,\lambda,\gamma)$ is indeed the
sphere of interest (recall that, in Figure~\ref{fig:sphereaut}, these are
$E_1$ for $\State$ and $E_2'$ for $\State'$ if we ignore active nodes and
colors).

It is not obvious that the above ideas really do work, all the less as the
construction will apply to nested words over two stacks, but no longer to
nested words over more than two stacks. After all, the key argument will be
provided by Proposition~\ref{prop:infinite}, stating an important property of
nested words over two stacks. Intuitively, it states the following: Suppose
that, in a nested word, there is an acyclic path from a node $i$ to another
node $i'$, and suppose this path is of a certain type $w$ (recording the
labelings and edges seen in the path). Then, applying the same path several
times will never lead back to $i$. This is finally the reason why a cycle in
an extended sphere that occurs in a run on a nested word $\tW$ is in fact
simulated by $\tW$.

Let us formally construct the generalized $\tNWA$
$\WA_\radius=(Q,\delta,\Init,F,C)$. An element of $Q$ is a subset $\State$ of
$\AlleSpheres_\radius(\crSigma)$ such that either $\State = \emptyset$, which
will be the only initial state, or the following conditions are satisfied:
\begin{enumerate}[(a)]
\item there is a unique extended sphere
  $(N,\succord,\mu,\lambda,\gamma,\alpha,\inst) \in \State$ such that $\gamma
  = \alpha$\\(we set $\core(\State) := (N,\succord,\mu,\lambda,\gamma)$)
\item there is $a \in \Sigma$ such that, for every
  $(N,\succord,\mu,\lambda,\gamma,\alpha,\inst) \in \State$,
  $\lambda(\alpha)=a$\\(so that we can assign a unique label $a$ to $\State$,
  denoted by $\labeling(\State)$)
\item for every two elements
  $\esphere=(N,\succord,\mu,\lambda,\gamma,\alpha,\inst)$ and $\esphere' =
  (N',\succord',\mu',\lambda',\gamma',\alpha',\inst')$ from $\State$, if
  $\core(\esphere) = \core(\esphere')$ and $\inst = \inst'$, then $\alpha =
  \alpha'$
\end{enumerate}
So let us turn to the transition relation $\delta = \langle \delta_1,\delta_2
\rangle$:
\begin{enumerate}[$\bullet$]\itemsep=1ex
\item For $\State,\State' \in Q$ and $a \in \Sigma$, we let
  $(\State,a,\State') \in \delta_1$ if $\State' \neq \emptyset$ and the
  following hold:\vspace{0.8ex}
\begin{enumerate}[(1)]\itemsep=0.8ex
\item for all $(N,\succord,\mu,\lambda,\gamma,\alpha,\inst) \in \State'$,
  $\alpha \not\in \dom(\mu^{-1})$ (i.e., $\mu^{-1}(\alpha)$ is not defined)
\item $\labeling(\State') = a$
\item for all $E=(N,\succord,\mu,\lambda,\gamma,\alpha,\inst) \in \State$
  and $i \in N$,\vspace{0.5ex}\\
  ${\parbox{14.5em}{\hfill$E[i] \in \State'$}} ~~\Longrightarrow~~ (\alpha,i)
  \in \succord$
\item for all $E=(N,\succord,\mu,\lambda,\gamma,\alpha,\inst) \in
  \State'$,\vspace{0.5ex}\\
  ${\parbox{14.5em}{\hfill$\State \neq \emptyset \mathrel{\wedge} \neg\exists
      i: (i,\alpha) \in \succord$}} ~~\Longrightarrow~~
  d_E(\gamma,\alpha)=\radius$
\item for all $E=(N,\succord,\mu,\lambda,\gamma,\alpha,\inst) \in
  \State$,\vspace{0.5ex}\\
  ${\parbox{14.5em}{\hfill$\neg\exists i: (\alpha,i) \in \succord$}}
  ~~\Longrightarrow~~ d_E(\gamma,\alpha)=\radius$
\item for all $E=(N,\succord,\mu,\lambda,\gamma,\alpha,\inst) \in
  \State'$ and $i \in N$,\vspace{0.5ex}\\ ${\parbox{14.5em}{\hfill$(i,\alpha)
      \in \succord$}} ~~\Longrightarrow~~ E[i] \in \State$
\item for all $E=(N,\succord,\mu,\lambda,\gamma,\alpha,\inst) \in \State$
  and $i \in N$,\vspace{0.5ex}\\
  ${\parbox{14.5em}{\hfill$(\alpha,i) \in \succord$}} ~~\Longrightarrow~~ E[i]
  \in \State'$
\end{enumerate}
\item For $\cState,\State,\State' \in Q$ and $a \in \rSigma$, we let
  $(\cState,\State,a,\State') \in \delta_2$ if $\cState,\State,\State' \neq
  \emptyset$ and (2)--(7) as above hold as well as the
  following:\vspace{0.8ex}
  \begin{enumerate}[(1')]\itemsep=1ex
  \item[(3')] for all $E=(N,\succord,\mu,\lambda,\gamma,\alpha,\inst) \in
    \cState$
    and $i \in N$,\vspace{0.5ex}\\
    ${\parbox{14.5em}{\hfill$E[i] \in \State'$}} ~~\Longrightarrow~~
    (\alpha,i) \in \mathord{\mu}$
  \item[(4')] for all $E=(N,\succord,\mu,\lambda,\gamma,\alpha,\inst) \in \State'$,\vspace{0.5ex}\\
    ${\parbox{14.5em}{\hfill$\alpha \not\in \dom(\mu^{-1})$}}
    ~~\Longrightarrow~~ d_E(\gamma,\alpha)=\radius$
  \item[(5')] for all $E=(N,\succord,\mu,\lambda,\gamma,\alpha,\inst) \in \cState$,\vspace{0.5ex}\\
    ${\parbox{14.5em}{\hfill$\alpha \not\in \dom(\mu)$}} ~~\Longrightarrow~~
    d_E(\gamma,\alpha)=\radius$
  \item[(6')] for all $E=(N,\succord,\mu,\lambda,\gamma,\alpha,\inst) \in
    \State'$,\vspace{0.5ex}\\ ${\parbox{14.5em}{\hfill$\alpha \in
        \dom(\mu^{-1})$}} ~~\Longrightarrow~~ E[\call(\alpha)] \in \cState$
  \item[(7')] for all $E=(N,\succord,\mu,\lambda,\gamma,\alpha,\inst) \in
    \cState$,\vspace{0.5ex}\\ ${\parbox{14.5em}{\hfill$\alpha
        \in \dom(\mu)$}} ~~\Longrightarrow~~ E[\ret(\alpha)] \in
    \State'$\vspace{0.5ex}
\end{enumerate}
\end{enumerate}
As already mentioned, the only initial state of $\WA_\radius$ is the empty
set, i.e., $\Init = \{\emptyset\}$. Moreover, $\State \in Q$ is a final state
if, for every extended sphere $(N,\succord,\mu,\lambda,\gamma,\alpha,\inst)
\in \State$, both $\alpha \not\in \dom(\mu)$ and there is no $i \in N$ such
that $(\alpha,i) \in \succord$. Finally, $\State$ is contained in $C$, the set
of calling states, if there is $(N,\succord,\mu,\lambda,\gamma,\alpha,\inst)
\in \State$ such that $\alpha \in \dom(\mu)$.

The mapping $\eta: Q \rightarrow \AllSpheres_\radius(\crSigma)$ as required in
Proposition~\ref{prop:main} is provided by $\core$. More precisely, we set
$\eta(\emptyset)$ to be some arbitrary sphere and $\eta(\State)=\core(\State)$
if $\State \neq \emptyset$.


Let us come back to the example in Figure~\ref{fig:sphereaut}, depicting two
states, $\State$ and $\State'$, of the sphere automaton for radius $r = 1$,
and a nested word that makes use of these states for being accepted. The
sphere automaton contains a transition $(\cState,\State,\oa,\State')$ for some
$\cState$.

We will verify in the following that conditions (2)--(7) are indeed satisfied.
The cases (3')--(7') as well as the construction of $\cState$ are left to the
reader.
\begin{enumerate}[(2)]
\item[(2)] All the active nodes in $\State'$ are labeled with $\oa$.
\item[(3)] Whenever a sphere from $\State$ is already present in $\State'$,
  then the corresponding active nodes are in the $\succrel$-relation. This
  applies to $E_1$ and $E_1'$ as well as to $E_2$ and $E_2'$.
\item[(4)] The extended sphere $E_4'$ is the only one in $\State'$ whose
  active node has no $\succrel$-predecessor. However, the distance between
  this active node and the sphere center equals $r=1$.
\item[(5)] There is one extended sphere in $\State$ without a
  $\succrel$-successor wrt.\ the active node, namely $E_4$. As required, the
  distance to the sphere center is $r=1$.
\item[(6)] There are three extended spheres in $\State'$ whose active nodes
  have a $\succrel$-predecessor: $E_1'$, $E_2'$, and $E_3'$. In fact, $\State$
  contains, in terms of $E_1$, $E_2$, and, respectively, $E_3$, all three
  extended spheres with the active node replaced by the respective
  $\succrel$-predecessor.
\item[(7)] Symmetrically to the case $(6)$, $E_1$, $E_2$, and $E_3$ from
  $\State$, where the active node is followed by a $\succrel$-successor, have
  their counterparts in $\State'$ in terms of $E_1'$, $E_2'$, and $E_3'$,
  respectively.
\end{enumerate}

\subsubsection{Every Nested Word Is Accepted}\label{subsect:anyaccepted}

Let $\tW = ([\tn],\tsuccord,\tmu,\tlambda)$ be an arbitrary nested word over
$\crSigma$. We show that $\tW \in \Lang(\WA_\radius)$. Let us first distribute
colors to each of the involved spheres. For this, we define the notion of an
overlap: for any $i,i' \in [\tn]$, $i$ and $i'$ are said to have an
$\radius$-\emph{overlap} in $\tW$ if $\Sph{\radius}{\tW}{i} \isom
\Sph{\radius}{\tW}{i'}$ and $d_\tW(i,i') \le 2r + 1$. For example, in
Figure~\ref{fig:embedding}, $i$ and $i'$ have a $2$-overlap.

\begin{myclaim}\label{cl:coloring}
  There is a mapping $\chi: [\tn] \rightarrow [\const]$ such that, for all
  $i,i' \in [\tn]$ with $i \neq i'$, the following holds: if $i$ and $i'$ have
  an $\radius$-overlap in $\tW$, then $\chi(i) \neq \chi(i')$.
\end{myclaim}

\begin{proof}
  The mapping is obtained as a graph coloring. Consider the graph
  $([\tn],\mathit{Arcs})$, $\mathit{Arcs} \mathrel{\subseteq} [\tn] \times
  [\tn]$, where, for $i,i' \in [\tn]$, we have $(i,i') \in \mathit{Arcs}$ iff
  $i \neq i'$ and $i$ and $i'$ have an $\radius$-overlap in $\tW$. Observe
  that $([\tn],\mathit{Arcs})$ cannot be of degree greater than $4 \cdot
  \maxN^2$. For each $i \in [\tn]$, there are at most four distinct events
  $i'$ such that $d_\tW(i,i') \le 1$. Now, if a position $j \in [\tn]$ wants
  to ``get in touch'' with $i$, it requires a position in its own sphere,
  another position in the sphere around $i$, and one of the four possibilities
  to relate these two positions. Hence, $([\tn],\mathit{Arcs})$ can be
  $\const$-\emph{colored} by a mapping $\chi: [\tn] \rightarrow [\const]$
  (i.e., $\chi(i) \neq \chi(i')$ for every $(i,i') \in \mathit{Arcs}$), which
  concludes the proof of Claim~\ref{cl:coloring}.
\end{proof}

We now specify $\rho: [\tn] \rightarrow Q$: for $i \in [\tn]$, we set $\rho(i)
= \{(\Sph{\radius}{\tW}{i'},i,\chi(i')) \mid i' \in [\tn]$ such that
$d_{\tW}(i,i') \le r\}$. With this definition, we can check that, for all $i
\in [\tn]$, $\rho(i)$ is a valid state of $\WA_\radius$, and that $\rho$ is
indeed an accepting run of $\WA_r$ on $\tW$.
%
%
So let $i \in [\tn]$ and let $E=(N,\succord,\mu,\lambda,\gamma,\alpha,\inst)$
and $E'=(N',\succord',\mu',\lambda',\gamma',\alpha',\inst')$ be contained in
$\rho(i)$.
\begin{enumerate}[(a)]
\item Assume that $\gamma = \alpha$ and $\gamma' = \alpha'$. Then,
  $(N,\succord,\mu,\lambda,\gamma,\gamma) \isom (\Sph{\radius}{\tW}{i},i)$ and
  $(N',\succord',\mu',\lambda',\gamma',\gamma') \isom
  (\Sph{\radius}{\tW}{i},i)$. Consequently, we have
  $(N,\succord,\mu,\lambda,\gamma,\gamma) \isom
  (N',\succord',\mu',\lambda',\gamma',\gamma')$. Moreover, $\col = \col' =
  \chi(i)$.
\item Of course, $\lambda(\alpha) = \lambda'(\alpha')$.
\item Assume $(N,\succord,\mu,\lambda,\gamma) \isom
  (N',\succord',\mu',\lambda',\gamma')$ and $\inst = \inst'$. There are
  $i_1,i_2 \in [\tn]$ with $d_\tW(i,i_1) \le \radius$, $d_\tW(i,i_2) \le
  \radius$, $(N,\succord,\mu,\lambda,\gamma,\alpha) \isom
  (\Sph{\radius}{\tW}{i_1},i)$, $(N,\succord,\mu,\lambda,\gamma,\alpha') \isom
  (\Sph{\radius}{\tW\!}{i_2},i)$, and $\inst = \chi(i_1) = \chi(i_2)$. Clearly,
  we have $\Sph{\radius}{\tW\!}{i_1} \isom \Sph{\radius}{\tW\!}{i_2}$.
  Furthermore, $i_1 = i_2$ and, therefore, $\alpha = \alpha'$. This is because
  $i_1$ and $i_2$ have an $\radius$-overlap in $\tW$ so that, according to
  Claim~\ref{cl:coloring}, $i_1 \neq i_2$ would imply $\chi(i_1) \neq
  \chi(i_2)$, which contradicts the premise.
\end{enumerate}


\noindent Now, for $i \in \{0,\ldots,\tn\}$ and $i' = i + 1$ with $i' \not\in
\dom(\tmu^{-1})$, we check that the triple $(\rho(i),\lambda(i'),\rho(i'))$ is
contained in $\delta_1$, where we let $\rho(0) = \emptyset$. Note first that,
of course, $\rho(i') \neq \emptyset$.

\begin{enumerate}[(1)]
\item Suppose $E=(N,\succord,\mu,\lambda,\gamma,\alpha,\inst) \in
  \rho(i')$. We have $E \isom (\Sph{\radius}{\tW}{i''},i',\chi(i''))$ for some
  $i'' \in [\tn]$ with $d_{\tW}(i',i'') \le r$. As $i' \not\in
  \dom(\tmu^{-1})$, we deduce $\alpha \not\in \dom(\mu^{-1})$.
\item Obviously, we have $\labeling(\rho(i')) = \tlambda(i')$.
\item Suppose $E=(N,\succord,\mu,\lambda,\gamma,\alpha,\inst) \in
  \rho(i)$ (we thus have $i \ge 1$) and $j \in N$ such that $E[j] \in
  \rho(i')$. Recall that we have to show that, then, $(\alpha,j) \in
  \succord$. There are $i_1,i_1' \in [\tn]$ such that $d_\tW(i_1,i) \le
  \radius$, $d_\tW(i_1',i') \le \radius$,
  $(N,\succord,\mu,\lambda,\gamma,\alpha) \isom (\Sph{\radius}{\tW}{i_1},i)$,
  $(N,\succord,\mu,\lambda,\gamma,j) \isom (\Sph{\radius}{\tW}{i_1'},i')$, and
  $\inst = \chi(i_1) = \chi(i_1')$. We easily see that $i_1$ and $i_1'$ have
  an $\radius$-overlap in $\tW$. We deduce, according to
  Claim~\ref{cl:coloring}, $i_1 = i_1'$. As, then,
  $(N,\succord,\mu,\lambda,\gamma,\alpha) \isom (\Sph{\radius}{\tW}{i_1},i)$,
  $(N,\succord,\mu,\lambda,\gamma,j) \isom (\Sph{\radius}{\tW}{i_1},i')$, and
  $(i,i') \in \tsuccord$, we can infer $(\alpha,j) \in \succord$.
\item Let $E=(N,\succord,\mu,\lambda,\gamma,\alpha,\inst) \in \rho(i')$,
  suppose $i' \ge 2$, and suppose that there is no $j \in N$ such that
  $(j,\alpha) \in \succord$. Recall that we have to show that
  $d_E(\gamma,\alpha)=\radius$. There is $i_1' \in [\tn]$ such that
  $d_\tW(i_1',i') \le \radius$ and $(N,\succord,\mu,\lambda,\gamma,\alpha)
  \isom (\Sph{\radius}{\tW}{i_1'},i')$. But if $d_E(\gamma,\alpha) < \radius$,
  then $d_\tW(i_1',i') < \radius$, and there must be a $\succord$-predecessor
  of $\alpha$, which is a contradiction. We therefore deduce that
  $d_E(\gamma,\alpha)=\radius$.
\item Let $E=(N,\succord,\mu,\lambda,\gamma,\alpha,\inst) \in \rho(i)$
  and suppose that there is no $j \in N$ such that $(\alpha,j) \in \succord$.
  Similarly to the case (4), we show that $d_E(\gamma,\alpha)=\radius$. In
  fact, there is $i_1 \in [\tn]$ such that $d_\tW(i_1,i) \le \radius$ and
  $(N,\succord,\mu,\lambda,\gamma,\alpha) \isom (\Sph{\radius}{\tW}{i_1},i)$.
  Again, if $d_E(\gamma,\alpha) < \radius$, then $d_\tW(i_1,i) < \radius$ so
  that there must be a $\succord$-successor of $\alpha$, which is a
  contradiction. We conclude that $d_E(\gamma,\alpha)=\radius$.
\item Let $E=(N,\succord,\mu,\lambda,\gamma,\alpha,\inst) \in \rho(i')$
  and $j \in N$ such that $(j,\alpha) \in \succord$. We show that, then, $E[j]
  \in \rho(i)$. There is $i_1' \in [\tn]$ such that $d_\tW(i_1',i') \le
  \radius$, $(N,\succord,\mu,\lambda,\gamma,\alpha) \isom
  (\Sph{\radius}{\tW}{i_1'},i')$, and $\inst = \chi(i_1')$. As $(j,\alpha) \in
  \succord$, $\alpha$ is not minimal so that we have $i \ge 1$. Since,
  furthermore, $d_E(\gamma,j) \le \radius$ implies $d_\tW(i_1',i) \le
  \radius$, and since we also have $(N,\succord,\mu,\lambda,\gamma,j) \isom
  (\Sph{\radius}{\tW}{i_1'},i)$ and $\inst = \chi(i_1')$, we deduce
  $E[j]=(N,\succord,\mu,\lambda,\gamma,j,\col) \in \rho(i)$.
\item Let $E=(N,\succord,\mu,\lambda,\gamma,\alpha,\inst) \in \rho(i)$
  and $j \in N$ such that $(\alpha,j) \in \succord$. We have to show that
  $E[j] \in \rho(i')$. There is $i_1 \in [\tn]$ such that $d_\tW(i_1,i) \le
  \radius$, $(N,\succord,\mu,\lambda,\gamma,\alpha) \isom
  (\Sph{\radius}{\tW}{i_1},i)$, and $\inst = \chi(i_1)$. Since $d_E(\gamma,j)
  \le \radius$ implies $d_\tW(i_1,i') \le \radius$, and since we have
  $(N,\succord,\mu,\lambda,\gamma,j) \isom (\Sph{\radius}{\tW}{i_1},i')$ and
  $\inst = \chi(i_1)$, we deduce $E[j]=(N,\succord,\mu,\lambda,\gamma,j,\col)
  \in \rho(i')$.
\end{enumerate}

\noindent Next, for $i_c,i,i' \in [\tn]$ with $i' = i + 1$ and
$(i_c,i') \in \tmu$, we check that the quadruple
$(\rho(i_c),\rho(i),\lambda(i'),\rho(i'))$ is contained in
$\delta_2$. Checking (2)--(7) proceeds as in the above cases. For
completeness, we present the cases (3')--(7'), which are shown
analogously.  First observe that, indeed, $\rho(i_c)$, $\rho(i)$, and
$\rho(i')$ are all nonempty.
\begin{enumerate}[(1')]
\item[(3')] Suppose $E=(N,\succord,\mu,\lambda,\gamma,\alpha,\inst) \in
  \rho(i_c)$ and $j \in N$ such that $E[j] \in \rho(i')$. We show that
  $(\alpha,j) \in \mu$. There are $i_1,i_1' \in [\tn]$ such that
  $d_\tW(i_1,i_c) \le \radius$, $d_\tW(i_1',i') \le \radius$,
  $(N,\succord,\mu,\lambda,\gamma,\alpha) \isom
  (\Sph{\radius}{\tW}{i_1},i_c)$, $(N,\succord,\mu,\lambda,\gamma,j) \isom
  (\Sph{\radius}{\tW}{i_1'},i')$, and $\inst = \chi(i_1) = \chi(i_1')$. Again,
  $i_1$ and $i_1'$ have an $\radius$-overlap in $\tW$. According to
  Claim~\ref{cl:coloring}, $i_1 = i_1'$. Then,
  $(N,\succord,\mu,\lambda,\gamma,\alpha) \isom
  (\Sph{\radius}{\tW}{i_1},i_c)$, $(N,\succord,\mu,\lambda,\gamma,j) \isom
  (\Sph{\radius}{\tW}{i_1},i')$, and $(i_c,i') \in \tmu$, so that we can
  deduce $(\alpha,j) \in \mu$.
\item[(4')] Let $E=(N,\succord,\mu,\lambda,\gamma,\alpha,\inst) \in \rho(i')$
  and suppose that there is no $j \in N$ such that $(j,\alpha) \in \mu$. We
  have to show that $d_E(\gamma,\alpha)=\radius$. There is $i_1' \in [\tn]$
  such that $d_\tW(i_1',i') \le \radius$ and
  $(N,\succord,\mu,\lambda,\gamma,\alpha) \isom
  (\Sph{\radius}{\tW}{i_1'},i')$. But if $d_E(\gamma,\alpha) < \radius$, then
  $d_\tW(i_1',i') < \radius$, so there must be a $\mu$-predecessor of
  $\alpha$, which is a contradiction. We deduce $d_E(\gamma,\alpha)=\radius$.
\item[(5')] Let $E=(N,\succord,\mu,\lambda,\gamma,\alpha,\inst) \in \rho(i_c)$
  and suppose that there is no $j \in N$ such that $(\alpha,j) \in \mu$. We
  show that, then, $d_E(\gamma,\alpha)=\radius$. There is $i_1 \in [\tn]$ such
  that $d_\tW(i_1,i_c) \le \radius$ and
  $(N,\succord,\mu,\lambda,\gamma,\alpha) \isom
  (\Sph{\radius}{\tW}{i_1},i_c)$. If $d_E(\gamma,\alpha) < \radius$, then
  $d_\tW(i_1,i_c) < \radius$ , so there must be a $\mu$-successor of $\alpha$,
  which is a contradiction. We conclude that $d_E(\gamma,\alpha)=\radius$.
\item[(6')] Let $E=(N,\succord,\mu,\lambda,\gamma,\alpha,\inst) \in \rho(i')$
  and $j \in N$ such that $(j,\alpha) \in \mu$. We show $E[j] \in \rho(i_c)$.
  There is $i_1' \in [\tn]$ such that $d_\tW(i_1',i') \le \radius$,
  $(N,\succord,\mu,\lambda,\gamma,\alpha) \isom
  (\Sph{\radius}{\tW}{i_1'},i')$, and $\inst = \chi(i_1')$. Due to
  $d_E(\gamma,j) \le \radius$, we also have $d_\tW(i_1',i_c) \le \radius$, and
  since $(N,\succord,\mu,\lambda,\gamma,j) \isom
  (\Sph{\radius}{\tW}{i_1'},i_c)$ and $\inst = \chi(i_1')$, we deduce $E[j]
  \in \rho(i_c)$.
\item[(7')] Let $E=(N,\succord,\mu,\lambda,\gamma,\alpha,\inst) \in \rho(i_c)$
  and $j \in N$ such that $(\alpha,j) \in \mu$. We have to show $E[j] \in
  \rho(i')$. There is $i_1 \in [\tn]$ such that $d_\tW(i_1,i_c) \le \radius$,
  $(N,\succord,\mu,\lambda,\gamma,\alpha) \isom
  (\Sph{\radius}{\tW}{i_1},i_c)$, and $\inst = \chi(i_1)$. From $d_E(\gamma,j)
  \le \radius$, it follows $d_\tW(i_1,i') \le \radius$. As, moreover,
  $(N,\succord,\mu,\lambda,\gamma,j) \isom (\Sph{\radius}{\tW}{i_1},i')$ and
  $\inst = \chi(i_1)$, we deduce $E[j]=(N,\succord,\mu,\lambda,\gamma,j,\col)
  \in \rho(i')$.
\end{enumerate}

\subsubsection{Every Run Keeps Track Of Spheres}\label{subsect:keepstrack}

We will now show that an accepting run reveals the sphere around any node.
This constitutes the more difficult part of the correctness proof.


We introduce some useful notation: By $\Dir$, we denote the set
$\{\rightarrow,\leftarrow,\rightone,\leftone,\righttwo,\lefttwo\}$ of
\emph{directions}. Now let $W = ([n],\succord,\mu,\lambda) \in \NW(\crSigma)$
be a nested word, $i,j \in [n]$, and let $w=e_1 \ldots e_m \in \Dir^\ast$
(where $e_k \in \Dir$ for all $k \in \{1,\ldots,m\}$).
We write $\Wleadstoeq{W}{w}{i}{j}$ if there are $i_0,i_1,\ldots,i_m \in [n]$
such that $i_0 = i$, $i_m = j$, and, for every $k \in \{0,\ldots,m-1\}$, one
of the following holds:
\begin{enumerate}[(a)]
\item $e_{k+1} = \mathord{\rightarrow}$ and $i_{k+1} = i_k + 1$
\item $e_{k+1} = \mathord{\leftarrow}$ and $i_{k+1} = i_k - 1$
\item $e_{k+1} = \mathord{\rightone}$ and $i_k \in \dom(\mu)$ and $\lambda(i_k) \in
  \cSigma^1$ and $i_{k+1} = \mu(i_k)$
\item $e_{k+1} = \mathord{\righttwo}$ and $i_k \in \dom(\mu)$ and $\lambda(i_k) \in
  \cSigma^2$ and $i_{k+1} = \mu(i_k)$
\item $e_{k+1} = \mathord{\leftone}$ and $i_k \in \dom(\mu^{-1})$ and
  $\lambda(i_k) \in \rSigma^1$, and $i_{k+1} = \mu^{-1}(i_k)$
\item $e_{k+1} = \mathord{\lefttwo}$ and $i_k \in \dom(\mu^{-1})$ and
  $\lambda(i_k) \in \rSigma^2$, and $i_{k+1} = \mu^{-1}(i_k)$
\end{enumerate}
Moreover, we write $\Wleadsto{W}{w}{i}{j}$ if there are pairwise distinct
$i_0,i_1,\ldots,i_{m-1} \in [n]$ and $i_m \in [n] \setminus
\{i_1,\ldots,i_{m-1}\}$ such that $i_0 = i$, $i_m = j$, and, for every $k \in
\{0,\ldots,m-1\}$, (a)--(f) as above hold.

We say that a string $w \in \Dir^+$ is \emph{circular} if
$\Wleadsto{W}{w}{i}{i}$ for some nested word $W \in \NW(\crSigma)$ and some
position $i$ of $W$. In other words, a circular string can produce a circle in
a nested word. For example, $\rightone ~\! \rightarrow ~\! \lefttwo ~\!
\rightarrow$ and $\rightone ~\! \rightarrow ~\! \righttwo ~\! \rightarrow ~\!
\leftone ~\! \rightarrow$ are circular (for an appropriate alphabet
$\crSigma$), whereas $\rightone ~\! \rightarrow ~\! \righttwo ~\! \rightarrow
~\! \leftone ~\! \leftarrow$ is not circular.

The following proposition is crucial for our project, and it fails when
considering nested words over more than two stacks.


\begin{prop}\label{prop:infinite}
  Let $w \in \Dir^+$ be circular. Then, for all $k \ge 2$, $w^k$ is not
  circular.
\end{prop}

\begin{figure}[h]
\begin{center}
  \scalebox{1}{
\begin{picture}(140,40)(-70,-17)
\gasset{Nh=4,Nw=4,Nadjustdist=1,AHangle=35,AHLength=1.0,AHlength=0.4,Nframe=n,Nfill=n,linewidth=0.08}
\unitlength=0.27em

\node(W)(0,0){$a \lra c \hspace{4em} a \lra c \hspace{2em} \overline{c} \lra  b \hspace{4em} \overline{c}
  \lra b \hspace{2em} \overline{b} \lra \overline{a} \hspace{4em} \overline{b}
  \lra \overline{a}$}

\node(A)(-38,-0.7){}
\node(B)(38,-0.2){}
\drawbcedge(A,-30,15,B,30,15){}

\node(A)(-64,-0.7){}
\node(B)(64,-0.2){}
\drawbcedge(A,-55,25,B,55,25){}

\node(A)(-27.8,-0.5){}
\node(B)(-17,-0.5){}
\drawbcedge(A,-25.3,-7,B,-19.5,-7){}

\node(A)(17,-0.5){}
\node(B)(28,-0.5){}
\drawbcedge(A,18.5,-7,B,25.5,-7){}

\node(A)(-54,-0.5){}
\node(B)(9,-0.6){}
\drawbcedge(A,-49,-18,B,5,-18){}

\node(A)(-9,-0.5){}
\node(B)(54,-0.6){}
\drawbcedge(A,-4,-18,B,51,-18){}

\node(B)(-36.5,-5){$i$}

\gasset{Nframe=n,Nfill=n}

\end{picture}
}
\caption{Proposition~\ref{prop:infinite} fails when considering three
  stacks\label{fig:threestacks}}
\end{center}
\end{figure}
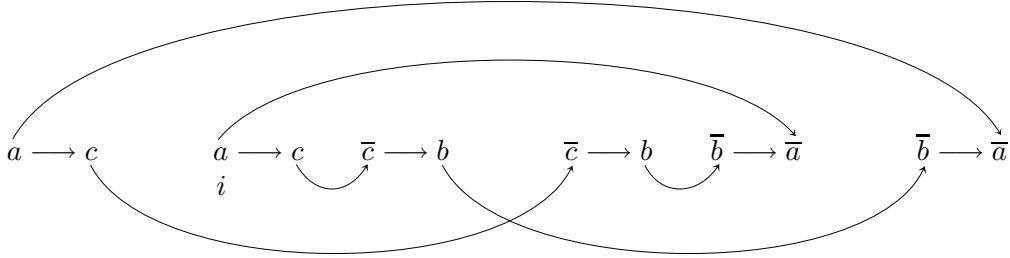

Before we prove Proposition~\ref{prop:infinite}, observe that it does not hold
as soon as a third stack comes into play. To see this, consider
Figure~\ref{fig:threestacks}, describing a part of a nested word $W$ over the
$3$-stack call-return alphabet $\langle
\{(\{a\},\{\overline{a}\}),(\{b\},\{\overline{b}\}),(\{c\},\{\overline{c}\})\},\emptyset
\rangle$. Suppose $w=~ \rightone ~\! \leftarrow ~\! \lefttwo ~\! \leftarrow
~\! \leftthree ~\! \leftarrow$ (where the meaning of $\leftthree$ is the
expected one), which is circular if we apply our definition to the framework
of three stacks. However, we have $\Wleadsto{W}{ww}{i}{i}$. It should be noted
that this does not imply that there is no sphere automaton or logical
characterization in the framework with more than two stacks. Indeed, we leave
as an open question if multiple stacks generally allow for a logical
characterization in terms of a fragment of MSO logic.

\begin{figure}[h]
\begin{center}
  \scalebox{1}{
\begin{picture}(140,40)(-70,-17)
\gasset{Nh=4,Nw=4,Nadjustdist=1,AHangle=35,AHLength=1.0,AHlength=0.4,Nframe=n,Nfill=n,linewidth=0.08}
\unitlength=0.27em

\node(W)(0,0){$a \lra b \hspace{4em} a \lra b \hspace{2em} \overline{b} \lra  a \hspace{4em} \overline{b}
  \lra a \hspace{2em} \overline{a} \lra \overline{a} \xrightarrow{\hspace{2.8em}} \overline{a}
  \lra \overline{a}$}

\node(A)(-38,-0.7){}
\node(B)(55,-0.2){}
\drawbcedge(A,-30,18,B,47,18){}

 \node(A)(-64.5,-0.7){}
 \node(B)(63.5,-0.2){}
 \drawbcedge(A,-55,25,B,67,25){}

 \node(A)(-28,-0.5){}
 \node(B)(-18,-0.5){}
 \drawbcedge(A,-25,-7,B,-21,-7){}

 \node(A)(17,-0.7){}
 \node(B)(28,-0.2){}
 \drawbcedge(A,19,7,B,25,7){}

 \node(A)(-54.5,-0.5){}
 \node(B)(9,-0.6){}
 \drawbcedge(A,-49,-18,B,5,-18){}

 \node(A)(-9.5,-0.7){}
 \node(B)(38,-0.2){}
 \drawbcedge(A,-6,12,B,35,12){}

 \node(B)(-37.3,-5){$i$}
 \node(B)(37.5,-5){$j_2$}
 \node(B)(54,-5){$j_1$}

\gasset{Nframe=n,Nfill=n}

\end{picture}
}
\caption{Intermediate positions need to be pairwise distinct\label{fig:distinct}}
\end{center}
\end{figure}
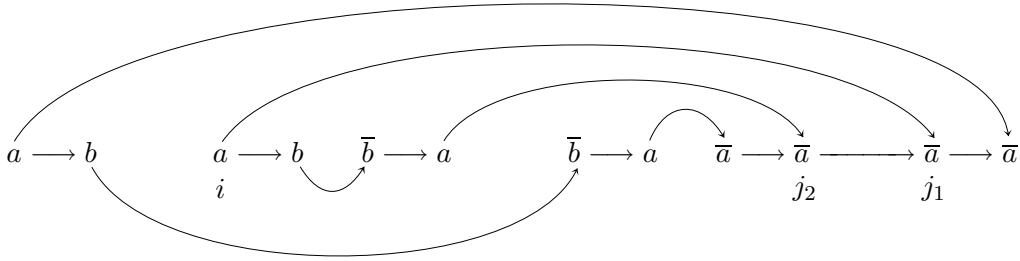

In the above definition of $\Wleadsto{W}{w}{i}{j}$, it is crucial to require
the elements $i_0,i_1,\ldots,i_{m-1} \in [n]$ to be pairwise distinct. This
can be seen considering a part of the nested word $W$ over the 2-stack
call-return alphabet $\langle
\{(\{a\},\{\overline{a}\}),(\{b\},\{\overline{b}\})\},\emptyset \rangle$ that
is depicted in Figure~\ref{fig:distinct}. Let $w=~\rightone ~\! \leftarrow ~\!
\leftarrow ~\! \leftone ~\! \leftarrow ~\! \lefttwo ~\! \leftarrow$, which is
a circular string. We have $\Wleadstoeq{W}{ww}{i}{i}$, i.e., starting from
$i$, we can follow the sequence of directions $w$ twice, arriving at $i$
again. However, apart from $i$, we have to visit $j_1$ and $j_2$ twice.
Indeed,
$i~\! \mathrel{{~~~\not\!\!\!\xhookrightarrow{ww~}}_{W}} i$.

\begin{Proof}{\it (of Proposition~\ref{prop:infinite}).} Let $W =
  ([n],\succord,\mu,\lambda) \in \NW(\crSigma)$, $w \in \Dir^+$, and $i \in
  [n]$. We have to show that, if $\Wleadsto{W}{w}{i}{i}$, then $w$ cannot be
  decomposed nontrivially into identical circular factors, i.e., there is no
  circular $u \in \Dir^+$ such that $w = u^k$ for some $k \ge
  2$.\footnote{Actually, one can even show that there is no $u \in \Dir^+$ at
    all (not even non-circular) such that $w = u^k$ for some $k \ge 2$.}

  To see this easily, we observe that a situation such as
  $\Wleadsto{W}{w}{i}{i}$ corresponds to a topological circle, as depicted in
  Figure~\ref{fig:topcircle}. A topological circle is a closed line in the
  two-dimensional plane that never crosses over itself. Let us construct
  topological circles according to the following procedure: We assume a
  straight (horizontal) line of the plane. Assume further a point $i$ on this
  line. Starting from $i$, we choose another two points as follows: Pick a
  symbol $\gamma$ from the alphabet
  $\{\rightright,\leftleft,\rightleft,\leftright\}$. According to this choice,
  we first draw a semicircle above the straight line ending somewhere on the
  line, and then, without interruption, a semicircle below the line, again
  resulting in a point on the line. Each semicircle is drawn in the direction
  indicated by $\gamma$, e.g., $\rightleft$ requires to draw the upper
  semicircle rightwards and the lower one leftwards, and $\rightright$
  requires both the upper and the lower semicircle to be drawn rightwards.
  This procedure is continued until we reach the original point $i$. We call a
  sequence from $\{\rightright,\leftleft,\rightleft,\leftright\}^+$ that
  allows us to draw a topological circle \emph{circular}. For example, in
  Figure~\ref{fig:topcircle}, we construct a topological circle by following
  the sequence $x=~\!\rightright~\! \rightright~\! \leftleft~ \rightleft~
  \leftleft~\! \leftleft$, starting in the left outermost point of
  intersection on the horizontal line. Thus, $x$ is circular, whereas
  $\leftright~\rightleft$ is not circular. Observe that we have $x \neq y^k$
  for all $y \in \{\rightright,\leftleft,\rightleft,\leftright\}^+$ and $k \ge
  2$.

\begin{figure}[h]
\begin{center}
\includegraphics[width=0.8\textwidth]{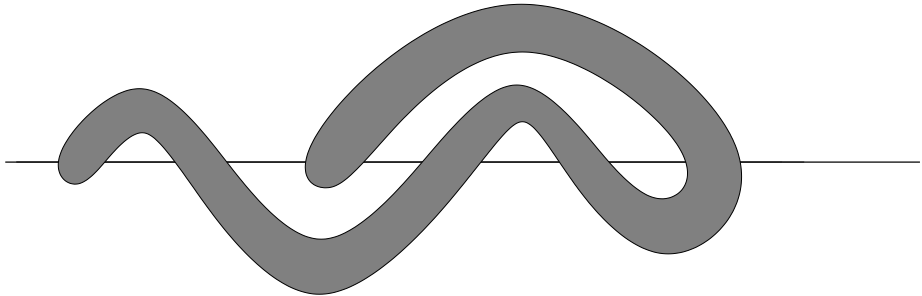}
\caption{Proof of Proposition~\ref{prop:infinite}\label{fig:topcircle}}
\end{center}
\end{figure}

  It is not hard to see that topological circles behave aperiodically in
  general, i.e., for any given $y \in
  \{\rightright,\leftleft,\rightleft,\leftright\}$, there is no $k \ge 2$ such
  that $y^k$ is circular. To show our proposition, we can even restrict to
  circular $y$. So let $y \in
  \{\rightright,\leftleft,\rightleft,\leftright\}^+$. But if $y$ is circular,
  then, for growing $k$, $y^k$ gives rise to a ``spiral'', and going back to
  the starting point would require to intersect the line that has been drawn
  hitherto.

  Let us relate our topological circles to the nested-word setting over two
  stacks. To this aim, we define a partial mapping $f: \Dir^+ \dashrightarrow
  \{\rightright,\leftleft,\rightleft,\leftright\}^+$ that associates with any
  circular string a sequence over
  $\{\rightright,\leftleft,\rightleft,\leftright\}$. This is done by reading a
  string from left to right and successively replacing every direction from
  $\Dir$ with a symbol from $\{\rightright,\leftleft,\rightleft,\leftright\}$,
  according to the following rules:
\begin{enumerate}[$\bullet$]
\item $\rightone$ is always replaced with $\rightright$

\item $\leftone$ is always replaced with $\leftleft$

\item $\rightarrow$ is replaced with
$\begin{cases}
  \leftright & \text{if the previous letter has been} \leftone\\
 \rightright & \text{otherwise}
\end{cases}$

\item $\leftarrow$ is replaced with
$\begin{cases}
\rightleft & \text{if the previous letter has been} \rightone\\
\leftleft & \text{otherwise}
\end{cases}$

\item $\righttwo$ is replaced with
$\begin{cases}
\leftright & \text{if the previous letter has been} \leftarrow\\
\rightright & \text{otherwise}
\end{cases}$

\item $\lefttwo$ is replaced with
$\begin{cases}
\rightleft & \text{if the previous letter has been} \rightarrow\\
\leftleft & \text{otherwise}
\end{cases}$
\end{enumerate}
For example, $f(\leftarrow~\! \leftone~\! \rightarrow~\! \righttwo~\!
\rightarrow~\! \leftone~\! \rightarrow~\! \lefttwo~\!) =~\! \leftleft~\!
\leftleft~ \leftright~ \rightright~\! \rightright~\! \leftleft~ \leftright~
\rightleft$\,. Let $w$ be circular. Clearly, $f(w)$ is circular as well, i.e.,
it allows us to draw a topological circle. We assume that the first letter of
$w$ stems from $\{\rightone,\leftone\}$. Other cases are either trivial or can
be reduced to that one. Then, if $w$ can be decomposed nontrivially into
identical circular factors, then this also applies to $f(w)$. Summarizing, the
power of a circular string is not circular anymore.



This concludes the proof of Proposition~\ref{prop:infinite}.
\end{Proof}

We will now show that, indeed, $\WA_\radius$ discovers the $\radius$-sphere
around any node of an input nested word.

Let $W = ([n],\succord,\mu,\lambda) \in \NW(\crSigma)$ be a nested word and
$\rho$ be a run of $\WA_\radius$ on $W$. Consider any $i \in [n]$, let
$(\Nat_i,\succord_i,\mu_i,\lambda_i,\gamma_i)$ refer to $\core(\rho(i))$, and
let $\inst_i$ be the unique element from $[\const]$ satisfying $E_i :=
(\Nat_i,\succord_i,\mu_i,\lambda_i,\gamma_i,\gamma_i,\inst_i) \in \rho(i)$.


 The following statement claims that an arbitrarily long path in $E_i$ is
 simulated by a corresponding path in $W$.

\begin{myclaim}\label{cl:simulate}
  Let $d \ge 0$ and suppose there are $j_0,\ldots,j_d \in \Nat_{i}$ such that
  $\gamma_{i} = j_0 \edge{E_i} j_1 \edge{E_i} \ldots \edge{E_i} j_d$. Then,
  there is a (unique) sequence of nodes $i_0,\ldots,i_d \in [n]$ such that
  \begin{enumerate}[$\bullet$]\itemsep=0.5ex
\item $i_0 = i$,
\item for each $k \in \{0,\ldots,d\}$, $E_i[j_k] \in \rho(i_k)$ (in particular,
  $\lambda(i_k) = \lambda_i(j_k)$), and
\item for each $k \in \{0,\ldots,d-1\}$,
  $\wsimulate{E_i}{W}{(j_k,j_{k+1})}{(i_k,i_{k+1})}$.
\end{enumerate}
\end{myclaim}

\begin{proof}
  The proof is by induction. Obviously, the statement holds for $d=0$. So
  assume $d \ge 0$ and suppose there are a sequence $j_0,\ldots,j_d,j_{d+1}
  \in \Nat_{i}$ such that $\gamma_{i} = j_0 \edge{E_i} j_1 \edge{E_i} \ldots
  \edge{E_i} j_d \edge{E_i} j_{d+1}$ and a unique sequence $i_0,i_1,\ldots,i_d
  \in [n]$ such that $i_0 = i$, $E_i[j_k] \in \rho(i)$ for each $k \in
  \{0,\ldots,d\}$, and $\wsimulate{E_i}{W}{(j_k,j_{k+1})}{(i_k,i_{k+1})}$ for
  each $k \in \{0,\ldots,d-1\}$. We consider four cases:
\begin{enumerate}[$\bullet$]
\item Assume $(j_d,j_{d+1}) \in \succord_i$. Then, $\rho(i_d)$ is not a final
  state so that $i_d < n$. We set $i_{d+1} = i_d + 1$. Due to (7), we have
  $E_i[j_{d+1}] \in \rho(i_{d+1})$.
\item Assume $(j_{d+1},j_d) \in \succord_i$. Then, according to (6), $i_d \ge
  2$. We set $i_{d+1} = i_d - 1$. Due to (6), we also have $E_i[j_{d+1}] \in
  \rho(i_{d+1})$.
\item Assume $(j_d,j_{d+1}) \in \mu_i$. Clearly, $\rho(i_d)$ is a calling
  state so that $\mu(i_d)$ is defined. Setting $i_{d+1} = \mu(i_d)$, we have,
  due to (7'), $E_i[j_{d+1}] \in \rho(i_{d+1})$.
\item Assume $(j_{d+1},j_d) \in \mu_i$. According to (1), $i_d \in
  \dom(\mu^{-1})$. With (6'), letting $i_{d+1} = \mu^{-1}(i_d)$, we have
  $E_i[j_{d+1}] \in \rho(i_{d+1})$.
\end{enumerate}
This concludes the proof of Claim~\ref{cl:simulate}.
\end{proof}


\begin{myclaim}\label{cl:forth}
  There is a homomorphism $h: \Sph{\radius}{W}{i} \rightarrow \core(\rho(i))$.
\end{myclaim}

\begin{proof}
  We show by induction the following statement:
  \vspace{1.5ex}\\
  \hspace*{\fill}
   \begin{minipage}{.8\linewidth}
     For every $d \in \{0,\ldots,\radius\}$, there is a homomorphism $h:
     \Sph{d}{W}{i} \rightarrow
     \Sph{d}{(\Nat_i,\succord_i,\mu_i,\lambda_i)}{\gamma_i}$ such that, for
     each $i' \in [n]$ with $d_W(i,i') \le d$, we have $E_i[h(i')] \in
     \rho(i')$.
   \end{minipage}
   \hspace{\fill}(*) \vspace{1.5ex}\\ Of course, (*) holds for $d=0$. So
   assume that (*) holds true for some natural number $d \in
   \{0,\ldots,\radius-1\}$, i.e., there is a homomorphism $h: \Sph{d}{W}{i}
   \rightarrow \Sph{d}{(\Nat_i,\succord_i,\mu_i,\lambda_i)}{\gamma_{i}}$ such
   that $E_i[h(i')] \in \rho(i')$ for each $i' \in [n]$ with $d_W(i,i') \le
   d$. We show that then (*) holds for $d+1$ as well. For this, let $i_1,i_2
   \in [n]$ such that $\dist_W(i,i_1) = d$ and $\dist_W(i,i_2) = d+1$.
\begin{enumerate}[$\bullet$]
\item Suppose $i_1 \succrel i_2$. Since $d_W(i,i_1) < r$, we also have
  $d_{E_i}(\gamma_i,h(i_1)) < r$. Due to (5), there is $j_2 \in \Nat_i$ such that
  $h(i_1) \succrel_i j_2$. Since $E_i[h(i_1)] \in \rho(i_1)$, we obtain, by (7)
  and (2), that $\lambda_i(j_2) = \lambda(i_2)$ and $E_i[j_2] \in \rho(i_2)$.
\item Similarly, we proceed if $i_2 \succrel i_1$. By
  $d_{E_i}(\gamma_i,h(i_1)) < r$ and (4), there is $j_2 \in \Nat_i$ such that
  $j_2 \succrel_i h(i_1)$. Since $E_i[h(i_1)] \in \rho(i_1)$, we obtain, by
  (6) and (2), that $\lambda_i(j_2) = \lambda(i_2)$ and $E_i[j_2] \in
  \rho(i_2)$.
\item If $(i_1,i_2) \in \mu$, then there exists, exploiting (5') and (7'),
  $j_2 \in \Nat_i$ such that $(h(i_1),j_2) \in \mu_i$, $\lambda_i(j_2) =
  \lambda(i_2)$, and $E_i[j_2] \in \rho(i_2)$.
 \item If $(i_2,i_1) \in \mu$, then we can find, due to (4') and (6'), $j_2
   \in \Nat_i$ such that $(j_2,h(i_1)) \in \mu_i$, $\lambda_i(j_2) =
   \lambda(i_2)$, and $E_i[j_2] \in \rho(i_2)$.
\end{enumerate}
Observe that $j_2$ is uniquely determined by $i_2$ and does not depend on the
choice of $i_1$ or on the relation between $i_1$ and $i_2$: If we obtained
distinct elements $j_2$ and $j_2'$, then the constraints $E_i[j_2] \in
\rho(i_2)$ and $E_i[j_2'] \in \rho(i_2)$ would imply that $\rho(i_2)$ is not a
valid state.

The above procedure extends the domain of the homomorphism $h$ by those
elements whose distance to $i$ is $d+1$. I.e., for $i_1,i_2 \in [n]$ with
$\dist_W(i,i_1)=\dist_W(i,i_2) = d+1$, we determined two \emph{unique}
elements $h(i_1),h(i_2) \in \Nat_i$, respectively. Let us show that
$\wsimulate{W}{\core(\rho(i))}{(i_1,i_2)}{(h(i_1),h(i_2)})$. Suppose $i_1
\succrel i_2$ (the case $i_2 \succrel i_1$ is symmetric). As $E_i[h(i_1)] \in
\rho(i_1)$ and $E_i[h(i_2)] \in \rho(i_2)$, we have, by (3), $h(i_1)
\succrel_i h(i_2)$. Similarly, with (3'), $(i_1,i_2) \in \mu$ implies
$(h(i_1),h(i_2)) \in \mu_i$.
\end{proof}
%

\begin{myclaim}\label{cl:back}
  There is a homomorphism $h': \core(\rho(i)) \rightarrow \Sph{\radius}{W}{i}$.
\end{myclaim}

\begin{proof}
  We show, again by induction, the following statement:
  \vspace{1.5ex}\\
  \hspace*{\fill}
   \begin{minipage}{.8\linewidth}
     For every natural number $d \in \{0,\ldots,\radius\}$, there is a
     homomorphism $h': \Sph{d}{(\Nat_i,\succord_i,\mu_i,\lambda_i)}{\gamma_i}
     \rightarrow \Sph{d}{W}{i}$ such that, for every $j \in \Nat_i$ with
     $d_{E_i}(\gamma_i,j) \le d$, we have $E_i[j] \in \rho(h'(j))$.
   \end{minipage}
   \hspace{\fill}(**) \vspace{1.5ex}\\ Clearly, (**) holds for $d=0$. Assume
   that (**) holds for some natural number $d \in \{0,\ldots,\radius-1\}$ and
   let $h': \Sph{d}{(\Nat_i,\succord_i,\mu_i,\lambda_i)}{\gamma_i} \rightarrow
   \Sph{d}{W}{i}$ be a corresponding homomorphism. Let $j_1,j_2 \in \Nat_i$
   such that $\dist_{E_i}(\gamma_i,j_1) = d$ and $\dist_{E_i}(\gamma_i,j_2) =
   d+1$.

   Suppose that $j_1 \succrel_i j_2$. As $E_i[j_1] \in \rho(h'(j_1))$,
   $\rho(h'(j_1))$ cannot be a final state of $\WA_\radius$ so that there is
   $i_2 \in [n]$ such that $h'(j_1) \succrel i_2$. Clearly, we have $E_i[j_2]
   \in \rho(i_2)$. Analogously, we proceed in the cases $j_2 \succrel_i j_1$,
   $(j_1,j_2) \in \mu_i$, and $(j_2,j_1) \in \mu_i$ to obtain such an element
   $i_2$. Note that $i_2$ is uniquely determined by $j_2$ and does not depend
   on the choice of $j_1$ or on the specific relation between $j_1$ and $j_2$.
   This is less obvious than the corresponding fact in the proof of
   Claim~\ref{cl:forth} but can be shown along the lines of the following
   procedure, proving that the extension of the domain of $h'$ by elements $j
   \in \Nat_i$ with $d_{E_i}(\gamma_i,j) = d+1$ is a homomorphism:


   We show that, for $j,j' \in N_i$ with $d_{E_i}(\gamma_i,j)
   =d_{E_i}(\gamma_i,j') = d+1$, we have
   $\wsimulate{E_i}{W}{(j,j')}{(h'(j),h'(j'))}$ (where the elements $h'(j)$
   and $h'(j')$ are obtained as indicated above). So suppose $j \edge{E_i}
   j'$. There are $\ell \in \{0,\ldots,d\}$ and pairwise distinct
   $j_0,\ldots,j_{2(d+1)-\ell} \in N_i$, such that
  \[\begin{array}{ccccccccccc} & & & & \!\!j_{\ell + 1} & \edge{E_i} & \ldots &
    \edge{E_i} & j_{d+1} & = & j\vspace{-1.2ex}\\ \gamma_i = j_0 \edge{E_i}
    & \ldots
    & \edge{E_i} j_\ell \!\!\!\!\! &
    \rotatebox{-45}{$\edge{E_i}$}\hspace{-1.5em}\rotatebox{45}{$\edge{E_i}$} &
    & & & & \rotatebox{-90}{\!\!\!\!\!\!\mbox{$\edge{E_i}$}} & &
    \vspace{-1.7ex}\\ & & & & \!\!j_{2(d+1)-\ell} & \edge{E_i} & \ldots & \edge{E_i}
    & j_{d+2} & = & j'
\end{array}\]
For ease of notation, set $D=2(d+1) - \ell$ and let, for $k \in \N$,
\[
\modit(k) = \left\{
    \begin{array}{cl}
      k & ~~ \text{~if~} k \le D\\
      ((k - \ell) \mod (D - \ell + 1)) + \ell & ~~ \text{~if~} k > D
    \end{array}
\right.
\]
I.e., the mapping $\modit$ counts until $D$ and afterwards modulo $D - \ell +
1$. According to Claim~\ref{cl:simulate}, there is a unique infinite sequence
$i_0,i_1,\ldots \in [n]$ such that
  \begin{enumerate}[$\bullet$]\itemsep=0.5ex
\item $i_0 = i$,
\item for each $k \in \N$, $E_i[j_{\modit(k)}] \in \rho(i_k)$, and
\item for each $k \in \N$,
  $\wsimulate{E_i}{W}{(j_{\modit(k)},j_{\modit(k+1)})}{(i_k,i_{k+1})}$.
\end{enumerate}
In what follows, we show that $i_{D+1}=i_\ell$, which implies
$\wsimulate{E_i}{W}{(j_{d+1},j_{d+2})}{(i_{d+1},i_{d+2})}$ so that
$\wsimulate{E_i}{W}{(j_{d+1},j_{d+2})}{(\hinv(j_{d+1}),\hinv(j_{d+2}))}$.
There is a circular string $w=e_\ell \ldots e_D \in \Dir^+$ such that
\begin{enumerate}[$\bullet$]
\item $\Wleadstoeq{E_i}{w}{j_\ell}{j_\ell}$,
\item $\longWleadstoeq{E_i}{e_\ell \ldots e_{\ell+k-1}}{j_\ell}{j_{\ell+k}}$
  for each $k \in \{1,\ldots,D-\ell\}$, and
\item $\Wleadstoeq{W}{w^k}{i_\ell}{i_{\ell + k(D - \ell + 1)}}$ for each $k \ge
  1$.
\end{enumerate}
We can obtain such a $w$ by setting, for each $k \in \{\ell,\ldots,D\}$,
\[
e_k = \left\{
    \begin{array}{cl}
      \rightarrow & ~~ \text{~if~} j_k \succrel_i j_{\modit(k+1)}\\
      \leftarrow & ~~ \text{~if~} j_{\modit(k+1)} \succrel_i j_k\\
      \rightone & ~~ \text{~if~} \lambda_i(j_k) \in \cSigma^1 \text{~and~}
      (j_k,j_{\modit(k+1)}) \in \mu_i
      \text{~and~} j_k \not\succrel_i j_{\modit(k+1)}\\
      \leftone & ~~ \text{~if~} \lambda_i(j_k) \in \rSigma^1 \text{~and~}
      (j_{\modit(k+1)},j_k) \in \mu_i
      \text{~and~} j_{\modit(k+1)} \not\succrel_i j_k\\
      \righttwo & ~~ \text{~if~} \lambda_i(j_k) \in \cSigma^2 \text{~and~}
      (j_k,j_{\modit(k+1)}) \in \mu_i
      \text{~and~} j_k \not\succrel_i j_{\modit(k+1)}\\
      \lefttwo & ~~ \text{~if~} \lambda_i(j_k) \in \rSigma^2 \text{~and~}
      (j_{\modit(k+1)},j_k) \in \mu_i
      \text{~and~} j_{\modit(k+1)} \not\succrel_i j_k
    \end{array}
\right.
\]
As $[n]$ is a finite set\footnote{\label{ftn:arg}In the context of infinite
  nested words, this argument can be replaced with the fact that, starting in
  $i$, there is no infinite sequence of \emph{pairwise distinct} nodes that
  follows the infinite sequence of directions $w^\omega$, i.e., the infinite
  repetition of $w$ (see Section~\ref{sec:buechi}).}, there are $p,q \in \N$
such that $\ell \le p < q$ and $i_p = i_q$. We choose $p$ and $q$ such that
$i_\ell,\ldots,i_{q-1}$ are pairwise distinct. We have both
$E_i[j_{\modit(p)}] \in \rho(i_p)$ and $E_i[j_{\modit(q)}] \in \rho(i_p)$.
According to the definition of the set of states of $\WA_\radius$, this
implies $j_{\modit(p)} = j_{\modit(q)}$. Let us distinguish three cases:
\begin{enumerate}[\hbox to6 pt{\hfill}]
\item\noindent{\hskip-11 pt\bf Case 1:}\ $p = \ell$ and $q = \ell + k(D - \ell + 1)$ for some $k \ge 1$.
  Then, $\Wleadsto{W}{w^k}{i_\ell}{i_{\ell + k(D - \ell + 1)}}$ so that,
  according to Proposition~\ref{prop:infinite}, we have $k=1$ and $i_\ell =
  i_{D+1}$, and we are done.
\item\noindent{\hskip-11 pt\bf Case 2:}\ $p > \ell$ and $q = p + k(D - \ell + 1)$ for some $k \ge 1$.
  Setting $e = e_{\modit(p-1)}$, we have both $\Wleadsto{W}{e}{i_{p-1}}{i_p}$
  and $\Wleadsto{W}{e}{i_{q-1}}{i_p}$, which is a contradiction, as $i_{p-1}
  \neq i_{q-1}$.
\item\noindent{\hskip-11 pt\bf Case 3:}\ $p \ge \ell$ and $q \neq p + k(D - \ell + 1)$ for every $k \ge
  1$. But this implies $\modit(p) \neq \modit(q)$ and, as the
  $j_\ell,\ldots,j_D$ are pairwise distinct, $j_{\modit(p)} \neq
  j_{\modit(q)}$, a contradiction.
\end{enumerate}
This concludes the proof of Claim~\ref{cl:back}.
\end{proof}
So let $h: \Sph{\radius}{W}{i} \rightarrow \core(\rho(i))$ and $h':
\core(\rho(i)) \rightarrow \Sph{\radius}{W}{i}$ be the unique homomorphisms
that we obtain following the constructive proofs of Claims~\ref{cl:forth} and
\ref{cl:back}, respectively. It is now immediate that $h$ is injective,
$h^{-1} = h'$, and $h: \Sph{\radius}{W}{i} \rightarrow \core(\rho(i))$ is an
isomorphism.

Recall that $\eta: Q \rightarrow \AllSpheres_\radius(\crSigma)$ shall map the
empty set to an arbitrary sphere and a nonempty set $\State \in Q$ onto
$\core(\State)$. Indeed, we constructed a generalized $\tNWA$
$\WA_\radius=(Q,\delta,\Init,F,C)$ together with a mapping $\eta: Q
\rightarrow \AllSpheres_\radius(\crSigma)$ such that
\begin{enumerate}[$\bullet$]
\item $\Lang(\WA_\radius)$ is the set of all nested words over $\crSigma$
  (cf.\ Section~\ref{subsect:anyaccepted}), and
\item for every nested word $W \in \NW(\crSigma)$, every accepting run $\rho$
  of $\WA_\radius$ on $W$, and every node $i$ of $W$, we have $\eta(\rho(i))
  \isom \Sph{\radius}{W}{i}$ (cf.\ Section~\ref{subsect:keepstrack}).
\end{enumerate}
This shows Proposition~\ref{prop:main}.





\section{Grids and Monadic Second-Order Quantifier
  Alternation}\label{sec:grids}

In this section, we show that the monadic second-order quantifier-alternation
hierarchy over nested words is infinite. In other words, the more alternation
of second-order quantification we allow, the more expressive formulas become.
From this, we can finally deduce that 2-stack visibly pushdown automata cannot
be complemented in general. In the proof, we use results that have been gained
in the setting of grids. By means of first-order reductions from grids into
nested words, we can indeed transfer expressiveness results for grids to the
nested-word setting. Let us first state a general result from \cite{MST02},
starting with the formal definition of a strong first-order reduction:

\begin{defi}[\cite{MST02}, Definition 32]\label{def:reduction} Let
  $\mathcal{C}$ and $\mathcal{C}'$ be classes of structures over relational
  signatures $\tau$ and $\tau'$, respectively. A \emph{strong first-order
    reduction} from $\mathcal{C}$ to $\mathcal{C}'$ with rank $m \ge 1$ is an
  injective mapping $\Phi: \mathcal{C} \rightarrow \mathcal{C}'$ such that the
  following hold:
\begin{enumerate}[(1)]
\item For every $G \in \mathcal{C}$, the universe of the structure $\Phi(G)$
  is $\bigcup_{k \in \{1,\ldots,m\}} (\{k\} \times \dom(G))$, i.e., the
  disjoint union of $m$ copies of $\dom(G)$, where $\dom(G)$ shall denote the
  universe of $G$.
\item There exists some $\psi(x_1,\ldots,x_m) \in \FO(\tau')$ such that, for every
  structure $G \in \mathcal{C}$, every $u_1,\ldots,u_m \in \dom(G)$, and every
  $k_1,\ldots,k_m \in [m]$, $\Phi(G) \models \psi[(k_1,u_1),\ldots,(k_m,u_m)]$
  iff $((k_1,u_1),\ldots,(k_m,u_m)) = ((1,u_1),\ldots,(m,u_1))$. (The
  intuition is that $u \in
  \dom(G)$ is represented by a model $((1,u),\ldots,(m,u))$ of $\psi$.)
\item For every relation symbol $r'$ from $\tau'$, say with arity $l$, and every
  $\kappa: [l] \rightarrow [m]$, there is $\phi^{r'}_\kappa(x_1,\ldots,x_l)
  \in \FO(\tau)$ such that, for each $G \in \mathcal{C}$ and each
  $u_1,\ldots,u_l \in \dom(G)$, $G \models \phi^{r'}_\kappa[u_1,\ldots,u_l]
  \text{~~iff~~} \Phi(G) \models r'[(\kappa(1),u_1),\ldots,(\kappa(l),u_l)]$.
\item For every relation symbol $r $ from $\tau$, say with arity $l$, there is
  $\phi^{r}(x_1,\ldots,x_l) \in \FO(\tau')$ such that, for each $G \in
  \mathcal{C}$ and each $u_1,\ldots,u_l \in \dom(G)$, $G \models
  r[u_1,\ldots,u_l] \text{~~iff~~} \Phi(G) \models
  \phi^r[(1,u_1),\ldots,(1,u_l)]$.
\end{enumerate}
\end{defi}

Once we have a strong first-order reduction from $\mathcal{C}$ to
$\mathcal{C}'$, logical definability carries over from $\mathcal{C}$ to
$\mathcal{C}'$:

\begin{thm}[\cite{MST02}, Theorem 33]\label{thm:sfor} Let $\mathcal{C}$ and
  $\mathcal{C}'$ be classes of structures over relational signatures $\tau$
  and $\tau'$, respectively. Let $\Phi: \mathcal{C} \rightarrow \mathcal{C}'$
  be a strong first-order reduction such that $\Phi(\mathcal{C})$ is
  $\LSigma_1(\tau')$-definable relative to $\mathcal{C}'$. Then, for every
  $\mathcal{L} \subseteq \mathcal{C}$ and $k \ge 1$, $\mathcal{L}$ is
  $\LSigma_k(\tau)$-definable relative to $\mathcal{C}$ iff
  $\Phi(\mathcal{L})$ is $\LSigma_k(\tau')$-definable relative to
  $\mathcal{C}'$.
\end{thm}

We proceed as follows. We first recall the notion of the class of grids, of
which we know that the monadic second-order quantifier-alternation hierarchy
is infinite. Then, we give a strong first-order reduction from the class of
grids to the class of nested words over a simple $2$-stack visibly pushdown
alphabet so that we can deduce that the monadic second-order
quantifier-alternation hierarchy over nested words is infinite, too. Note that
we will add to ordinary grids some particular labeling in terms of $a$ and
$b$, which will simplify the upcoming constructions. It is, however, easy to
see that well-known results concerning ordinary grids extend to these extended
grids (cf.\ Theorem~\ref{thm:mqah} below).

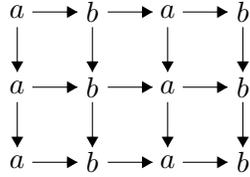
\begin{figure}[h]
\begin{center}
  \scalebox{1}{
    \begin{picture}(30,24)(0,-2)
      \gasset{Nw=4,Nh=4,AHangle=30,AHLength=1.6,Nframe=n,Nfill=n}
\node(A1)(0,20){$a$}\node(A2)(0,10){$a$}
\node(A3)(0,0){$a$}\node(B1)(10,20){$b$}
\node(B2)(10,10){$b$}\node(B3)(10,0){$b$}
\node(C1)(20,20){$a$}\node(C2)(20,10){$a$}
\node(C3)(20,0){$a$}\node(D1)(30,20){$b$}
\node(D2)(30,10){$b$}\node(D3)(30,0){$b$}

\drawedge(A1,A2){}\drawedge(A2,A3){}
\drawedge(B1,B2){}\drawedge(B2,B3){}
\drawedge(C1,C2){}\drawedge(C2,C3){}
\drawedge(D1,D2){}\drawedge(D2,D3){}
\drawedge(A1,B1){}\drawedge(B1,C1){}
\drawedge(C1,D1){}\drawedge(A2,B2){}
\drawedge(B2,C2){}\drawedge(C2,D2){}
\drawedge(A3,B3){}\drawedge(B3,C3){}
\drawedge(C3,D3){}
\end{picture}
}
\caption{The (3,4)-grid\label{fig:examplegrid}}
\end{center}
\end{figure}

We fix a signature $\signG=\{P_a,P_b,\succone,\succtwo\}$ with $P_a,P_b$ unary
and $\succone,\succtwo$ binary relation symbols. Let $n,m \ge 1$ be natural
numbers. The $(n,m)$-\emph{grid} is the $\signG$-structure $G(n,m) = ([n]
\times [m],\gdown,\gright,P_a,P_b)$ such that $\gdown = \{((i,j),(i+1,j)) \mid
i \in [n-1]$, $j \in [m]\}$, $\gright = \{((i,j),(i,j+1)) \mid i \in [n]$, $j
\in [m-1]\}$, $P_a = \{(i,j) \in [n] \times [m] \mid j$ is odd$\}$, and $P_b =
\{(i,j) \in [n] \times [m] \mid j$ is even$\}$. The $(3,4)$-grid is
illustrated in Figure~\ref{fig:examplegrid}. By $\Grids$, we denote the set of
all the grids.

\begin{thm}[\cite{MST02}]\label{thm:mqah} The monadic second-order
  quantifier-alternation hierarchy over grids is infinite. I.e., for every $k
  \ge 1$, there is a set of grids that is $\LSigma_{k+1}(\signG)$-definable
  relative to $\Grids$ but not $\LSigma_{k}(\signG)$-definable relative to
  $\Grids$.
\end{thm}

For the rest of this section, we suppose that $\crSigma$ is the $2$-stack
call-return alphabet given by $\cSigma^1=\{a\}$, $\rSigma^1=\{\overline{a}\}$,
$\cSigma^2=\{b\}$, $\rSigma^2=\{\overline{b}\}$, and $\intSigma = \emptyset$.
In particular, the following results assume all alphabets apart from
$\intSigma$ to be nonempty.

\newcommand{\pos}{\mathit{pos}}
\newcommand{\nodemap}{\chi_{n,m}}
\newcommand{\fullmap}{\overline{\chi}_{n,m}}

We now describe an encoding $\Phi: \Grids \rightarrow \NW(\crSigma)$ of grids
into nested words over $\crSigma$. Given $n,m \ge 1$, we let
\[\Phi(G(n,m)) := \left\{
 \begin{array}{ll}
   \nested\Bigl(a^n \bigl[(\overline{a} b)^n (\overline{b}
   a)^n\bigr]^{(m-1)/2~}\overline{a}^n\Bigr) & \text{~if~}  m
   \text{~is~odd}\vspace{0.5ex}\\
   \nested\Bigl(a^n \bigl[(\overline{a} b)^n (\overline{b} a)^n\bigr]^{m/2-1}
   (\overline{a} b)^n \overline{b}~\!\!^n\Bigr)  & \text{~if~}  m \text{~is~even}
 \end{array}
\right.
\]
The idea is that the first $n$ $a$'s (and, as explained below, the
corresponding return events) in a nested word represent the first column of
$G(n,m)$ seen from top to bottom; the first $n$ $b$'s represent the second
column, where the column is seen from bottom to top; the second $n$ $a$'s
stand for the third column, again considered from top to bottom, and so on.
The encoding $\Phi(G(3,4))$ of the (3,4)-grid as a nested word is depicted in
Figure~\ref{fig:encoding}. We claim that $\Phi$ is indeed a strong first-order
reduction from the set of grids to the set $\NW(\crSigma)$ of nested words
over $\crSigma$. Observe that $\Phi(G(n,m))$ does not have as domain the set
$\{1,2\} \times ([n] \times [m])$ as required in the definition of a strong
first-order reduction. However, below, we will introduce a bijection
$\fullmap: \{1,2\} \times ([n] \times [m]) \rightarrow [2 \cdot n \cdot m]$ to
identify every element in the domain of $\Phi(G(n,m))$ with some element in
$\{1,2\} \times ([n] \times [m])$.

\begin{figure}
\begin{center}
  \scalebox{0.69}{
\begin{picture}(150,50)(-75,-25)
\gasset{Nh=4,Nw=4,Nadjustdist=1,AHangle=35,AHLength=1.0,AHlength=0.4,Nframe=n,Nfill=n,linewidth=0.08}
\unitlength=0.27em

\newcommand{\ra}{\rightarrow}

\node(W)(0,0){$a \xrightarrow{~\hspace{2.4em}} a \xrightarrow{~\hspace{2.4em}}
  a \ra \overline{a} \ra b \ra \overline{a} \ra b \ra \overline{a} \ra b \ra
  \overline{b} \ra a \ra \overline{b} \ra a \ra \overline{b} \ra a \ra
  \overline{a} \ra b \ra \overline{a} \ra b \ra \overline{a} \ra b \ra
  \overline{b} \xrightarrow{~\hspace{2.4em}} \overline{b}
  \xrightarrow{~\hspace{2.4em}} \overline{b}$}

\node(A)(-26.5,-0.5){}
\node(B)(-18,-0.5){}
\drawbcedge(A,-24.5,-6,B,-20,-6){}

\node(A)(-42,-0.5){}
\node(B)(-3,-0.5){}
\drawbcedge(A,-37,-14,B,-8,-14){}

\node(A)(-56,-0.7){}
\node(B)(13,-0.7){}
\drawbcedge(A,-46,-25,B,3,-25){}

\node(A)(64,-0.5){}
\node(B)(73,-0.5){}
\drawbcedge(A,66,-6,B,71,-6){}

\node(A)(48.5,-0.5){}
\node(B)(88,-0.5){}
\drawbcedge(A,53.5,-14,B,83,-14){}

\node(A)(34,-0.7){}
\node(B)(104,-0.7){}
\drawbcedge(A,44,-25,B,94,-25){}

\put(-49,-2){
\node(A)(-23,1){}
\node(B)(-13,1.2){}
\drawbcedge(A,-21,6,B,-15,6){}

\node(A)(-39.5,0.5){}
\node(B)(1,1){}
\drawbcedge(A,-34.5,14,B,-4,14){}

\node(A)(-56,0.7){}
\node(B)(16,1){}
\drawbcedge(A,-46,25,B,6,25){}
}

\put(44,-2){
\node(A)(-26,1){}
\node(B)(-16,1.2){}
\drawbcedge(A,-24,6,B,-18,6){}

\node(A)(-40,0.5){}
\node(B)(-1,1.2){}
\drawbcedge(A,-35,14,B,-6,14){}

\node(A)(-56,0.7){}
\node(B)(14,1.2){}
\drawbcedge(A,-46,25,B,4,25){}
}

\gasset{Nframe=n,Nfill=n}

\end{picture}
}
\caption{The encoding $\Phi(G(3,4))$ of the (3,4)-grid as a nested word\label{fig:encoding}}
\end{center}
\end{figure}

\begin{prop}\label{prop:reduction}
  We have that $\Phi: \Grids \rightarrow \NW(\crSigma)$ is a strong
  first-order reduction with rank $2$. Moreover, $\Phi(\Grids)$ is
  $\LSigma_1(\signNW)$-definable relative to $\NW(\crSigma)$.
\end{prop}

\begin{proof}
  Let us first introduce a useful notation. Given a nested word
  $W=([n],\succord,\mu,\lambda)$ and $c \in \Sigma$ such that $W$ contains at
  least $k$ positions labeled with $c$, we let $\pos_c(W,k)$ denote the least
  position $i$ in $W$ such that $|\{j \in [i] \mid \lambda(j) = c\}| = k$
  (i.e., $\pos_c(W,k)$ denotes the position of the $k$-th $c$ in $W$).

  Let $n,m \ge 1$ and let $([2 \cdot n \cdot m],\succord,\mu,\lambda)$ refer
  to $\Phi(G(n,m))$. Recall that $\lambda$ can be seen as the collection of
  unary relations $\lambda_c = \{i \in [2 \cdot n \cdot m] \mid \lambda(i) =
  c\}$ for $c \in \Sigma$.
  Let us map any node in the $(n,m)$-grid (i.e., any element from $[n] \times
  [m]$) to a position of $\Phi(G(n,m))$ by defining a function $\nodemap: [n]
  \times [m] \rightarrow [2 \cdot n \cdot m]$ as follows:
\[
\nodemap(i,j) = \left\{
  \begin{array}{ll}
      \pos_a(\Phi(G(n,m)),n \cdot [(j+1)/2 - 1] + i) & ~~ \text{~if~} j \text{~is~odd}\\
      \pos_b(\Phi(G(n,m)),n \cdot [j/2 - 1] + (n + 1 - i)) & ~~ \text{~if~} j \text{~is~even}
    \end{array}
\right.
\]
for any $(i,j) \in [n] \times [m]$. Intuitively, $\nodemap(i,j) \in [2 \cdot n
\cdot m]$ represents the node $(i,j)$ in the $(n,m)$-grid. This mapping is
further extended towards a bijection $\fullmap: \{1,2\} \times ([n] \times
[m]) \rightarrow [2 \cdot n \cdot m]$ as required by
Definition~\ref{def:reduction} (item (1)). Namely, we map $\fullmap(1,(i,j))$
onto $\nodemap(i,j)$ and $\fullmap(2,(i,j))$ onto $\mu(\nodemap(i,j))$.

We are  prepared to specify the first-order formulas as supposed in
Definition~\ref{def:reduction}:
Let \[\psi(x_1,x_2) = \mu(x_1,x_2)~.\tag{2}\] Indeed, for every $n,m \ge 1$,
$k_1,k_2 \in \{1,2\}$, and $u_1,u_2 \in [n] \times [m]$, we have
\[\Phi(G(n,m)) \models \psi[\fullmap(k_1,u_1),\fullmap(k_2,u_2)]
\text{~iff~} ((k_1,u_1),(k_2,u_2)) = ((1,u_1),(2,u_1))~.\]
We will identify a map $\kappa: [l] \rightarrow \{1,2\}$ with the tuple
$(\kappa(1),\ldots,\kappa(l))$. Let, for $c \in \Sigma$ and $\kappa \in
\{1,2\}$,
\[
\phi^{\lambda_c}_\kappa(x) = \left\{
    \begin{array}{cl}
      P_c(x) & ~~ \text{~if~} c \in \{a,b\} \text{~and~} \kappa = 1\\
      P_{\overline{c}}(x) & ~~ \text{~if~} c \in \{\overline{a},\overline{b}\} \text{~and~} \kappa = 2\\
      \mathit{false} & ~~ \text{~otherwise}
    \end{array}
  \right.\tag{3}
\]
where we let $\overline{\overline{a}} = a$ and $\overline{\overline{b}} = b$.
For every $n,m \ge 1$, $\kappa \in \{1,2\}$, and $u\in [n] \times [m]$, we
have
\[G(n,m) \models \phi^{\lambda_c}_\kappa(x)[u] \text{~iff~} \Phi(G(n,m))
\models (\lambda(x) = c)[\fullmap(\kappa,u)]~.\]
Further, let, for $\kappa \in \{1,2\} \times \{1,2\}$,
\[
\phi^{\succord}_\kappa(x_1,x_2) = \left\{
    \begin{array}{cl}
      \succone(x_1,x_2) \mathrel{\wedge} \neg \exists z~ \succtwo(z,x_1) & ~~ \text{~if~} \kappa = (1,1)\\
      \left(\begin{array}{rl} & P_a(x_1) \mathrel{\wedge} \succone(x_2,x_1) \mathrel{\wedge} \neg
          \exists z~ \succtwo(x_1,z)\\
          \vee & P_b(x_1) \mathrel{\wedge} \succone(x_1,x_2) \mathrel{\wedge} \neg
          \exists z~ \succtwo(x_1,z)
        \end{array}\right) & ~~ \text{~if~} \kappa = (2,2)\\
      \left(\begin{array}{rl} & (x_1 = x_2) \mathrel{\wedge} P_a(x_1) 
          \mathrel{\wedge} \neg \exists z~ \succone(x_1,z)\\
          \vee & (x_1 = x_2) \mathrel{\wedge} P_b(x_1)
          \mathrel{\wedge} \neg \exists z~ \succone(z,x_1)\\
          \vee & P_a(x_1) \mathrel{\wedge} P_b(x_2)
          \mathrel{\wedge} \exists z~ (\succone(z,x_1) \mathrel{\wedge}  \succtwo(z,x_2))\\
          \vee & P_b(x_1) \mathrel{\wedge} P_a(x_2)
          \mathrel{\wedge} \exists z~ (\succone(z,x_1) \mathrel{\wedge}  \succtwo(x_2,z))
        \end{array}\right) & ~~ \text{~if~} \kappa = (1,2)\\
      \left(\begin{array}{rl} & P_a(x_1) \mathrel{\wedge} P_b(x_2)
          \mathrel{\wedge} \succtwo(x_1,x_2)\\ \vee & P_b(x_1) \mathrel{\wedge} P_a(x_2)
          \mathrel{\wedge} \succtwo(x_1,x_2)\end{array}\right) & ~~ \text{~if~} \kappa = (2,1)
    \end{array}
  \right.
\]
For every $n,m \ge 1$, $\kappa \in \{1,2\} \times \{1,2\}$, and $u_1,u_2 \in
[n] \times [m]$, we have
\[G(n,m) \models \phi^{\succord}_\kappa(x_1,x_2)[u_1,u_2] \text{~iff~} \Phi(G(n,m))
\models (x_1 \succrel
x_2)[\fullmap(\kappa(1),u_1),\fullmap(\kappa(2),u_2)]~.\]
Finally, to complete step (3), let, for $\kappa \in \{1,2\} \times \{1,2\}$,
\[
\phi^{\mu}_\kappa(x_1,x_2) = \left\{
    \begin{array}{cl}
      x_1 = x_2 & ~~ \text{~if~} \kappa = (1,2)\\
      \mathit{false} & ~~
      \text{~otherwise}
    \end{array}
  \right.
\]
Then, for every $n,m \ge 1$, $\kappa \in \{1,2\} \times \{1,2\}$ and $u_1,u_2
\in [n] \times [m]$,
\[G(n,m) \models \phi^{\mu}_\kappa(x_1,x_2)[u_1,u_2] \text{~iff~} \Phi(G(n,m))
\models (\mu(x_1,x_2))[\fullmap(\kappa(1),u_1),\fullmap(\kappa(2),u_2)]~.\]
Let
\[\phi^{P_a}(x) = (\lambda(x) = a) \text{~~and~~} \phi^{P_b}(x) = (\lambda(x)
= b)~.\tag{4}\] Of course, we have, for each $n,m \ge 1$, $c \in \{a,b\}$, and
$u \in [n] \times [m]$, \[G(n,m) \models P_c(x)[u] \text{~iff~} \Phi(G(n,m))
\models (\phi^{P_c})[\fullmap(1,u)]~.\]
Let
\[\phi^{\succone}(x_1,x_2) = \left(\begin{array}{rl} & \lambda(x_1) = a
    \mathrel{\wedge} \lambda(x_2) = a
    \mathrel{\wedge} (x_1 \succrel x_2 \mathrel{\vee} \exists z~ (x_1 \succrel z
    \mathrel{\wedge} z \succrel x_2))\\ \vee & \lambda(x_1) = b
    \mathrel{\wedge} \lambda(x_2) = b
    \mathrel{\wedge} \exists z~ (x_2 \succrel z
    \mathrel{\wedge} z \succrel x_1) \end{array}\right)\]
and let furthermore
\[\phi^{\succtwo}(x_1,x_2) = \exists z~(\mu(x_1,z) \mathrel{\wedge} z \succrel
x_2)~.\] Then, for each $n,m \ge 1$, $u_1,u_2 \in [n] \times [m]$, and $k \in
\{1,2\}$, it holds
\[G(n,m) \models \succk(x_1,x_2)[u_1,u_2] \text{~iff~} \Phi(G(n,m)) \models
(\phi^{\succk})[\fullmap(1,u_1),\fullmap(1,u_2)]~.\] With the above formulas,
it is now immediate to verify that $\Phi$ is indeed a strong first-order
reduction.

Now observe that $\Phi(\Grids)$ is the ``conjunction'' of
\begin{enumerate}[$\bullet$]\itemsep=1.2ex
\item the regular expression
  $
  \Bigl(a^+ \bigl[(\overline{a} b)^+ (\overline{b}
  a)^+\bigr]^{\ast~}\overline{a}^+\Bigr) + \Bigl(a^+ \bigl[(\overline{a} b)^+
  (\overline{b} a)^+\bigr]^\ast (\overline{a} b)^+
  \overline{b}~\!\!^+\Bigr),$
\item the first-order formula $\forall x \exists y~\bigl(\mu(x,y)
  \mathrel{\vee} \mu(y,x)\bigr),~\text{and}$
\item the first-order property (written in shorthand)
\[\begin{array}{rrl} \forall x_1,x_2,y_1,y_2~ \Bigl( & & \lambda(x_1) =
  \lambda(x_2) \mathrel{\wedge} \mu(x_1,y_1) \mathrel{\wedge} \mu(x_2,y_2)\\
  & \rightarrow &
\begin{array}[t]{rl}
  &  \bigl(\lambda(x_1) = a \mathrel{\wedge} x_2 - x_1 = 1 ~\mathrel{\rightarrow}~ y_1 - y_2 \in \{1,2\} \bigr)\\
  \wedge &  \bigl(\lambda(y_1) = \overline{a} \mathrel{\wedge} y_1 - y_2 = 1
  ~\mathrel{\rightarrow}~ x_2 - x_1 \in \{1,2\} \bigr)\\
  \wedge &  \bigl(\lambda(y_1) = \overline{b} \mathrel{\wedge} y_1 - y_2 = 1
  ~\mathrel{\rightarrow}~ x_2 - x_1 = 2 \bigr)\\
  \wedge &  \bigl(x_2 - x_1 = 2 \mathrel{\wedge} \lambda(x_1 + 1) \neq \lambda(x_1)
  ~\mathrel{\rightarrow}~ y_1 - y_2 \in \{1,2\} \bigr)\\
  \wedge &  \bigl(y_1 - y_2 = 2 \mathrel{\wedge} \lambda(y_2 + 1) \neq \lambda(y_2)
  ~\mathrel{\rightarrow}~ x_2 - x_1 \in \{1,2\} \bigr)\Bigr)
\end{array}
\end{array}\]
\end{enumerate}
As the regular expression represents a $\LSigma_1(\signNW)$-definable
property, $\Phi(\Grids)$ is $\LSigma_1(\signNW)$-definable relative to
$\NW(\crSigma)$, which concludes the proof of
Proposition~\ref{prop:reduction}.
\end{proof}

Combining Theorem~\ref{thm:sfor}, Theorem~\ref{thm:mqah}, and
Proposition~\ref{prop:reduction}, we obtain the following:

\begin{thm}\label{thm:hierarchyNW}
  $\!$The monadic second-order quantifier-alternation hierarchy over nested words
  is infinite. I.e., for all $k \ge 1$, there is a set of nested words over
  $\crSigma$ (with $\crSigma$ as specified above) that is
  $\LSigma_{k+1}(\signNW)$-definable relative to $\NW(\crSigma)$ but not
  $\LSigma_{k}(\signNW)$-definable relative to $\NW(\crSigma)$.
\end{thm}
Recall that Theorem~\ref{thm:hierarchyNW} relies on a particularly simple
call-return alphabet and the presence of at least two stacks. Indeed, its
proof is based on the $2$-stack call-return alphabet $\crSigma$, which is
given by $\cSigma^1=\{a\}$, $\rSigma^1=\{\overline{a}\}$, $\cSigma^2=\{b\}$,
$\rSigma^2=\{\overline{b}\}$, and $\intSigma = \emptyset$.

Finally, Theorems~\ref{thm:equiv} and \ref{thm:hierarchyNW} give rise to the
following theorem:

\begin{thm}
  The class of nested-word languages that are recognized by \tVPA is, in
  general, not closed under complementation. More precisely, there is a set
  $\Lang$ of nested words over $\crSigma$ (with $\crSigma$ as specified above)
  such that the following hold:
\begin{enumerate}[\em(1)]
\item There is a $\tVPA$ $\PA$ over $\crSigma$ such that $\Lang(\PA) = \Lang$.
\item There is no $\tVPA$ $\PA$ over $\crSigma$ such that $\Lang(\PA) =
  \NW(\crSigma) \setminus \Lang$.
\end{enumerate}
\end{thm}

This implies that the deterministic model of a \tVPA (see
\cite{Madhusudan2007} for its formal definition) is strictly weaker than the
general model. This fact was, however, already shown in \cite{Madhusudan2007}:
Consider the language $L = \{(ab)^mc^nd^{m-n}x^ny^{m-n} \mid m \in \N$, $n \in
[m]\}$ and the 2-stack call-return alphabet $\crSigma$ given by
$\cSigma^1=\{a\}$, $\rSigma^1=\{c,d\}$, $\cSigma^2=\{b\}$,
$\rSigma^2=\{x,y\}$, and $\intSigma = \emptyset$. Then, $L$ is accepted by
some \tVPA over $\crSigma$ but not by any deterministic \tVPA over $\crSigma$.

\vfill\eject

\section{\Buchi Multi-Stack Visibly Pushdown Automata}\label{sec:buechi}

We now transfer some fundamental notions and results from the finite case into
the setting of infinite (nested) words.

\subsection{\Buchi Multi-Stack Visibly Pushdown Automata}

Let $\nstack \ge 1$, and let $\crSigma =
\langle\{(\cSigma^\stack,\rSigma^\stack)\}_{\stack \in
  [\nstack]},\intSigma\rangle$ be a $\nstack$-stack call-return alphabet.

\begin{defi}
  A \emph{\Buchi multi-stack visibly pushdown automaton} (\Buchi \MVPA) over
  $\crSigma$ is a tuple $\PA=(Q,\Gamma,\delta,\Init,F)$ whose components agree
  with those of an ordinary \MVPA, i.e., $Q$ is its finite set of
  \emph{states}, $\Init \subseteq Q$ is the set of \emph{initial states}, $F
  \subseteq Q$ is the set of \emph{final states}, $\Gamma$ is the finite
  \emph{stack alphabet} containing the special symbol $\bot$, and $\delta$ is
  a triple $\crdelta$ with $\cdelta \mathrel{\subseteq} Q \times \cSigma
  \times (\Gamma \setminus \{\bot\}) \times Q$, $\rdelta \mathrel{\subseteq} Q
  \times \rSigma \times \Gamma \times Q$, and $\intdelta \mathrel{\subseteq} Q
  \times \intSigma \times Q$.

  A \emph{\Buchi 2-stack visibly pushdown automaton} (\Buchi \tVPA) is a
  \Buchi $\MVPA$ that is defined over a 2-stack alphabet.
\end{defi}

Consider an infinite string $w = a_1 a_2 \ldots \in \Sigma^\omega$. A run of
the \Buchi \MVPA $\PA$ on $w$ is a sequence $\rho = (q_0,\sigma_0^1, \ldots,
\sigma_0^\nstack) (q_1,\sigma_1^1, \ldots, \sigma_1^\nstack) \ldots \in {(Q
  \times \Contents^{[K]})}^\omega$ (recall that $\Contents={(\Gamma \setminus
  \{\bot\})}^\ast \cdot \{\bot\}$) such that $q_0 \in \Init$, $\sigma_0^\stack
= \bot$ for every stack $\stack \in [\nstack]$, and {\bf [Push]}, {\bf [Pop]},
and {\bf [Internal]} as specified in the finite case hold for every $i \in
\infDom$. We call the run accepting if $\{q \mid q = q_i$ for infinitely many
$i \in \N\} \mathrel{\cap} F \neq \emptyset$. A string $w \in \Sigma^\omega$
is accepted by $\PA$ if there is an accepting run of $\PA$ on $w$. The such
defined (string) language of $\PA$ is denoted by $L^\omega(\PA)$.

For the infinite case, we can likewise establish a relational structure of
\emph{infinite} nested words:

\begin{defi}
  An \emph{infinite nested word} over $\crSigma$ is a structure
  $(\infDom,\succord,\mu,\lambda)$ where $\succord=\{(i,i+1) \mid i \in
  \infDom\}$, $\lambda: \infDom \rightarrow \Sigma$, and $\mu =
  \bigcup_{\stack \in [\nstack]} \mu^\stack \subseteq \infDom \times \infDom$
  where, for every $\stack \in [\nstack]$ and $(i,j) \in \infDom \times
  \infDom$, $(i,j) \in \mu^\stack$ iff $i < j$, $\lambda(i) \in
  \cSigma^\stack$, $\lambda(j) \in \rSigma^\stack$, and $\lambda(i+1) \ldots
  \lambda(j-1)$ is $\stack$-well formed.
\end{defi}

The set of infinite nested words over $\crSigma$ is denoted by
$\infNW(\crSigma)$.
Again, given infinite nested words $\nword=(\infDom,\succord,\mu,\lambda)$ and
$\nword'=(\infDom,\succord',\mu',\lambda')$, $\lambda=\lambda'$ implies
$\nword = \nword'$ so that we can represent $\nword$ as $\str(\nword) :=
\lambda(1) \lambda(2) \ldots \in \Sigma^\omega$.
Vice versa, given a string $w \in \Sigma^\omega$, there is exactly one
infinite nested word $\nword$ over $\crSigma$ such that $\str(\nword) = w$,
which we denote $\nested(w)$.


\begin{defi}
  A \emph{generalized \Buchi multi-stack nested-word automaton} (generalized
  \Buchi \MNWA) over $\crSigma$ is a tuple $\WA=(Q,\delta,\Init,F,C)$ where
  $Q$, $\delta$, $\Init$, $F$, and $C$ are as in a generalized \MNWA. Recall
  that, in particular, $\delta$ is a pair $\langle \delta_1,\delta_2 \rangle$
  with $\delta_1 \mathrel{\subseteq} Q \mathrel{\times} \Sigma
  \mathrel{\times} Q$ and $\delta_2 \mathrel{\subseteq} Q \mathrel{\times} Q
  \mathrel{\times} \rSigma \mathrel{\times} Q$.

  We call $\WA$ a \emph{generalized \Buchi 2-stack nested-word automaton}
  (generalized \Buchi \tNWA) if it is defined over a 2-stack alphabet.

  If $C=\emptyset$, then we call $\WA$ a \Buchi \MNWA (\Buchi \tNWA, if
  $K=2$).
\end{defi}

A run of $\WA$ on an infinite nested word
$\nword=(\infDom,\succord,\mu,\lambda) \in \infNW(\crSigma)$ is a mapping
$\rho: \infDom \rightarrow Q$ such that $(q,\lambda(1),\rho(1)) \in \delta_1$
for some $q \in \Init$, and, for all $i \ge 2$, we have
 \[\left\{
 \begin{array}{rll}
   (\rho(\mu^{-1}(i)),\rho(i-1),\lambda(i),\rho(i)) & \!\!\!\in \delta_2 &~~ \text{~if~}
   \lambda(i) \in \rSigma \text{~and~} \mu^{-1}(i) \text{~is~defined}\\
   (\rho(i-1),\lambda(i),\rho(i)) & \!\!\!\in \delta_1 &~~ \text{~otherwise}
 \end{array}
\right.
 \]
 The run $\rho$ is accepting if $\rho(i) \in F$ for infinitely many $i \in
 \infDom$ and, for all $i \in \infDom$ with $\rho(i) \in C$, both $\lambda(i)
 \in \cSigma$ and $\mu(i)$ is defined. The language of $\WA$, denoted by
 $\Lang^\omega(\WA)$, is the set of infinite nested words over $\crSigma$ that
 allow for an accepting run of $\WA$.

 As we still have a one-to-one correspondence between strings and nested
 words, we may let $\Lang^\omega(\PA)$ with $\PA$ a \Buchi \MVPA stand for the
 set $\{\nested(w) \mid w \in L^\omega(\PA)\}$.

 It is now straightforward to adapt Lemma~\ref{lem:generalized} and
 Lemma~\ref{lem:PAWA} to the infinite setting:

\begin{lem}\label{lem:Bgen}
  For every generalized \Buchi \MNWA $\WA$, there is a \Buchi \MNWA $\WA'$
  such that $\Lang^\omega(\WA') = \Lang^\omega(\WA)$.
\end{lem}

\begin{lem}\label{lem:BPAWA}
  Let $\Lang \subseteq \infNW(\crSigma)$. The following are equivalent:
\begin{enumerate}[\em(1)]
\item There is a \Buchi \MVPA $\PA$ such that $\Lang^\omega(\PA) = \Lang$.
\item There is a \Buchi \MNWA $\WA$ such that $\Lang^\omega(\WA) = \Lang$.
\end{enumerate}
\end{lem}

\subsection{\Buchi 2-Stack Visibly Pushdown Automata vs.\ Logic}

In this section, we will again restrict to two stacks. Unfortunately, EMSO
logic over infinite nested words turns out to be too weak to capture all the
behaviors of \Buchi \tVPA. Given that EMSO logic considers a successor
relation instead of an order relation, one cannot even express that one
particular action occurs infinitely often. To overcome this deficiency, one
can introduce a first-order quantifier $\exists^\infty x \phi$ meaning that
there are infinitely many positions $x$ to satisfy the property $\phi$
\cite{LSV:06:11}.

So let us fix a 2-stack call-return alphabet
$\crSigma=\langle\{(\cSigma^1,\rSigma^1),(\cSigma^2,\rSigma^2)\},\intSigma\rangle$
for the rest of the paper. We introduce the logic
$\MSO^\infty(\tau_{\crSigma})$, which is given by the following grammar:
\begin{align*}
  \phi ::= &~ \lambda(x)=a ~\mid~ \succ{x}{y} ~\mid~ \mu(x,y) ~\mid~ x = y ~\mid~ x \in X ~\mid~ \\
  &~ \neg \phi ~\mid~ \phi_1 \vee \phi_2 ~\mid~ \exists x \phi ~\mid~
  \exists^\infty x \phi ~\mid~ \exists X \phi
\end{align*}
where $a \in \Sigma$. The fragments $\EMSO^\infty(\signNW)$ and
$\FO^\infty(\signNW)$ are defined as one would expect.
The satisfaction relation is as usual concerning the familiar fragment
$\MSO(\signNW)$. Moreover, given a formula
$\phi(y,x_1,\ldots,x_m,X_1,\ldots,X_n) \in \MSO^\infty(\signNW)$, an infinite
nested word $W$, $(i_1,\ldots,i_m) \in (\infDom)^m$, and $(I_1,\ldots,I_n) \in
(2^{\infDom})^n$, we set $W \models (\exists^\infty y
\phi)[i_1,\ldots,i_m,I_1,\ldots,I_n]$ iff $W \models
\phi[i,i_1,\ldots,i_m,I_1,\ldots,I_n]$ for infinitely many $i \in \infDom$.
Given a sentence $\phi \in \MSO^\infty(\signNW)$, we denote by
$\Lang^\omega(\phi)$ the set of infinite nested words over $\crSigma$ that
satisfy $\phi$.


To establish a connection between the extended logic and our \Buchi automata
models, we have to provide an extension of Hanf's Theorem.


\begin{thm}[cf.\ \cite{LSV:06:11}]\label{thm:infhanf} Let $\phi \in
  \FO^\infty(\tau_{\crSigma})$ be a sentence. There is a positive Boolean
  combination $\psi$ of formulas of the form \[\exists^{=t}x\, \chi(x)
  \text{~~~and~~~} \exists^{>t}x\, \chi(x) \text{~~~and~~~}
  \exists^{<\infty}x\, \chi(x) \text{~~~and~~~} \exists^{=\infty}x\, \chi(x)\]
  where $t \in \N$ and $\chi(x) \in \FO(\tau_{\crSigma})$ is local such that,
  for every nested word $W \in \infNW(\crSigma)$, we have
  \[W \models \phi \text{~~~iff~~~} W \models \psi.\]
\end{thm}

Unfortunately, we do not know if $\psi$ can be computed effectively in this
extended setting.






We observe that the \tNWA $\WA_\radius$ constructed in the proof of
Proposition~\ref{prop:main} can be easily adapted to obtain its counterpart
for infinite nested words:

\begin{prop}\label{prop:mainBuchi}
  Let $\radius \in \N$ be any natural number. There are a generalized \Buchi
  $\tNWA$ $\WA_\radius^\omega=(Q,\delta,\Init,F,C)$ over $\crSigma$ and a
  mapping $\eta: Q \rightarrow \AllSpheres_\radius(\crSigma)$ such that
\begin{enumerate}[$\bullet$]
\item $\Lang^\omega(\WA_\radius^\omega) = \infNW(\crSigma)$ and
\item for every $W \in \infNW(\crSigma)$, every accepting run $\rho$ of
  $\WA_\radius^\omega$ on $W$, and every node $i \in \posN$ of $W$, we have
  $\eta(\rho(i)) \isom \Sph{\radius}{W}{i}$.
\end{enumerate}
\end{prop}

\begin{proof}
  First, note that Proposition~\ref{prop:infinite} and the crucial argument
  stated in the proof of Claim~\ref{cl:back} (see Footnote~\ref{ftn:arg}) hold
  for infinite nested words just as well. Now, we look at the generalized
  \tNWA $\WA_\radius=(Q,\delta,\Init,F,C)$ as constructed in the proof of
  Proposition~\ref{prop:main}. As the only purpose of the set $F$ of final
  states is to ensure progress in some states where progress is required in
  terms of spheres with a non-maximal active node, we can set
  $\WA_\radius^\omega$ to be $(Q,\delta,\Init,Q,C)$, and we are done.
\end{proof}

With this, we can easily extend Lemma~\ref{lem:counting} and determine a
\Buchi \tNWA to detect if a particular sphere occurs infinitely often in an
infinite nested word:

\begin{lem}\label{lem:infcounting}
  Let $\radius \in \N$ and let $S \in \Spheres_\radius(\crSigma)$. There is a
  generalized \Buchi $\tNWA$ $\WA$ over $\crSigma$ such that
  $\Lang^\omega(\WA) = \{W \in \NW^\omega(\crSigma) \mid$ there are infinitely
  many $i \in \infDom$ such that $\Sph{\radius}{W}{i} \isom S\}$.
\end{lem}

\begin{proof}
  We start from the generalized \Buchi \tNWA
  $\WA_\radius^\omega=(Q,\delta,\Init,Q,C)$ and the mapping $\eta: Q
  \rightarrow \AllSpheres_\radius(\crSigma)$ from
  Proposition~\ref{prop:mainBuchi}. To obtain $\WA$ as required in the
  proposition, we simply set the set of final states to be $\{q \in Q \mid
  \eta(q) \isom S\}$.
\end{proof}

\begin{thm}
  Let $\Lang \mathrel{\subseteq} \infNW(\crSigma)$ be a set of infinite nested
  words over the 2-stack call-return alphabet $\crSigma$. Then, the following
  are equivalent:
\begin{enumerate}[\em(1)]
\item There is a \Buchi \tVPA $\PA$ over $\crSigma$ such that
  $\Lang^\omega(\PA) = \Lang$.
\item There is a sentence $\phi \in \EMSO^\infty(\tau_{\crSigma})$ such that
  $\Lang^\omega(\phi) = \Lang$.
\end{enumerate}
\end{thm}

\begin{proof}
  To prove $(1) \rightarrow (2)$, one again uses standard methods. Basically,
  second-order variables $X_q$ for $q \in Q$ encode an assignment of states to
  positions in a nested word. Then, the first-order part of the formula
  expresses that this assignment is actually an accepting run. To take care of
  the acceptance condition, we add the disjunction of formulas
  $\exists^{=\infty} x~(x \in X_q)$ with $q$ a final state.

  For the direction $(2) \rightarrow (1)$, we make use of
  Lemmas~\ref{lem:Bgen}, \ref{lem:BPAWA}, \ref{lem:infcounting}, (a simple
  variation of) Lemma~\ref{lem:counting}, and the easy fact that the class of
  languages of infinite nested words that are recognized by generalized \Buchi
  \tNWA is closed under union and intersection. With this, the proof proceeds
  exactly as in the finite case.
\end{proof}


\section{Open Problems}\label{sec:openproblems}

We leave open if visibly pushdown automata still admit a logical
characterization in terms of EMSO logic once they are equipped with more than
two stacks.

We conjecture that every first-order definable set of nested words over two
stacks is recognized by some unambiguous \tVPA, i.e., by a \tVPA in which an
accepting run is unique. To achieve such an automaton, the coloring of spheres
as performed by $\WA_\radius$ by simply \emph{guessing} and subsequently
verifying it has to be done unambiguously.

We do not know if EMSO logic over nested words becomes more expressive if we
allow atomic formulas $x \mathrel{<} y$ with the obvious meaning. For this
logic, it is no longer possible to apply Hanf's theorem as the degree of the
resulting structures is not bounded anymore.

Our method might lead to logical characterizations for concurrent queue
systems, where several visibly pushdown automata communicate with each other
via channels \cite{TACAS08}. In this extended setting, we deal with both
multiple stacks and channels. A corresponding logic then has to provide an
additional matching predicate $\textup{msg}(x,y)$ to relate the sending and
reception of a message (see, for example, \cite{BolligJournal}). It remains to
identify channel architectures for which a logical characterization is
possible. Using results from \cite{TACAS08}, this might lead to partial
results concerning the decidability of corresponding satisfiability problems.

Finally, it might be worthwhile to study if our technique leads to a logical
characterization of \tVPA for more general 2-stack call-return alphabets as
introduced in \cite{Murano2007}.

\noindent {\bf Acknowledgment}~ We thank the anonymous referees for their
careful reading and many useful remarks.










\begin{thebibliography}{10}

\bibitem{AM2004}
R.~Alur and P.~Madhusudan.
\newblock Visibly pushdown languages.
\newblock In {\em Proceedings of the 36th Annual ACM Symposium on Theory of
  Computing (STOC 2004)}, pages 202--211. ACM Press, 2004.

\bibitem{AlurM06}
R.~Alur and P.~Madhusudan.
\newblock Adding nesting structure to words.
\newblock In {\em Proceedings of the 10th International Conference on
  Developments in Language Theory (DLT 2006)}, volume 4036 of {\em Lecture
  Notes in Computer Science}, pages 1--13. Springer, 2006.

\bibitem{ABH-dlt08}
M.~F. Atig, B.~Bollig, and P.~Habermehl.
\newblock Emptiness of multi-pushdown automata is \(2\){ETIME}-complete.
\newblock In {\em {P}roceedings of the 12th {I}nternational {C}onference on
  {D}evelopments in {L}anguage {T}heory (DLT 2008)}, volume 5257 of {\em
  Lecture Notes in Computer Science}, pages 121--133. Springer, 2008.

\bibitem{LSV:06:11}
B.~Bollig and D.~Kuske.
\newblock {M}uller message-passing automata and logics.
\newblock {\em Information and Computation}, 206(9-10):1084--1094, 2008.

\bibitem{BolligJournal}
B.~Bollig and M.~Leucker.
\newblock Message-passing automata are expressively equivalent to {EMSO} logic.
\newblock {\em Theoretical Computer Science}, 358(2-3):150--172, 2006.

\bibitem{multi96}
L.~Breveglieri, A.~Cherubini, C.~Citrini, and S.~{Crespi Reghizzi}.
\newblock Multi-push-down languages and grammars.
\newblock {\em International Journal of Foundations of Computer Science},
  7(3):253--292, 1996.

\bibitem{Buechi:60}
J.~B{\"u}chi.
\newblock Weak second order logic and finite automata.
\newblock {\em Z. Math. Logik Grundlag. Math.}, 5:66--62, 1960.

\bibitem{Murano2007}
D.~Carotenuto, A.~Murano, and A.~Peron.
\newblock 2-visibly pushdown automata.
\newblock In {\em Proceedings of the 11th International Conference on
  Developments in Language Theory (DLT 2007)}, volume 4588 of {\em Lecture
  Notes in Computer Science}, pages 132--144. Springer, 2007.

\bibitem{Droste2000}
M.~Droste, P.~Gastin, and D.~Kuske.
\newblock Asynchronous cellular automata for pomsets.
\newblock {\em Theoretical Computer Science}, 247(1-2):1--38, 2000.

\bibitem{Elgot1961}
C.~C. Elgot.
\newblock Decision problems of finite automata design and related arithmetics.
\newblock {\em Trans. Amer. Math. Soc.}, 98:21--52, 1961.

\bibitem{Hanf1965}
W.~Hanf.
\newblock Model-theoretic methods in the study of elementary logic.
\newblock In J.~W. Addison, L.~Henkin, and A.~Tarski, editors, {\em The Theory
  of Models}. North-Holland, Amsterdam, 1965.

\bibitem{Hopcroft2000}
J.~E. Hopcroft, R.~Motwani, and J.~D. Ullman.
\newblock {\em Introduction to Automata Theory, Languages and Computability}.
\newblock Addison-Wesley, 2000.

\bibitem{Madhusudan2007}
S.~{La Torre}, P.~Madhusudan, and G.~Parlato.
\newblock A robust class of context-sensitive languages.
\newblock In {\em Proceedings of the 22nd IEEE Symposium on Logic in Computer
  Science (LICS 2007)}, pages 161--170. IEEE Computer Society Press, 2007.

\bibitem{TACAS08}
S.~{La Torre}, P.~Madhusudan, and G.~Parlato.
\newblock Context-bounded analysis of concurrent queue systems.
\newblock In {\em Proceedings of the 14th International Conference on Tools and
  Algorithms for the Construction and Analysis of Systems (TACAS 2008)},
  Lecture Notes in Computer Science, pages 299--314. Springer, 2008.

\bibitem{Schwentick94}
C.~Lautemann, Th. Schwentick, and D.~Therien.
\newblock Logics for context-free languages.
\newblock In {\em {Proceedings of the 1994 Annual Conference of the European
  Association for Computer Science Logic (CSL 1994)}}, volume 933 of {\em
  Lecture Notes in Computer Science}, pages 205--216, 1995.

\bibitem{Libkin2004}
L.~Libkin.
\newblock {\em {E}lements of {F}inite {M}odel {T}heory}.
\newblock Springer, 2004.

\bibitem{MST02}
O.~Matz, N.~Schweikardt, and W.~Thomas.
\newblock The monadic quantifier alternation hierarchy over grids and graphs.
\newblock {\em Information and Computation}, 179(2):356--383, 2002.

\bibitem{ThoPOMIV96}
W.~Thomas.
\newblock Elements of an automata theory over partial orders.
\newblock In {\em Proceedings of Workshop on Partial Order Methods in
  Verification (POMIV 1996)}, volume~29 of {\em DIMACS}. AMS, 1996.

\bibitem{Tho-automata-ttp}
W.~Thomas.
\newblock Automata theory on trees and partial orders.
\newblock In {\em Proceedings of Theory and Practice of Software Development
  (TAPSOFT 1997), 7th International Joint Conference CAAP/FASE}, volume 1214 of
  {\em Lecture Notes in Computer Science}, pages 20--38. Springer, 1997.

\bibitem{Tho97handbook}
W.~Thomas.
\newblock Languages, automata and logic.
\newblock In A.~Salomaa and G.~Rozenberg, editors, {\em Handbook of Formal
  Languages}, volume 3, Beyond Words, pages 389--455. Springer, 1997.

\end{thebibliography}
\end{document}